\documentclass[aps,onecolumn,superscriptaddress,showpacs,floatfix,nofootinbib]{revtex4-2}
\usepackage{color}
\usepackage{xcolor}
\usepackage{subfiles}
\usepackage{amsmath}
\usepackage{amssymb}
\usepackage{amsthm}
\usepackage{caption}
\usepackage{subcaption}
\usepackage{mathrsfs}
\usepackage{graphicx}
\usepackage{epstopdf}
\usepackage{fancyhdr}
\usepackage{array}
\usepackage{physics}
\usepackage[all]{xy}
\usepackage{eufrak}
\usepackage{float} 
\usepackage{euscript}
\usepackage{enumerate}
\usepackage{slashed}
\usepackage{hyperref}
\usepackage{caption}
\usepackage{epstopdf} % COMPILE EPS (AND PDF) FIGURES USING PDFLATEX!!!

\usepackage{bm}
\usepackage{cancel}
\usepackage{subcaption}

\hypersetup{pdftex,colorlinks=true,linkcolor=blue,citecolor=blue,menucolor=black,urlcolor=blue,filecolor=blue}

\definecolor{mygreen}{rgb}{0.1, 0.6, 0.1}
\definecolor{jblue}{rgb}{0 0.4470 0.7410}

\begin{document}

\title{Uncertainty quantification of holographic transport and energy loss for \\ the hot and baryon-dense QGP}

\author{Musa R. Khan}
\email{mkhan97@uh.edu}
\affiliation{Department of Physics, University of Houston, Houston, TX 77204, USA}

\author{Ayrton Nascimento}
\email{ayrton@pos.if.ufrj.br}
\affiliation{Instituto de Física, Universidade Federal do Rio de Janeiro, 21941-909, Rio de Janeiro, RJ, Brazil}

\author{Yumu~Yang}
\email{yumuy2@illinois.edu}
\affiliation{Illinois Center for Advanced Studies of the Universe, Department of Physics,
University of Illinois Urbana-Champaign, Urbana, IL 61801, USA}

\author{Joaquin~Grefa}
\email{vgrefaju@central.uh.edu}
\affiliation{Center for Nuclear Research, Department of Physics, Kent State University, Kent, OH 44243, USA}
\affiliation{Department of Physics, University of Houston, Houston, TX 77204, USA}

\author{Mauricio~Hippert}
\email{hippert@cbpf.br}
\affiliation{Centro Brasileiro de Pesquisas F\'isicas, Rua Dr. Xavier Sigaud 150, Rio de Janeiro, RJ, 22290-180, Brazil}

\author{Jorge~Noronha}
\email{jn0508@illinois.edu}
\affiliation{Illinois Center for Advanced Studies of the Universe, Department of Physics,
University of Illinois Urbana-Champaign, Urbana, IL 61801, USA}

\author{Claudia~Ratti}
\email{cratti@central.uh.edu}
\affiliation{Department of Physics, University of Houston, Houston, TX 77204, USA}

\author{Romulo~Rougemont}
\email{rougemont@ufg.br}
\affiliation{Instituto de F\'isica, Universidade Federal de Goi\'as, Av. Esperan\c{c}a - Campus Samambaia, CEP 74690-900, Goi\^{a}nia, Goi\'{a}s, Brazil}

\date{\today}

\begin{abstract}
We investigate several transport coefficients across the phase diagram of a holographic Einstein-Maxwell-Dilaton (EMD) model of hot and dense QCD with $N_f=2+1$ flavors. 
Our results are obtained from an open-source implementation of this model in C++, publicly available as a module within the MUSES Framework. 
This code includes a new numerical method to extract thermodynamic quantities from near-boundary asymptotics in holographic models, introduced here for the first time, 
which greatly improves numerical stability and performance in comparison to earlier implementations. 
Thanks to this improved technique, we are able to compute results for many realizations of our holographic model, sampled from a Bayesian posterior distribution constrained by lattice QCD results at zero chemical potential. 
This allows us to propagate lattice QCD error bars to predictions of transport coefficients in a wide window of temperature and baryon chemical potential, covering the crossover region, the neighborhood of the predicted critical point, and the line of first order phase transition. The physical observables include baryon and thermal conductivities, baryon diffusion, shear and bulk viscosities, the jet-quenching parameter, the heavy-quark drag force, and Langevin diffusion coefficients.
At vanishing baryon density, we compare our results to estimates extracted by the JETSCAPE Collaboration from heavy-ion data, with which we find good agreement.
\end{abstract}

\maketitle
\tableofcontents
%%%%%%%%%%%%%%%%%%%%%%%%%%%%%%%%%
\section{Introduction}
\label{sec:intro}

Understanding the properties of strongly interacting matter across the QCD phase diagram is a central goal of contemporary nuclear physics \cite{Fukushima:2010bq,Lovato:2022vgq,MUSES:2023hyz}. 
Temperatures of $T\gtrsim 150$ MeV can be explored in relativistic heavy-ion collisions, where a tiny and short-lived quark-gluon plasma (QGP) is formed \cite{Heinz:2000bk,Gyulassy:2004zy,PHENIX:2004vcz,BRAHMS:2004adc,PHOBOS:2004zne,STAR:2005gfr,Muller:2012zq,Busza:2018rrf}. 
More than two decades of experimental and theoretical efforts have been dedicated to exploring and characterizing this system both at RHIC and at CERN, leading to substantial progress in understanding its properties and evolution, including precise constraints on collective flow and transport properties \cite{MUSES:2023hyz,Bernhard:2019bmu,JETSCAPE:2020shq,Nijs:2020ors,JETSCAPE:2020mzn}. 
Among these advances, it has been established that the QGP found in these collisions exhibits an exceptionally low shear viscosity to entropy ratio, characteristic of strongly coupled systems \cite{Kovtun:2004de,Heinz:2013th,Luzum:2013yya}.

While collisions at top energies probe low values of the baryon chemical potential, there is increasing interest in investigating the QGP in the baryon-rich regime   
by exploring lower beam energies, of the order of a few to a few tens of  GeV per nucleon. 
A major goal of experimental programs such as the RHIC Beam Energy Scan and the CBM experiment planned at FAIR (GSI) is that of mapping the QCD phase diagram to determine whether and under which conditions the smooth crossover to the QGP becomes a discontinuous phase transition---namely, a first-order transition line starting from a second-order critical end point (CEP)~\cite{Almaalol:2022xwv,Lovato:2022vgq,Cebra:2014sxa,HADES:2019auv,Friese:2006dj,Tahir:2005zz,Lutz:2009ff,Durante:2019hzd,Kekelidze:2017tgp,Kekelidze:2016wkp}. 
Nonetheless, a realistic theoretical description of the baryon-dense QGP formed in these collisions still faces major challenges. 
First-principles lattice QCD (LQCD) simulations cannot yet reliably determine the baryon-dense equation of state \cite{Philipsen:2012nu,Ratti:2018ksb}, and  transport properties also remain poorly constrained \cite{MUSES:2023hyz}. 
In this context, one must rely on surrogate descriptions such as effective models of QCD, which should be constructed so as to reproduce current knowledge---for instance, the low-density equation of state and the nearly inviscid behavior of the QGP. 

An approach capable of capturing strongly coupled results and reproducing several features of QCD can be found in some holographic models, based on the gauge-gravity duality \cite{Maldacena:1997re,Gubser:1998bc,Witten:1998qj,Witten:1998zw}---see, for instance~\cite{Erdmenger:2007cm,Brodsky:2014yha,Gursoy:2010fj,DeWolfe:2013cua,Hoyos:2021uff,Rougemont:2023gfz}. 
The Einstein--Maxwell--Dilaton (EMD) class of models represents a particularly flexible bottom-up holographic approach, and can 
be tuned to reproduce LQCD results at vanishing baryon density~\cite{Gubser:2008ny,Gubser:2008yx,Gubser:2008sz,DeWolfe:2010he,DeWolfe:2011ts,Finazzo:2014cna,Rougemont:2015oea,Rougemont:2015wca,Rougemont:2015ona,Finazzo:2015xwa,Finazzo:2016mhm,Critelli:2016cvq,Knaute:2017opk,Rougemont:2017tlu,Critelli:2017oub,Rougemont:2018ivt,Rougemont:2020had,Grefa:2021qvt,Zollner:2021stb,Cai:2022omk,Grefa:2022sav,Chen:2022goa,Li:2023mpv,Hippert:2023bel,Zhao:2023gur,Chen:2024ckb,Fu:2024wkn,Chen:2024epd,Jokela:2024xgz,Chen:2024mmd,Cai:2024eqa,Cao:2024jgt,Zhu:2025gxo,Li:2025ugv,Chen:2025fpd,Shen:2025zkj,Zeng:2025tcz}. 
Recently, in Ref.~\cite{Hippert:2023bel}, this procedure was made more systematic by applying Bayesian inference tools to constrain the EMD model with $2+1$-flavors LQCD results for the entropy density and baryon susceptibility, both evaluated at zero baryon chemical potential. 
This Bayesian analysis showed that LQCD results at vanishing baryon density can place stringent constraints on features of the holographic model phase diagram and equation of state at higher baryon densities, revealing a strong preference for the existence of a critical point within two different functional realizations of the EMD model~\cite{Hippert:2023bel}. 
Moreover, results were shown to be in good quantitative agreement with state-of-the-art LQCD results at finite baryon density from the $T'$--expansion~\cite{Borsanyi:2021sxv}, even though they were not used to construct the model, lending further support to its phenomenological applicability \cite{Hippert:2023bel,Rougemont:2023gfz}.

Unlike other later analyses \cite{Zhu:2025gxo,Chen:2025fpd}, the Bayesian analysis of Ref.~\cite{Hippert:2023bel} employed fully numerical solutions to the EMD equations of motion, instead of relying on analytical solutions applicable to simpler realizations of the model. 
For this reason, improvements in numerical performance and stability were essential for carrying out the Bayesian sampling. 
In particular, that work benefited from 
a novel numerical strategy
which replaces least-squares fits to near-boundary data by a relaxational procedure to isolate the ultraviolet asymptotic coefficients used to extract thermodynamic observables. 
This approach leads to substantial gains in numerical stability and computational efficiency and was crucial for enabling the exploration of Bayesian posterior distributions. 
It has been implemented in a \texttt{C++} code developed within the MUSES Framework~\cite{yang_2025_14695243}, which was employed in~\cite{Hippert:2023bel}.

Here, we introduce this new method for the first time in the context of holographic models.
We also take advantage of the Bayesian posterior samples from \cite{Hippert:2023bel} to propagate LQCD uncertainties  to predictions for a broad set of transport coefficients throughout the phase diagram, and find them to be tightly constrained within the EMD model. 
This is achieved with a new version of the MUSES holographic module that incorporates the calculation of baryon and thermal conductivities, shear and bulk viscosities, the jet-quenching parameter, the heavy-quark drag force, and Langevin diffusion coefficients, which is described below and released publicly~\cite{yang_2025_14695243}. 
Our results for the conductivities and viscosities are directly relevant for heavy-ion collisions at Beam Energy Scan and FAIR energies, while results for heavy-quark transport and jet quenching---although not expected to play a central phenomenological role at such low beam energies---are presented for completeness.

Finally, we note that, during the development of the present work, two interesting papers~\cite{Li:2025ugv,Chen:2025fpd} appeared with which our manuscript has some overlap. In~\cite{Li:2025ugv}, the jet-quenching parameter and the energy loss of heavy and light quarks have
been investigated in the framework of holographic EMD models at finite temperature and baryon density, adjusted to LQCD data with $N_f = 2$, $2+1$, and $2+1+1$ flavors. They analyzed these observables in the crossover region, and also in the phase transition region comprising the CEP and the first order phase transition line of their EMD models, but without quantifying the statistical uncertainties in their predictions through Bayesian inference. In~\cite{Chen:2025fpd}, a Bayesian analysis of an EMD model with $N_f=2+1$ flavors was employed to investigate the heavy quark drag force, jet-quenching parameter,
heavy quark diffusion coefficient, and the bulk and shear viscosities strictly at zero chemical potential. On the other hand, in the present work, we investigate a broader set of physical observables, including the baryon susceptibility, baryon and thermal conductivities, baryon diffusion, shear and bulk viscosities, jet-quenching parameter, heavy quark drag force and Langevin diffusion coefficients across the phase diagram of the EMD model using Bayesian analysis with two different sets of functional forms for the free functions of the holographic setup, further asserting the robustness of the predictions provided here. Indeed, our approach provides the first in-depth analysis in the literature regarding the behavior of all these physical observables across the crossover and phase transition regions at finite baryon density with Bayesian uncertainty bands fully included.

The present manuscript is organized as follows. In section~\ref{sec:holo}, we review the basics of the holographic EMD setup and the Bayesian inference procedure, while also presenting our new numerical scheme for a more efficient, fast, and numerically stable extraction of the ultraviolet expansion coefficients required for the holographic computation of several physical observables. In section~\ref{sec:baryon}, we employ the aforementioned methods to compute the baryon susceptibility, the baryon and thermal conductivities, and the baryon diffusion coefficient at finite temperature and baryon density. In section~\ref{sec:bulk}, we compute the shear and bulk viscosities, while in section~\ref{sec:loss} we evaluate the heavy quark drag force, the Langevin diffusion coefficients, and the jet-quenching parameter. In section~\ref{sec:comparison}, as a way of assessing theoretical uncertainties, we compare results obtained within two different parametric forms, or ansatze, for the holographic potentials in our EMD model. We present our main conclusions in section~\ref{sec:conclusion}. In appendix~\ref{sec:app}, we further check the internal robustness of our holographic predictions for the QGP transport coefficients by performing a Bayesian analysis with different functional forms for the free functions of the EMD setup, as  was done for thermodynamic observables in~\cite{Hippert:2023bel}.
The numerical implementation developed in this work is released through the MUSES Collaboration and is publicly available at Ref.~\cite{yang_2025_14695243}.
The Bayesian posterior samples employed in the analysis are likewise publicly available in a separate repository \cite{Hippert2024_ZenodoDataset}.

Throughout this work we use natural units with $c=\hbar=k_B=1$ and the mostly plus metric signature.

%%%%%%%%%%%%%%%%%%%%%%%%%%%%%%%%%
%%%%%%%%%%%%%%%%%%%%%%%%%%%%%%%%%
\section{Holographic EMD setup at finite temperature and baryon chemical potential}
\label{sec:holo}

One of the earliest realizations of the holographic gauge-gravity duality (then known as AdS-CFT correspondence) corresponds to the dual relationship between pure gravity on (asymptotically) AdS$_5$ spacetimes and a supersymmetric and conformal field theory (CFT) called $\mathcal{N}=4$ Supersymmetric Yang-Mills (SYM) theory, living on the four-dimensional conformally flat boundary of the asymptotically AdS bulk geometry~\cite{Maldacena:1997re}. Since the QGP has no supersymmetry and is highly nonconformal around the hadronization/deconfinement crossover region, one needs to break supersymmetry and conformal symmetry in a rather specific way in order to quantitatively reproduce the behavior of QCD thermodynamics at finite temperature and baryon density using holographic gauge-gravity models. One of the most successful approaches in this regard was originally proposed in \cite{Gubser:2008ny,Gubser:2008yx,Gubser:2008sz,DeWolfe:2010he,DeWolfe:2011ts} by making use of bottom-up holographic models comprising a dilaton field with an adequately engineered potential. We refer the reader to the comprehensive review in Ref.~\cite{Rougemont:2023gfz} for refinements and extensions of such approach, applications for computing several observables relevant for characterizing the QGP, as well as a discussion of different formulations of other bottom-up holographic models that are used in the literature.

In the holographic approach used here, supersymmetry and conformal invariance are broken by introducing a real tachyonic scalar dilaton field $\phi$ in the gravity action for the bulk theory, whose boundary value sources a QFT operator triggering a relevant deformation of the renormalization group (RG) flow of the dual gauge theory at the boundary. Moreover, a baryon chemical potential at the boundary QFT is associated with the boundary value of a bulk Maxwell field $A^\mu$, which promotes the global $U(1)_B$ symmetry associated to baryon number conservation at the boundary QFT, to a local gauge symmetry in the bulk gravity theory. This allows one to consider charged black hole duals carrying a conserved charge, which we identity with baryon number.

The EMD action for the bulk gravity theory may be written as follows \cite{DeWolfe:2010he},
\begin{align}
S=\int_{\mathcal{M}_5} d^5x\,\mathcal{L} &= \frac{1}{2\kappa_5^2}\int_{\mathcal{M}_5} d^5x\,\sqrt{-g}\left[R-\frac{(\partial_\mu\phi)^2}{2} -V(\phi) -\frac{f(\phi)F_{\mu\nu}^2}{4}\right],
\label{eq:EMDaction}
\end{align}
where $\mathcal{M}_5$ is the five-dimensional bulk manifold, $g$ is the determinant of the metric tensor $g_{\mu\nu}$, $R$ is the Ricci curvature scalar, $F_{\mu\nu}=\partial_\mu A_\nu-\partial_\nu A_\mu$ is the Maxwell strength tensor, $\kappa_5^2\equiv 8\pi G_5$, where $G_5$ is the five-dimensional Newton's constant, and $V(\phi)$ is the dilaton potential. Since the dilaton is a real scalar field, there is no minimal coupling with the Maxwell field. Instead, the $F_{\mu\nu}^2$ term in the EMD action is scaled by the Maxwell-dilaton coupling function, $f(\phi)$, so that the RG flow is coupled to the baryon current without breaking $U(1)_B$. We remark that along with $\kappa_5^2$, $V(\phi)$, and $f(\phi)$, there is another free parameter in the EMD setup which is the asymptotic AdS radius, $L$. As in previous works, we set here $L=1$, and exchange it with an energy scale $\Lambda$ as a free parameter of the holographic model. In this way, the number of free parameters of the bottom-up setup is preserved, and $\Lambda$ is then interpreted as a characteristic energy scale of the non-conformal holographic model (playing a role analogous to the QCD dimensional transmutation scale, $\Lambda_\textrm{QCD}$). In practice, we use powers of $\Lambda$ to express dimensionful observables of the dual QFT at the boundary in powers of MeV. These observables are originally computed on the gravity side of the holographic gauge-gravity correspondence in powers of the inverse of the asymptotic AdS radius $L$ of the bulk geometry.

The general EMD field equations are obtained by extremizing the EMD bulk action \eqref{eq:EMDaction} with respect to the fields $g_{\mu\nu}$, $A^\mu$, and $\phi$, respectively,
\begin{align}
R_{\mu\nu}-\frac{g_{\mu\nu}}{3}\left[V(\phi)-\frac{f(\phi)}{4}F_{\alpha\beta}^2\right]-\frac{1}{2}\partial_\mu\phi\partial_\nu\phi-\frac{f(\phi)}{2}g^{\alpha\beta}F_{\mu\alpha}F_{\nu\beta}&=0,\label{eq:EinsteinEqs}\\
\partial_\mu\left(\sqrt{-g}f(\phi)g^{\mu\alpha}g^{\nu\beta}F_{\alpha\beta}\right)&=0,\label{eq:MaxwellEqs}\\
\frac{1}{\sqrt{-g}}\partial_\mu\left(\sqrt{-g}g^{\mu\nu}\partial_\nu\phi\right)-\frac{\partial V(\phi)}{\partial\phi}-\frac{F_{\mu\nu}^2}{4}\frac{\partial f(\phi)}{\partial\phi}&=0.\label{eq:DilatonEq}
\end{align}

In order to describe isotropic and homogeneous states at finite temperature and baryon charge density in thermodynamic equilibrium, we consider a charged black hole Ansatz for the EMD bulk fields depending solely on the holographic radial coordinate, $r$, which is the fifth dimension perpendicular to the spacetime directions parallel to the boundary. This fifth dimension corresponds to a geometrization of the RG flow of the dual QFT at the boundary, with ultraviolet information being encoded close to the boundary, while infrared information is encoded close to the black hole event horizon. Then, with a particular choice of the gauge fixing associated to diffeomorphism invariance, the Ansatz for the EMD fields can be written as follows~\cite{DeWolfe:2010he,Rougemont:2015wca,Critelli:2017oub,Grefa:2021qvt,Rougemont:2023gfz},
\begin{align}
ds^2 = g_{\mu\nu}dx^\mu dx^\nu = e^{2A(r)}[-h(r)dt^2+d\vec{x}^2]+\frac{dr^{2}}{h(r)}, \qquad
\phi = \phi(r), \qquad A_{\mu}=\Phi(r)\delta_\mu^0,
\label{eq:ansatz}
\end{align}
where the boundary lies at $r\to\infty$ and the radial location of the black hole horizon $r_H$ is given by the largest root of the blackening function, $h(r_H)=0$. By substituting the above Ansatz into the EMD field equations, one obtains,
\begin{align}
\phi''(r)+\left[\frac{h'(r)}{h(r)}+4A'(r)\right]\phi'(r)-\frac{1}{h(r)}\left[\frac{\partial V(\phi)}{\partial\phi}-\frac{e^{-2A(r)}\Phi'(r)^{2}}{2}\frac{\partial f(\phi)}{\partial\phi}\right]&=0,\label{eq:EoM1}\\
\Phi''(r)+\left[2A'(r)+\frac{d[\ln{f(\phi)}]}{d\phi}\phi'(r)\right]\Phi'(r)&=0,\label{eq:EoM2}\\
A''(r)+\frac{\phi'(r)^{2}}{6}&=0,\label{eq:EoM3}\\
h''(r)+4A'(r)h'(r)-e^{-2A(r)}f(\phi)\Phi'(r)^{2}&=0,\label{eq:EoM4}\\
h(r)[24A'(r)^{2}-\phi'(r)^{2}]+6A'(r)h'(r)+2V(\phi)+e^{-2A(r)}f(\phi)\Phi'(r)^{2}&=0,\label{eq:EoM5}
\end{align}
which is a set of coupled ordinary differential equations discussed in detail in~\cite{DeWolfe:2010he,Rougemont:2015wca,Critelli:2017oub,Grefa:2021qvt}. Below we briefly review some of the main points involved and, in subsection~\ref{sec:new}, we present a new and very efficient way of numerically solving these differential equations.

The set of dynamical equations of motion for the unknown radial functions $\phi(r)$, $\Phi(r)$, $A(r)$, and $h(r)$, given by Eqs.~\eqref{eq:EoM1} --- \eqref{eq:EoM4}, together with the constraint Eq.~\eqref{eq:EoM5}, are solved numerically. Two different sets of coordinates are usually employed to analyze the structure of these differential equations and the asymptotic behavior of the bulk fields near the black hole horizon, in the deep infrared, and close to the boundary, corresponding to a nontrivial ultraviolet (UV) fixed point of the holographic RG flow. One of these sets of coordinates corresponds to the so-called \textit{standard coordinates}, denoted here with a tilde, in terms of which the blackening function goes to unity at the boundary, $\tilde{h}(\tilde{r}\to\infty)=1$, and also $\tilde{A}(\tilde{r}\to\infty)\to\tilde{r}$, so that holographic formulas for physical observables are written in standard form in these coordinates. The other set of coordinates are the so-called \textit{numerical coordinates}, denoted here without a tilde, which are defined in terms of convenient rescalings of the standard coordinates. These rescalings are done with the purpose of fixing definite numerical values for the radial position of the event horizon, $r_H=0$, and also for some of the initially undetermined infrared Taylor-expansion coefficients of the bulk fields close to the event horizon, since for starting the numerical integration of the bulk equations of motion at the black hole horizon (which then proceeds up to the boundary), all the infrared Taylor-expansion coefficients must have definite numerical values. In fact, after those rescalings, all the near-horizon coefficients are expressed as functions of just two initially undetermined infrared coefficients, $\phi_0\equiv\phi(r_H=0)$ and $\Phi_1\equiv\Phi'(r_H=0)$. These are the two near-horizon conditions which one needs to choose in order to construct different numerical solutions for the above specified field equations. In that way, each pair of event horizon conditions $(\phi_0,\Phi_1)$ numerically generates a charged black hole solution which is dual to a given thermodynamic equilibrium state at the boundary QFT, with definite values of temperature ($T$), baryon chemical potential ($\mu_B$), entropy density ($s$), and baryon charge density ($\rho_B$)~\cite{DeWolfe:2010he}. By generating a large ensemble of charged black hole solutions one can therefore construct the holographic phase diagram of the model in the $(T,\mu_B)$-plane.

In the numerical coordinates, one can show from Eqs.~\eqref{eq:EoM1} --- \eqref{eq:EoM4} that the EMD fields scale in the UV limit $r\to \infty$, near the boundary and far from the event horizon, according to \cite{DeWolfe:2010he,Rougemont:2015wca,Critelli:2017oub,Grefa:2021qvt},
\begin{align}
         A(r)&= \alpha(r)+O\left(e^{-2\nu\alpha(r)} \right),\label{UVA}\\
         h(r)&= h_{0}^{\textrm{far}}+O\left(e^{-4\alpha(r)} \right),\label{UVh}\\
         \phi(r)&= \phi_{A}e^{-\nu\alpha(r)}+O\left(e^{-2\nu\alpha(r)}\right),\label{UVphi}\\
         \Phi(r)&= \Phi_{0}^{\textrm{far}}+ \Phi_{2}^{\textrm{far}}e^{-2\alpha(r)}+O\left(e^{-(2+\nu)\alpha(r)}\right),\label{UVPhi}
\end{align}
where $\alpha(r) = A^{\textrm{far}}_{-1} \, r + A^{\textrm{far}}_0$.  The UV coefficient $A^{\textrm{far}}_{-1} = 1/h_{0}^{\textrm{far}}$ in $\alpha(r)$ can be found by using the constraint Eq.~\eqref{eq:EoM5}, while the UV coefficient $\Phi_{2}^{\textrm{far}}$ can be found by using the radially conserved Gauss charge \cite{DeWolfe:2010he,Critelli:2017oub},
\begin{equation}\label{eq:Phi2far}
    \Phi_{2}^{\textrm{far}} = - \frac{\sqrt{h_{0}^{\textrm{far}}}}{2f(0)}f(\phi_0)\,\Phi_1\,.
\end{equation}
The remaining UV coefficients, $A^{\textrm{far}}_0$, $\phi_A$, $h_{0}^{\textrm{far}}$, and $\Phi_0^{\textrm{far}}$, are obtained by matching the UV expansions~\eqref{UVA} --- \eqref{UVPhi} to the near-boundary behavior of the numerical solutions of the EMD equations of motion. In particular, the UV coefficients $\phi_A$, $h_{0}^{\textrm{far}}$, $\Phi_0^{\textrm{far}}$, and $\Phi_2^{\textrm{far}}$ are relevant for the calculation of thermodynamic observables. Finally, the exponent in the leading UV contribution for $\phi(r\to\infty)$ in Eq.~\eqref{UVphi} involves $\nu \equiv d - \Delta$, where,
\begin{equation}\label{eq:nu}
    \Delta \equiv \frac{d + \sqrt{d^2 + 4\,m^2}}{2},
\end{equation}
is the scaling dimension of the boundary QFT operator dual to the bulk dilaton field $\phi$, with $-d^2/4=-4\le m^2 \equiv d^2V(0)/d\phi^2 <0$ for a relevant operator satisfying the Breitenlohner-Freedman bound for massive scalar fields on stable asymptotically AdS$_5$ spacetimes~\cite{Breitenlohner:1982jf,Breitenlohner:1982bm} (see also~\cite{Ramallo:2013bua}).

The holographic formulas for the thermodynamic observables referred above are given by~\cite{DeWolfe:2010he,Rougemont:2015wca,Critelli:2017oub,Grefa:2021qvt,Rougemont:2023gfz},
\begin{align}
T &= \left.\frac{\sqrt{-g'_{\tilde{t}\tilde{t}}g^{\tilde{r}\tilde{r}}\,'}}{4\pi}\right|_{\tilde{r}=\tilde{r}_{H}} \!\!\!\!\!\!\!\!\!\!\!\!\Lambda=\frac{e^{\tilde{A}(\tilde{r}_{H})}}{4\pi}|\tilde{h}'(\tilde{r}_{H})|\Lambda = \frac{1}{4\pi\phi_{A}^{1/\nu}\sqrt{h_{0}^{\textrm{far}}}}\Lambda, \label{eq:T}\\
\mu_B &= \lim_{\tilde{r}\rightarrow\infty}\tilde{\Phi}(\tilde{r})\Lambda = \frac{\Phi_{0}^{\textrm{far}}}{\phi_{A}^{1/\nu}\sqrt{h_{0}^{\textrm{far}}}}\Lambda, \label{eq:muB}\\
s &= \frac{S}{V}\Lambda^3=\frac{A_{H}}{4G_{5}V}\Lambda^3=\frac{2\pi}{\kappa_{5}^{2}}e^{3\tilde{A}(\tilde{r}_{H})}\Lambda^{3} = \frac{2\pi}{\kappa_{5}^{2}\phi_{A}^{3/\nu}}\Lambda^{3}, \label{eq:s}\\
\rho_B &= \lim_{\tilde{r}\rightarrow\infty}\frac{\partial\mathcal{L}}{\partial(\partial_{\tilde{r}}\tilde{\Phi})}\Lambda^{3} = -\frac{\Phi_{2}^{\textrm{far}}}{\kappa_{5}^{2}\phi_{A}^{3/\nu}\sqrt{h_{0}^{\textrm{far}}}}\Lambda^{3}, \label{eq:rhoB}
\end{align}
where $A_H$ is the area of the black hole event horizon. Above, we first wrote the holographic formulas in the standard coordinates and next converted them to the corresponding expressions in the numerical coordinates, where the numerical solutions are made available. As mentioned before, we note the use of adequate powers of the energy scale $\Lambda$ employed to express the corresponding observables in powers of MeV.

The dimensionless ratio,
\begin{equation}
\hat{\chi}_2^B\equiv\frac{\chi_2^B}{T^2}\equiv \frac{\partial^2(P/T^4)}{\partial(\mu_B/T)^2} = \frac{\partial(\rho_B/T^3)}{\partial(\mu_B/T)},
\end{equation}
is called the reduced second order baryon susceptibility, where $P$ is the plasma pressure. By considering this expression in the limit of zero chemical potential, one may derive the following holographic formula \cite{DeWolfe:2010he,Rougemont:2015wca},
\begin{equation}
\hat{\chi}_{2}^B(T,\mu_{B}=0)=\frac{1}{16\pi^{2}}\frac{s}{T^{3}}\frac{1}{f(0)\int_{r_{H}}^{\infty}dr\ e^{-2A(r)}f(\phi(r))^{-1}},
\label{eq:chi2B0}
\end{equation}
which is to be evaluated on top of the numerical EMD solutions generated with the horizon condition $\Phi_{1}=0$, corresponding to setting $\mu_B=0$. As discussed in \cite{Rougemont:2015wca,Critelli:2017oub,Grefa:2021qvt}, in order to numerically evaluate the integral \eqref{eq:chi2B0} one must make the substitutions $r_{H}\rightarrow r_{\textrm{start}}$ and $\infty\rightarrow r_{\textrm{max}}$ in the limits of integration, with $r_{\textrm{start}}$ being a small infrared cutoff, typically of $\mathcal{O}(10^{-8})$, which is employed to avoid the singular point of the EMD field equations at the rescaled numerical horizon $r_H=0$, and $r_{\textrm{max}}$ corresponding to a numerical parametrization of the radial location of the boundary of the bulk geometry, which ideally should lie at $r\to\infty$. However, because it is not possible to use infinity as a number in numerical computations, one needs to stop the radial integration of the equations of motion at $r_{\textrm{max}}$, which is some number typically around $2 - 10$. In practice, this number is chosen so that $R(r_{\textrm{max}})\approx R_{\textrm{AdS}_5}=-20$ (in units of $L=1$) within some small numerical tolerance, where $R_{\textrm{AdS}_5}$ is the Ricci scalar curvature of the AdS$_5$ spacetime corresponding to the strongly coupled UV fixed point of the holographic RG flow. 

It is important to remark that classical gauge-gravity models have a nontrivial strongly coupled UV fixed point, contrary to the trivial UV fixed point of QCD, which implies that differently from QCD, classical gauge-gravity models are not asymptotically free, but just asymptotically safe. This phenomenological limitation of classical holographic plasmas becomes clearly manifest by the constant value of shear viscosity over entropy density, $\eta/s=1/4\pi$, which is valid for all values of temperature and baryon chemical potential~\cite{Kovtun:2004de,Policastro:2001yc,Buchel:2003tz}. This is in striking contrast to the temperature-dependent profile for $\eta/s$ in QCD, which increases with temperature in the asymptotically free limit of very high temperatures \cite{Arnold:2000dr} (see \cite{McLaughlin:2021dph,Danhoni:2024kgi} for the phenomenological implications of this in heavy-ion collisions). In fact, as it is well-known, classical gauge-gravity models can only possibly provide a good description of some target system when such a system is operating in a strongly coupled regime. However, we also remark that, based on the theoretical foundations of the holographic duality, one expects that higher curvature corrections to the bulk gravity action should drive the running coupling towards lower values, and it has been indeed shown in \cite{Cremonini:2012ny} that $\eta/s$ may acquire a temperature-dependent profile in higher curvature holographic models with a dilaton field employed to break conformal symmetry. In fact, this has been recently explored in~\cite{Chen:2025fpd}, where a holographic Bayesian profile was provided for the shear viscosity of the QGP at zero baryon density using higher derivative corrections to the EMD setup. This approach provided a qualitative improvement in the agreement with the JETSCAPE result for the shear viscosity to entropy density ratio~\cite{JETSCAPE:2020shq,JETSCAPE:2020mzn}, relatively to the constant value $\eta/s=1/4\pi$ valid for holographic gauge-gravity models with just two derivatives of the metric field in the bulk action. Such a qualitative improvement refers to the production of a minimum for $\eta/s$ in the crossover region. However, the obtained numerical variation on the temperature profile for $\eta/s$ is so small that, quantitatively, the approach followed in~\cite{Chen:2025fpd} is almost indistinguishable from the usual constant holographic result $\eta/s=1/4\pi$. This means that one still needs to devise another approach to try quantitatively reproducing the temperature profile for $\eta/s$ in QCD by using holographic models.

%%%%%%%%%%%%%%%%%%%%%%%%%%%%%%%%%
\subsection{New numerical scheme for UV near-boundary asymptotics}
\label{sec:new}

Regarding the set of UV coefficients $\{\phi_A, h_{0}^{\textrm{far}}, \Phi_0^{\textrm{far}}, \Phi_2^{\textrm{far}}\}$ relevant for the computation of thermodynamic observables in Eqs.~\eqref{eq:T} --- \eqref{eq:rhoB}, we saw that $\Phi_2^{\textrm{far}}$ can be evaluated according to Eq.~\eqref{eq:Phi2far}, once $h_{0}^{\textrm{far}}$ is known. Two of the remaining three UV coefficients can be easily obtained from the numerical solutions and from Eqs.~\eqref{UVh} and~\eqref{UVPhi} by setting $h_{0}^{\textrm{far}}=h(r_\textrm{max})$ and $\Phi_0^{\textrm{far}}=\Phi(r_\textrm{max})$. However, the extraction of $\phi_A$ is more challenging, as it appears in Eq.~\eqref{UVphi} scaled by an exponentially suppressed factor near the boundary. Naively dividing $\phi(r_\textrm{max})$ by this exponentially small factor near the boundary to obtain $\phi_A^{(\textrm{naive})}=\phi(r_\textrm{max}) /e^{-\nu A(r_\textrm{max})}$ would involve a ratio between two very small values, raising numerical issues with precision and noise.

This issue has been previously addressed by fitting the UV scaling form for $\phi(r)$ in Eq.~\eqref{UVphi} to the numerical solutions within some adaptive radial region~\cite{Critelli:2017oub,Grefa:2021qvt}. However, performing this fit: i) demands the judicious choice of the fitting region; ii) is computationally costly; and iii) often leads to numerical noise, which may be particularly large in the phase transition regions of the phase diagram of the model~\cite{Grefa:2021qvt}. Here we discuss a new method, already successfully employed in Ref.~\cite{Hippert:2023bel} (although not discussed there), which we found to be much more stable and computationally efficient.

The UV coefficient $\phi_A$ can be extracted via the inclusion of an extra field $C(r)$, to which corresponds an extra differential equation of relaxational form,
\begin{align} \label{eq:relaxation}
    \frac{dC(r)}{dr}=-\Gamma_{C}\left[C(r) - \phi(r)\, e^{\nu A(r)}\right].
\end{align}
The solution to Eq.~\eqref{eq:relaxation} can be written as,
\begin{align} \label{eq:relaxsol}
    C(r) = C(r_{H})\, e^{-\Gamma_C (r-r_{H})} + \Gamma_C \int_{r_{H}}^r dr'\, e^{-\Gamma_C(r-r')}\, \phi(r) \, e^{\nu A(r)}\,.
\end{align}
For $r-r_{H}\gg 1/\Gamma_C$, the dependence on the horizon conditions is damped away and $C \approx \langle \phi(r) \, e^{\nu A(r)} \rangle_{[\exp(-\Gamma_C\Delta r)]}$ becomes the average of $\phi_A^{(\textrm{naive})}(r)=\phi(r) \, e^{\nu A(r)}$ with an exponential weight.  While $\phi_A^{(\textrm{naive})}(r)$ is subject to numerical noise, noisy features are eliminated by averaging over a range $\Delta r \sim 1/\Gamma_C$, with exponentially decaying weight.

The convergence of the method can be evaluated by checking that $|C - \phi\, e^{\nu A}|/C\ll 1$. In practice, we take the method to have converged whenever $|C - \phi\, e^{\nu A}|/C$ is smaller than a precision parameter $\epsilon_C\ll 1$. The UV coefficient $\phi_A$ can then be directly estimated by setting $\phi_A = C(r_\textrm{max})$. %(r\gg 1/\Gamma_{C})$.

This new method has the following advantages:
\begin{enumerate}
    \item Efficiency: $\phi_A$ can be extracted directly from the final value of $C(r)$, with no need for a costly fitting procedure on a judicious choice of radial interval;
    \item Stability: the resulting values of $\phi_A$ turn out to be less subject to numerical noise --- we believe this is so because $\phi_A$ in this new method depends linearly on the values of $\phi(r)$, which is not the case for a least squares fitting;
    \item Computational time: with the improved stability against numerical noise, the application of filters to the numerical results is rendered unnecessary, drastically reducing the computational cost of the final results.
\end{enumerate}
\begin{figure}[h!]
    \centering
    \begin{subfigure}{0.45 \linewidth}
        \centering
    \includegraphics[width=\linewidth]{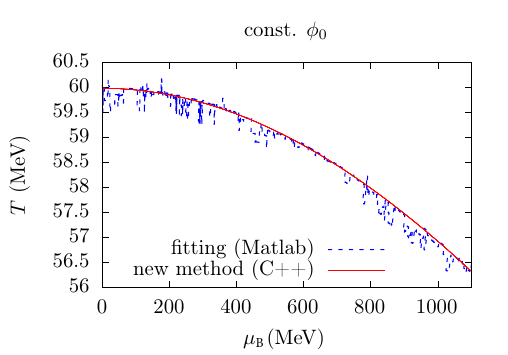}
    \end{subfigure}
    \hfill
    \begin{subfigure}{0.45\linewidth}
        \centering
    \includegraphics[width=\linewidth]{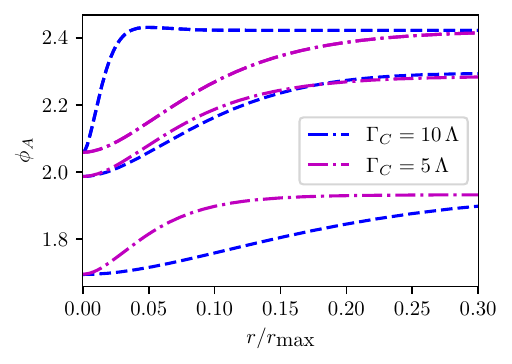}
    \end{subfigure}
\caption{Left panel: A constant $\phi_0$-line starting at $(T,\mu_B)= (60,0)$ MeV, computed with a prefactor $\phi_A$  extracted by different means.  The dot-dashed blue curve shows results obtained using the methods presented in Ref.~\cite{Grefa:2021qvt}, while the solid red curve represents results extracted with the present relaxational approach. Right panel: Ultraviolet scaling coefficient $\phi_A$, obtained from Eq.~\eqref{eq:relaxsol}, with $\Gamma_{C}/\Lambda = 10$ (dashed blue curves) and $\Gamma_{C}/\Lambda = 5$ (magenta dot-dashed curves), for ramdomly selected values of $115$ MeV$\leq T\leq 250$ MeV and $\mu_{B} \leq 3.5 \,T$.}
    \label{fig:numerical}
\end{figure}

Although there is no need for a judicious choice of fitting range, the rate $\Gamma_{C}$ must be carefully chosen. If $\Gamma_{C}$ is too large, the value of $\phi_A$ will be averaged over just a few points and the result will be susceptible to numerical noise. On the other hand, if $\Gamma_{C}$ is too small, relaxation to the correct value will take longer, and convergence will become slow. Moreover, requiring an improved precision on $\phi_A$ and on the Ricci scalar $R$ will lead to convergence at larger values of the AdS radius, where the extraction of $\phi_A$ becomes increasingly challenging. Therefore, the value of $\Gamma_C$ must be adjusted according to precision. The new numerical method is not only computationally robust but it also solves many issues related to noise associated with the previously applied numerical method, as can be clearly seen in the left panel of Fig.~\ref{fig:numerical}. 

In order to test how the convergence of this new method depends on the choice of $\Gamma_{C}$, the right panel of Fig.~\ref{fig:numerical} shows how our new approach to extract $\phi_A$ evolves with increasing radius, for different values of the relaxation rate $\Gamma_{C}$ and randomly selected values of $115$ MeV$\leq T\leq 250$ MeV  and $\mu_{B} \leq 3.5 \,T$. Notice that, in the right panel of  Fig.~\ref{fig:numerical}, we do not show results up to the maximum value of $r=r_{\textrm{max}}$ where numerical convergence is found. The reason behind this is that the results being compared become harder to distinguish when values of the radius up to $r=r_{\textrm{max}}$ are shown.

%%%%%%%%%%%%%%%%%%%%%%%%%%%%%%%%%
\subsection{New algorithm to locate the critical point}

\begin{figure}[h!]
    \centering
    \begin{subfigure}{0.45\linewidth}
        \centering
    \includegraphics[width=\linewidth]{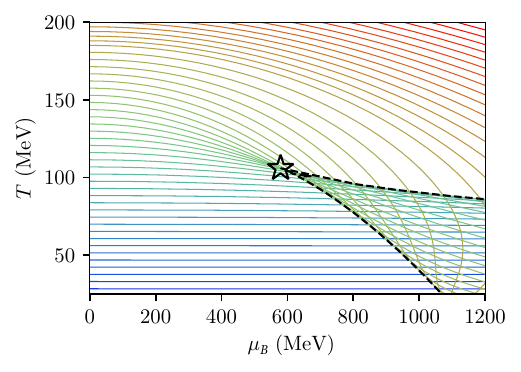}
    \end{subfigure}
    \hfill
    \begin{subfigure}{0.52\linewidth}
        \centering
    \includegraphics[width=\linewidth]{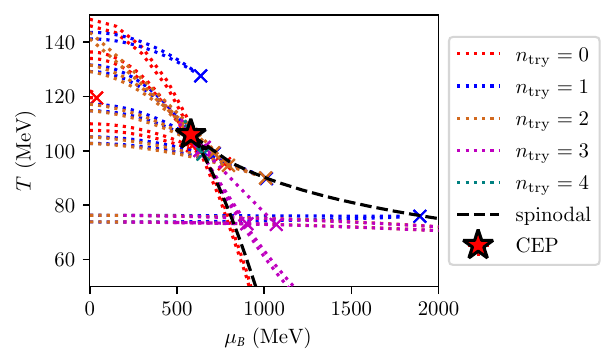}
    \end{subfigure}
\caption{Left panel: Lines of constant $\phi_0$ are displayed as thin solid lines across the phase diagram, colored according to the value of $\phi_0$. Stable, metastable and unstable states are realized as different dual black holes corresponding to the same value of $(\mu_{B}, T)$. Thick dashed lines correspond to the spinodal lines, where metastable solutions become unstable. The star marks the position of the critical endpoint, where these lines end. Right panel: The process of finding the critical point at which constant $\phi_0$-lines cross each other. Parabolas interpolating previously found values of $(\mu_B,T)$ on a constant $\phi_0$ line are used to find the critical point.}
    \label{fig:cep_determination}
\end{figure}

Our goal in the Bayesian analysis of the EMD model presented in Ref~\cite{Hippert:2023bel} was to draw predictions for the critical point from our posterior distribution. Here, we describe the CEP-locating algorithm that was devised to extract the location of the conjectured CEP in the prior and posterior samples of the Bayesian analysis. In this manuscript, we use this algorithm for a fast determination of the CEP location for each Bayesian posterior we utilize to construct our uncertainty bands.

The phase diagram in our model is of the form shown in the left panel of Fig.~\ref{fig:cep_determination}, where lines of constant $\phi_0$ bend and often intersect each other. This clearly indicates the existence of several dual black-hole solutions corresponding to the same $\mu_B$ and $T$, and therefore to different phases. In the left panel of Fig.~~\ref{fig:cep_determination}, the region within which constant $\phi_0$ lines cross each other is delimited by the spinodal lines, shown as dashed black lines. In this region there are three black hole duals for each point, corresponding to thermodynamically stable, metastable and unstable solutions. We observe that the two spinodal lines shown in the left panel of Fig.~~\ref{fig:cep_determination} meet at a cusp where multiple $\phi_0$ lines all cross each other, so that $\phi_0$ is undefined. As a consequence, the EoS becomes non-analytic, indicating the location of the second-order CEP, marked with a star. This crossing of multiple constant $\phi_0$ lines can be used to quickly locate the CEP.

To find crossings between two lines of constant $\phi_0$, we interpolate these lines with quadratic functions $T^{(\phi_0)}(\mu_B) = a^{(\phi_0)}\,\mu_B^2 + b^{(\phi_0)}\mu_B + c^{(\phi_0)}$. From the interpolated curves, checking for crossings at real $(\mu_B,T)$ pairs becomes trivial. By also interpolating $\Phi_1^{(\phi_0)}(\mu_B)$ for each of these lines, we can estimate the values of $(\phi_0,\Phi_1)$ where the two lines intersect. Using these values as initial conditions to solve Eqs.~\eqref{eq:EoM1} to \eqref{eq:EoM4}, we can find the actual values of $(\mu_B,T)$ and check if they are indeed the same within specified precision and accuracy goals. If they are not the same within the requested errors, we can use the new point to update our quadratic interpolation of  $T^{(\phi_0)}(\mu_B)$ and $\Phi_1^{(\phi_0)}(\mu_B)$ and repeat the process up to a maximum number of trials, until either a crossing between the two lines is found, the two lines move away from each other, or the crossing enters unphysical values of $\phi_0$, $\Phi_1$ and $T$ and $\mu_B$ (e.g., $T<0$ or $\Phi_1> \Phi_1^{\textrm{max}}$). 

We start by interpolating $T^{(i)}(\mu_B)$ and ${\Phi_1}^{i}(\mu_B)$ for different values of $\phi_0 = \phi_0^{(i)}$, $i = 0,1,\ldots,N_{\phi_0}$, within a given range, and looking for crossings between neighboring lines $T^{(i)}(\mu_B)$, $T^{(i+1)}(\mu_B)$. 
To increase the number of crossings we are able to find, we select values of $\phi_0^{(i)}$ for which an intersection between neighboring lines was found, and search for additional intersections between every pair of non-neighboring lines. Again, this is done by iteratively interpolating these lines until the process converges or fails.

The process of finding the points at which constant $\phi_0$ lines cross each other is illustrated in the right panel of Fig.~\ref{fig:cep_determination}, where different line styles correspond to different iterations, $n_{\textrm{try}}=0,1,\ldots,4$, in the search for the intersection points, marked by a `$\times$' symbol. After just a few trials, the remaining intersection points converge to the region between the spinodal lines, as should be expected. Finally, we take the crossing point at the lowest value of $\mu_B$ as the location of the critical point. This point is marked as an unfilled star in the left panel, and as a red star in the right panel of Fig.~\ref{fig:cep_determination}. The entire process can be performed in a few seconds, which is sufficiently fast for use in our Bayesian analysis.

%%%%%%%%%%%%%%%%%%%%%%%%%%%%%%%%%
\subsection{Bayesian inference and lattice QCD results}
\label{sec:bayes}

In this section, we present the functional forms chosen for the free functions $V(\phi)$ and $f(\phi)$ and review how to use Bayesian inference to fix all the free parameters of the EMD model by taking continuum extrapolated LQCD data with $N_f=2+1$ flavors and physical quark masses as first principles inputs for the equation of state and baryon susceptibility of the QGP at zero chemical potential.

With the holographic formulas~\eqref{eq:T},~\eqref{eq:muB},~\eqref{eq:s}, and~\eqref{eq:rhoB} describing the basic thermodynamic observables defining the equation of state for the strongly coupled plasma, one may fix $\{V(\phi),\kappa_5^2,\Lambda\}$ by imposing that the equation of state of the holographic model at zero baryon chemical potential matches the corresponding LQCD data~\cite{Borsanyi:2013bia}. Moreover, by numerically solving the integral \eqref{eq:chi2B0} over the background EMD solutions at $\mu_B=0$ and by requiring that the baryon susceptibility at zero chemical potential in the holographic model matches the corresponding LQCD result~\cite{Borsanyi:2021sxv}, one can fix $f(\phi)$.

Here we make use of two ansatze for $V(\phi)$ and $f(\phi)$. The first functional form is a polynomial-hyperbolic Ansatz (PHA), given by
\begin{align}
    \label{PHA_Vofphi}V(\phi)&=-12\cosh({\gamma\,\phi})+b_2\,\phi^2+b_4\,\phi^4+b_6\,\phi^6,\\
    \label{PHA_fofphi}
f(\phi)&=\frac{\text{sech}(c_1\,\phi+c_2\,\phi^2+c_3\,\phi^3)}{1+d_1}+\frac{d_1}{1+d_1}\,\text{sech}(d_2\,\phi).
\end{align}
The second one is a parametric Ansatz (PA), with the following functional form
\begin{align}
    \label{PA_Vphi}
    V(\phi)&=-12\,\cosh\left[\left(\frac{\gamma_1\,\Delta\,\phi^2_V+\gamma_2\,\phi^2}{\Delta\,\phi^2_V+\phi^2}\right)\phi\right],\\
    \label{PA_fphi}
    f(\phi)&=1-(1-A_1)\left[\frac{1}{2}+\frac{1}{2}\tanh\left(\frac{\phi-\phi_1}{\delta\phi_1}\right)\right]-A_2\,\left[\frac{1}{2}+\frac{1}{2}\tanh\left(\frac{\phi-\phi_2}{\delta\phi_2}\right)\right].
\end{align}
A few considerations are taken into account in order to propose the above ansatze for $V(\phi)$ and $f(\phi)$ (see the recent review~\cite{Rougemont:2023gfz} for detailed discussions). Regarding the dilaton potential $V(\phi)$, it is expected by the standards of the holographic dictionary that in the UV regime $(\phi\to 0)$ one recovers in the bulk action, from $V(\phi)$, twice the value of the cosmological constant of the $\text{AdS}_5$ spacetime. Therefore, one requires that, $V(0)=-12$ (in units of $L=1$). In this way, the functional forms \eqref{PHA_Vofphi} and \eqref{PA_Vphi} for $V(\phi)$ are such that, close to the boundary at $\phi(r\to\infty)\to 0$, the Ricci scalar approaches $R=-20$ (in units of $L=1$), which allows us to have asymptotically $\text{AdS}_5$ background solutions, as prescribed by the holographic dictionary. Besides that, the functional forms proposed for $V(\phi)$ are engineered so as to have the holographic entropy density at zero chemical potential matching the corresponding LQCD result from Ref.~\cite{Borsanyi:2013bia}. As for the Maxwell-dilaton coupling function $f(\phi)$, it is constrained by matching to the LQCD result from Ref.~\cite{Borsanyi:2021sxv} for the baryon susceptibility at zero chemical potential, the corresponding holographic formula given by Eq.~\eqref{eq:chi2B0}. An appealing feature of $\hat{\chi}_2^B(T,\mu_B=0)$ in Ref.~\cite{Borsanyi:2021sxv} is its fast inflection around $T=140\,\text{MeV}$. The freedom in $A_\mu$ allows us to set a normalization such that $f(0)=1$, as originally proposed in Ref.~\cite{DeWolfe:2010he}. Then, to reproduce this feature of the baryon susceptibility from LQCD, we must consider a functional $f(\phi)$ that presents a fast variation close to the boundary $\phi(r\to\infty)\to 0$~\cite{Rougemont:2023gfz}. These features can be achieved with Eqs.~\eqref{PHA_Vofphi} and \eqref{PHA_fofphi} corresponding to the PHA ansatz. The PA ansatz, represented by Eqs.~\eqref{PA_Vphi} and \eqref{PA_fphi}, reproduce the main features of Eqs.~\eqref{PHA_Vofphi} and \eqref{PHA_fofphi} in a more transparent way, in the sense that it is easier to tune specific features of $V(\phi)$ and $f(\phi)$ by adjusting the free parameters in the PA ansatz~\cite{Hippert:2023bel}. We note that both ansatze lead to mostly similar results, at least in the regime of applicability of the present model. Results for the transport coefficients obtained within the two ansatze are compared in section~\ref{sec:comparison} below.

\begin{figure}%[H]
    \centering
    % ===== Left panel =====
    \begin{subfigure}[t]{0.49\linewidth}
        \centering
        \includegraphics[width=\linewidth]
        {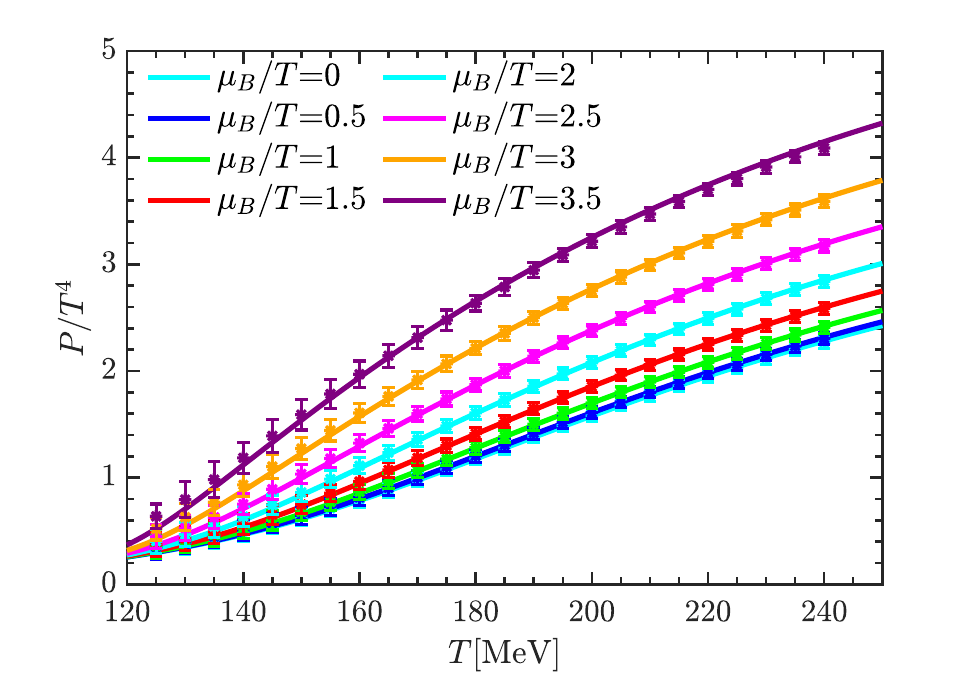}
    \end{subfigure}
    \hfill
    % ===== Right panel =====
    \begin{subfigure}[t]{0.49\linewidth}
        \centering
\includegraphics[width=\linewidth]
{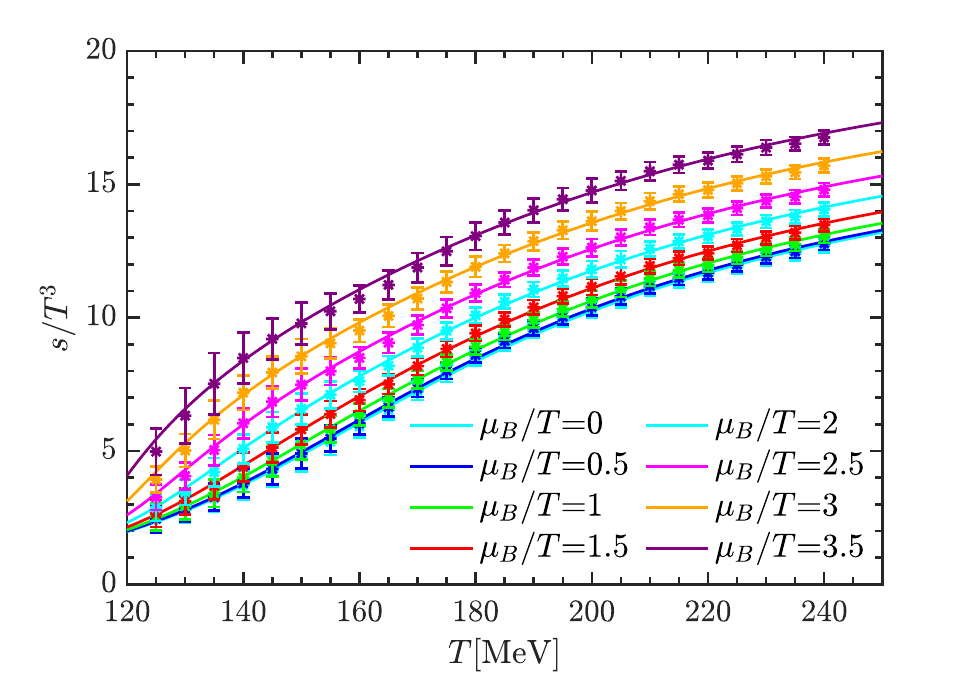}
    \end{subfigure}
\caption{Comparison between the Equation of State at finite density obtained from the best fit to lattice results at zero chemical potential for the PHA Ansatz, and the state-of-the-art lattice QCD results from \cite{Borsanyi:2021sxv}. For detailed comparison between the lattice results and PA, PHA models, see Ref.~\cite{Hippert:2023bel}.}
    \label{fig:Lattice}
\end{figure}

These specific ansatze were introduced in Ref.~\cite{Hippert:2023bel}, where Bayesian inference was employed to determine the free parameters in the functions that yield optimal agreement with lattice QCD results for each chosen functional form of $ V(\phi)$ and $f(\phi)$.
The prior distributions for $V(\phi)$ and $f(\phi)$ led to critical points widely scattered across the QCD phase diagram, and in some cases resulted in no critical point at all. However, posterior distributions constrained by lattice QCD data at zero chemical potential revealed holographic critical points tightly clustered within the ranges $T=101$ --- 108 MeV and $\mu_{B}=560$ --- 625 MeV, at 95\% confidence level. In Ref.~\cite{Hippert:2023bel}, the Bayesian analysis focused primarily on thermodynamic quantities.
In this work, our primary objective is to use the posterior distributions obtained from the previous analysis to examine transport coefficients and energy-loss observables in the QGP. Specifically, we aim to investigate the behavior of these physical observables near the critical end point and across the first order phase transition line and to assess its robustness against the Bayesian analysis uncertainty band. All figures presented in the main text correspond to the PHA Ansatz. Results for the PA Ansatz are shown in the Appendix~\ref{sec:app} for completeness. As a reminder of the thermodynamic predictions of our model at finite baryon chemical potential, Fig.~\ref{fig:Lattice} shows the results based on our best-fit parameter for the PHA Ansatz, for $s/T^3$ and $P/T^4$ at finite ($T,\mu_B$) compared to the corresponding LQCD results. We note that since finite $\mu_B$ lattice results were not used to fix the free parameters of the holographic model, such results follow as true predictions of the holographic setup. Comparing such predictions with LQCD results at finite baryon density, as done in Fig.~\ref{fig:Lattice}, is a clear way of accessing the phenomenological reliability of the present holographic EMD framework. Further comparisons involving holographic predictions for transport coefficients, like the bulk viscosity and the jet quenching parameter, and the profiles favored in multistage Bayesian models simultaneously matching several heavy ion data are presented in Fig.~\ref{fig:JetScape}, to be discussed in the upcoming sections.

As a final remark regarding the physical relevance of the bulk free parameters $\kappa_5^2\equiv 8\pi G_5$ and $\Lambda$, their values can be fixed, together with the form of the dilaton potential $V(\phi)$, by matching the holographic formulas~\eqref{eq:T} and~\eqref{eq:s} to the LQCD result for the entropy density as a function of temperature at zero chemical potential~\cite{Hippert:2023bel}. In top-down holographic models (which is not the case of our bottom-up EMD setup), one can explicitly show that $\kappa_5^2\equiv 8\pi G_5 \sim 1/N_c^2$, where $N_c$ is the number of colors of the gauge group $SU(N_c)$ of the dual QFT at the boundary. For instance, in the most well-known case corresponding to the strongly interacting $\mathcal{N}=4$ Supersymmetric Yang-Mills (SYM) plasma with gauge group $SU(N_c)$, the relation (in units of the AdS radius $L = 1$) is $\kappa_5^2=4\pi^2/N_c^2$. In the standard holographic gauge-gravity dictionary, $N_c$ is taken to be very large, so that $\kappa_5^2$ should be very small, implying weakly-coupled gravity in the bulk. On the other hand, in our approach we do not match our holographic entropy density at zero chemical potential to a large $N_c$ QCD result, but to the physical $SU(N_c=3)$ LQCD result extrapolated to the continuum limit with physical values of the quark masses. Although the precise relation between the bulk gravitational constant $\kappa_5^2\equiv 8\pi G_5$ and the rank $N_c$ of the dual QFT gauge group $SU(N_c)$ is unknown in bottom-up models, if we heuristically take into consideration the $\mathcal{N}=4$ SYM relation $\kappa_5^2=4\pi^2/N_c^2$ and use $N_c = 3$, we obtain $\kappa_5^2 \approx 4.4$, which is not a small number. In fact, this heuristic estimate has the same order of magnitude of $\kappa_5^2 \approx 11.4$, which is the value favored by the Bayesian analyses of LQCD data within the holographic EMD setup performed by considering both, the PHA and PA ansatze --- see the Maximum A Posteriori (MAP) values in Table II of Ref.~\cite{Hippert:2023bel}. The bulk gravitational constant itself is not directly observable at the dual boundary QFT, although it appears in holographic formulas for several physical observables of the boundary QFT.

Regarding the energy scale factor $\Lambda$ (which replaces $1/L$ as a free parameter in our gauge-gravity setup), it plays a role analogous to the $\Lambda_\textrm{QCD}$ scale within the holographic EMD model. Although the value of $\Lambda_\textrm{QCD}$ is not directly observable and depends on the number of active quarks and the particular renormalization scheme considered, a commonly quoted value (for 3 active quark flavors in the $\overline{\textrm{MS}}$ scheme) is $\Lambda_\textrm{QCD} \sim 330$ MeV. In Table II of Ref.~\cite{Hippert:2023bel}, the MAP values of the energy scale factor $\Lambda$ in the holographic EMD setup are $\Lambda_\textrm{EMD}(\textrm{PHA}) = 1129$ MeV $\approx 3.4\, \Lambda_\textrm{QCD}$ and $\Lambda_\textrm{EMD}(\textrm{PA}) = 955$ MeV $\approx 2.9\, \Lambda_\textrm{QCD}$. Like the dimensional transmutation scale $\Lambda_\textrm{QCD}$ in QCD, the energy scale factor $\Lambda$ in the holographic EMD model is not directly observable, although its value is employed to express gravity results in physical units at the dual boundary QFT.

%%%%%%%%%%%%%%%%%%%%%%%%%%%%%%%%%
%%%%%%%%%%%%%%%%%%%%%%%%%%%%%%%%%
\section{Holographic transport coefficients and energy loss}
\label{sec:results}

We revisit here the holographic calculation of several transport coefficients~\cite{Rougemont:2015wca,Rougemont:2015ona,Rougemont:2017tlu,Grefa:2022sav} in view of the Bayesian inference procedure discussed in the previous section. The merits of such approach are twofold: a) the Bayesian analysis allows for a much more systematic and precise determination of the best-fit model given some choice for the functional forms of the free functions $V(\phi)$ and $f(\phi)$ of the EMD setup; and b) the Bayesian procedure allows for the construction of uncertainty bands for the theoretical predictions of the EMD setup based on statistical posterior distributions, which in turn provides a way of checking the internal robustness of such holographic predictions against variations of the free parameters of the model.

In the EMD setup, the holographic formulas for computing the baryon and thermal conductivities as well as the baryon diffusion coefficient were derived in~\cite{DeWolfe:2011ts,Rougemont:2015ona}, while the holographic formulas for the shear and bulk viscosities were obtained in~\cite{DeWolfe:2011ts} --- see also~\cite{Rougemont:2017tlu,Grefa:2022sav}. In terms of parton energy loss, the original holographic formulas for the heavy quark drag force, the Langevin diffusion coefficients, and the jet-quenching parameter were derived for the conformal SYM plasma in~\cite{Gubser:2006bz,Herzog:2006gh,Casalderrey-Solana:2006fio,Gubser:2006qh,Gubser:2006nz,Liu:2006ug,Liu:2006he,Casalderrey-Solana:2007ahi,DEramo:2010wup}, later extended for non-conformal dilatonic models in~\cite{Gursoy:2009kk,Gursoy:2010aa}, and then adapted for EMD setups in~\cite{Rougemont:2015wca,Finazzo:2016mhm,Rougemont:2020had}, see also~\cite{Grefa:2022sav}. Since those formulas are well-known, we will not review their derivations here and refer the interested reader to consult the above references for detailed discussions. However, before presenting the relevant holographic formulas and discussing the corresponding numerical results, we make some brief comments regarding the general ideas involved.

For the isotropic EMD model considered here, there is a spatial $SO(3)$ rotation symmetry which separates the diffeomorphism and gauge invariant combinations of linearized perturbations of the bulk fields into different irreducible representations. The $SO(3)$ triplet channel is related to the baryon conductivity to be discussed in subsection~\ref{sec:baryon}. The $SO(3)$ quintuplet and singlet channels are related, respectively, to the shear and bulk viscosities, to be analyzed in subsection~\ref{sec:bulk}. The different irreducible representations do not mix at the linear level, so that one needs to solve decoupled differential equations for the corresponding linear perturbations on top of the EMD black hole backgrounds~\cite{DeWolfe:2011ts}. One can show that the perturbation in the quintuplet channel leads to $\eta/s = 1/4\pi$ for all values of $(T>0,\mu_B\ge 0)$, as expected for any holographic model which is homogeneous, isotropic, and with at most two derivatives of the bulk metric field in the gravity action~\cite{Kovtun:2004de,Policastro:2001yc,Buchel:2003tz}. We remark, however, that the dimensionless combination entering the hydrodynamic expression for the viscous part of the boundary gauge theory energy-momentum tensor at finite baryon chemical potential is $\eta T/(\epsilon+P)$ instead of $\eta/s$ \cite{Liao:2009gb,Denicol:2013nua}, where $P$ and $\epsilon$ are, respectively, the pressure and the energy density of the plasma. The dimensionless combination $\eta T/(\epsilon+P)$ reduces to $\eta/s$ when $\mu_B=0$. However, $\eta T/(\epsilon+P)$ displays a nontrivial dependence on $T$ and $\mu_B$ at finite baryon density, as we are going to see in subsection~\ref{sec:bulk}.

Regarding physical observables describing energy loss for partons traversing the strongly coupled medium, they are typically computed in holography by using the Nambu-Goto (NG) action for probe strings defined over the background field solutions. This probe string action involves the square root of the `t Hooft coupling, $\sqrt{\lambda_t}=L^2/\alpha'=L^2/\ell_s^2=1/\ell_s^2$ (in units of $L=1$), where $\ell_s$ is the fundamental string length, which is typically taken as a free parameter in bottom-up constructions. However, by the standards of the holographic dictionary, one expects that $\ell_s$ should be very small and, consequently, $\lambda_t$ should be very large in the classical limit corresponding to the gauge-gravity duality. Therefore, as done e.g.~in~\cite{Rougemont:2015wca,Critelli:2016cvq,Rougemont:2020had,Grefa:2022sav}, by considering infinitely heavy probe quarks traversing the plasma we neglect both the coupling between the background dilaton field and the Ricci scalar induced on the probe string worldsheet, and also the minimal coupling between the background Maxwell field and the string endpoint attached to a probe quark at the boundary, since both couplings are of order zero in the `t Hooft coupling, while the NG action is of order $1/2$~\cite{Gursoy:2007er,Kiritsis:2011ha}. Thus, both terms do not contribute to leading order in the `t Hooft coupling for the calculations of the heavy quark drag force and the Langevin diffusion coefficients. In the case of the jet-quenching parameter related to the energy loss of light partons, we remark that the minimal coupling term is automatically zero because the contributions from each string endpoint at the boundary in this computation cancel out for such a coupling.

%%%%%%%%%%%%%%%%%%%%%%%%%%%%%%%%%
\subsection{Baryon transport: baryon susceptibility, baryon and thermal conductivities and baryon diffusion}
\label{sec:baryon}

In the relativistic Navier-Stokes approximation \cite{Rocha:2023ilf}, the baryon current for a hot and baryon dense fluid at the flat four-dimensional boundary gauge theory with Minkowski metric $\eta_{\mu\nu}$ is given by~\cite{Kapusta:2014dja}
\begin{align}
J_B^\mu = \rho_B u^\mu + \Delta J_B^\mu,\qquad \Delta J_B^\mu = T\sigma_B \left(\eta^{\mu\nu}-u^\mu u^\nu\right) \partial_\nu\left(\frac{\mu_B}{T}\right),
\label{eq:Jmu}
\end{align}
where $u^\mu$ is the local velocity field of the fluid and $\Delta J_B^\mu$ is the dissipative part of the baryon current, which is characterized by the baryon conductivity $\sigma_B(T,\mu_B)$. This transport coefficient therefore quantifies how the baryon current responds to gradients of $\mu_B/T$ in the medium. By the Nernst-Einstein relation~\cite{Kapusta:2014dja,Iqbal:2008by}, it is also proportional to the diffusion of baryon charge in the medium, $D_B(T,\mu_B)$, with the proportionality factor being the baryon susceptibility $\chi_2^B(T,\mu_B)\equiv(\partial^2P/\partial\mu_B^2)_T=(\partial\rho_B/\partial\mu_B)_T$,
\begin{align}
T D_B(T,\mu_B) = T \frac{\sigma_B(T,\mu_B)}{\chi_2^B(T,\mu_B)} = \frac{\hat{\sigma}_B(T,\mu_B)}{\hat{\chi}_2^B(T,\mu_B)},
\label{eq:DB}
\end{align}
where $\hat{\sigma}_B\equiv \sigma_B/T$ and $\hat{\chi}_2^B\equiv\chi_2^B/T^2$ are dimensionless ratios. On the other hand, the thermal conductivity $\sigma_T(T,\mu_B)$, which measures the ability of the medium to conduct heat, is also related to the baryon conductivity as follows~\cite{Kapusta:2012zb},
\begin{align}
\sigma_T = \frac{\chi_2^B D_B}{T} \left(\frac{\epsilon+P}{\rho_B}\right)^2 = T\sigma_B \left(\frac{s}{\rho_B}+\frac{\mu_B}{T}\right)^2,
\label{eq:sigmaT1}
\end{align}
where we made use of the fundamental thermodynamic relation, $\epsilon=Ts+\mu_B\rho_B-P$, and the Nernst-Einstein relation. Since $\rho_{B}(T,\mu_{B}\rightarrow0)\rightarrow\mu_{B}\chi_{2}^{B}(T,\mu_{B}\rightarrow0)\to 0$, in order to construct a well-defined dimensionless combination in the limit of zero baryon chemical potential, one may consider as in~\cite{Jain:2009pw},
\begin{align}
    \left[\frac{\mu_{B}^{2}\,\sigma_{T}}{\eta\, T}\right](T,\mu_{B})&=4\pi\frac{\sigma_{B}}{T}\frac{\mu_B^2}{Ts}\left(\frac{Ts}{\rho_{B}}+\mu_B\right)^{2},\label{eq:sigmaT2}\\
    &\to \frac{4\pi}{\left(\chi_2^B/T^2\right)^2} \frac{\sigma_{B}}{T} \frac{s}{T^{3}} = \frac{4\pi\hat{\sigma}_B}{(\hat{\chi}_2^B)^2} \frac{s}{T^3},\,\,\, \textrm{as}\,\,\, \mu_B\to 0,\label{eq:sigmaT2-0}
\end{align}
where we used the holographic relation $\eta/s=1/4\pi$ in writing the RHS of~\eqref{eq:sigmaT2}. While the limit~\eqref{eq:sigmaT2-0} is finite, the LHS of~\eqref{eq:sigmaT2} also depends on the shear viscosity $\eta$ besides the thermal conductivity $\sigma_T$. Since it is the combination $\mu_B^2\sigma_T$ which remains finite in the $\mu_B\to 0$ limit, a simpler dimensionless combination which does not depend on any other transport coefficient besides the thermal conductivity may be constructed as done in~\cite{Rougemont:2015ona,Grefa:2022sav},
\begin{align}
\left[\frac{\mu_B^2\,\sigma_T}{T^4}\right](T,\mu_B) &= \frac{\sigma_B}{T}\left(\frac{\mu_B}{\rho_B}\,\frac{s}{T}+\frac{\mu_B^2}{T^2}\right)^2, \label{eq:sigmaT}\\
&\to \frac{\sigma_B}{T}\left(\frac{s}{T^3}\,\frac{T^2}{\chi_2^B}\right)^2 =  \frac{\hat{\sigma}_B}{(\hat{\chi}_2^B)^2} \left(\frac{s}{T^3}\right)^2, \,\,\, \textrm{as} \,\,\, \mu_B\to 0.\label{eq:sigmaT-0}
\end{align}
In Ref.~\cite{Denicol:2018wdp}, a hybrid phenomenological approach based on $3+1$ hydrodynamics and hadronic transport was used to simulate relativistic heavy-ion collisions producing a hot and baryon dense QGP with subsequent hadronization. The baryon conductivity influences the difference between proton and antiproton mean transverse momentum, and also the difference between proton and antiproton elliptic flow. Therefore, the investigation of the dissipative dynamics of baryon charge is relevant for the theoretical modeling of heavy-ion collisions at lower beam energies, when finite baryon chemical potential effects become important.

Notice from the above discussion that in order to obtain the baryon diffusion coefficient~\eqref{eq:DB} and the thermal conductivity~\eqref{eq:sigmaT}, besides the basic thermodynamic quantities given in Eqs.~\eqref{eq:T} ---~\eqref{eq:rhoB}, one needs to compute the numerical derivative corresponding to the baryon susceptibility, $\chi_2^B(T,\mu_B)=(\partial\rho_B/\partial\mu_B)_T$, and the baryon conductivity $\sigma_B(T,\mu_B)$. The latter can be obtained from the analysis of the diffeomorphism and gauge invariant perturbation $a(r,\omega)$ in the $SO(3)$ triplet channel of the homogeneous and isotropic holographic EMD model, corresponding to any of the three spatial components $a^i$ of the Maxwell field perturbation $a^\mu$. The equation of motion for this perturbation is given by~\cite{DeWolfe:2011ts},
\begin{align}\label{eq:condEOM}
a''(r,\omega)+\left[2A'(r)+\frac{h'(r)}{h(r)}+\frac{\partial_\phi f(\phi)}{f(\phi)}\phi'(r)\right] a'(r,\omega)
+\frac{e^{-2A(r)}}{h(r)}\left[\frac{\omega^2}{h(r)}-f(\phi)\Phi'(r)^{2}\right] a(r,\omega)=0,
\end{align}
where the prime denotes derivative with respect to $r$, and $\omega$ is the (angular) frequency of the homogeneous plane wave Ansatz used for the bulk perturbation.
One needs to solve~\eqref{eq:condEOM} on top of the numerical EMD backgrounds by using the infalling wave condition at the event horizon, $r_{\textrm{start}}$, which corresponds to solving for the retarded thermal correlator of the boundary QFT baryon current operator, with the additional requirement that the perturbation is normalized to unity at the boundary, $r_\textrm{max}$~\cite{DeWolfe:2011ts}.
These two boundary conditions are systematically implemented via~\cite{Grefa:2022sav},
 \begin{equation}\label{eq:condWaveCond}
     a(r,\omega)\equiv\frac{r^{-i\omega}G(r,\omega)}{r_{\textrm{max}}^{-i\omega}\,G(r_{\textrm{max}},\omega)},
 \end{equation}
where $G(r,\omega)$ must be a regular function at the event horizon, with its equation of motion being obtained by substituting~\eqref{eq:condWaveCond} into~\eqref{eq:condEOM}. The holographic formula for the baryon conductivity expressed in physical units of MeV is~\cite{DeWolfe:2011ts,Rougemont:2015ona,Rougemont:2017tlu},\\
\begin{equation}\label{eq:conduct}
    \sigma_{B}(T,\mu_{B})=-\frac{\Lambda}{2\kappa_{5}^{2}\phi_{A}^{1/\nu}}\lim_{\omega\rightarrow0}\frac{1}{\omega}\left(e^{2A(r)}\,h(r) f(\phi)\,\textrm{Im}[a^{*}(r,\omega)a'(r,\omega)]\right)\biggr|_\textrm{on-shell}.
\end{equation}
The term between brackets in Eq.~\eqref{eq:conduct} is a radially conserved flux which can then be evaluated at any value of $r$~\cite{DeWolfe:2011ts}. In practice, as discussed in~\cite{Grefa:2022sav}, the limit of zero frequency is approximated in numerical computations by evaluating~\eqref{eq:conduct} at some small but nonzero value $\omega=\omega_{\textrm{eval}}$, which is typically chosen to be $\mathcal{O}(10^{-5})$. In order to check whether this is in fact a good approximation to the zero frequency limit, we tested for several different backgrounds at finite $(T,\mu_B)$ if $\sigma_B(T,\mu_B)$ remains unchanged, within some small numerical tolerance, when computed using different small values of $\omega_{\textrm{eval}}$. This is satisfied for $\omega_{\textrm{eval}} \sim 10^{-7}$ --- $10^{-2}$. On the other hand, for very small frequencies there are spurious divergences related to issues with numerical precision, while for $\omega_{\textrm{eval}}\gtrsim 10^{-1}$ the evaluation frequency is not small enough to serve as a good approximation to the limit of zero frequency in~\eqref{eq:conduct}.

\begin{figure}%[H]
    \centering
    % ===== Left panel =====
    \begin{subfigure}[t]{0.49\linewidth}
        \centering
        \includegraphics[width=\linewidth]
        {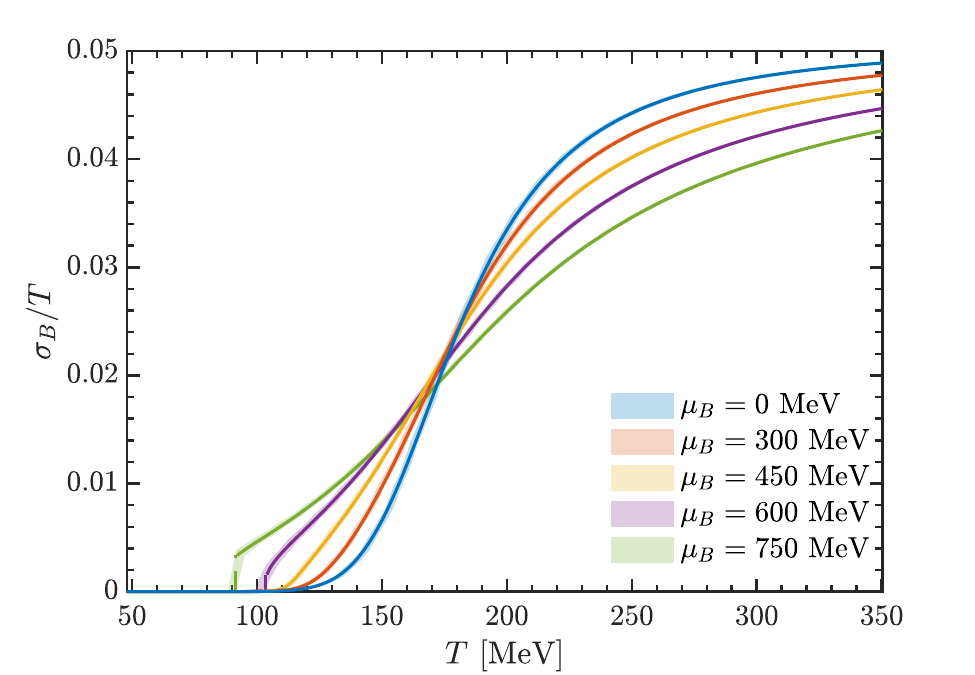}
    \end{subfigure}
    \hfill
    % ===== Right panel =====
    \begin{subfigure}[t]{0.49\linewidth}
        \centering
\includegraphics[width=\linewidth]
{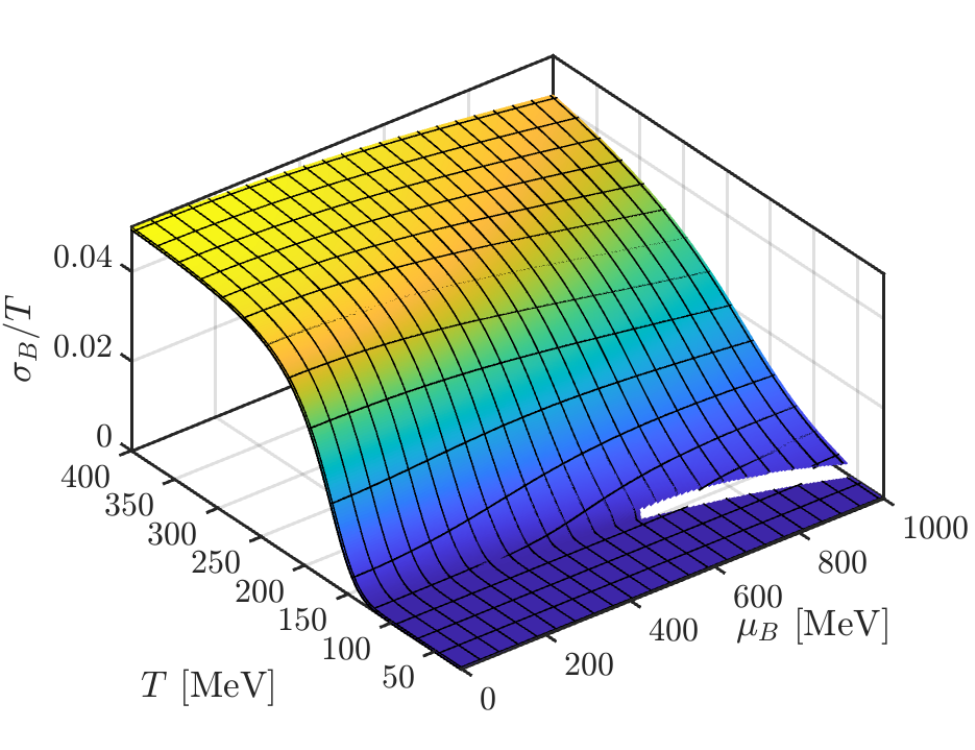}
    \end{subfigure}
\caption{
The left panel shows posterior bands at 95\% confidence level, and corresponding best-fit curves (solid lines) of $\sigma_B/T$ as a function of $T$ for different values of $\mu_B$ while the right panel displays the $\sigma_B/T$ as a function of $T$ and $\mu_B$ computed from the best-fit parameterization.}
    \label{fig:baryon_conductivity}
\end{figure}

\begin{figure}%[H]
    \centering
    % ===== Left panel =====
    \begin{subfigure}[t]{0.49\linewidth}
        \centering
        \includegraphics[width=\linewidth]
        {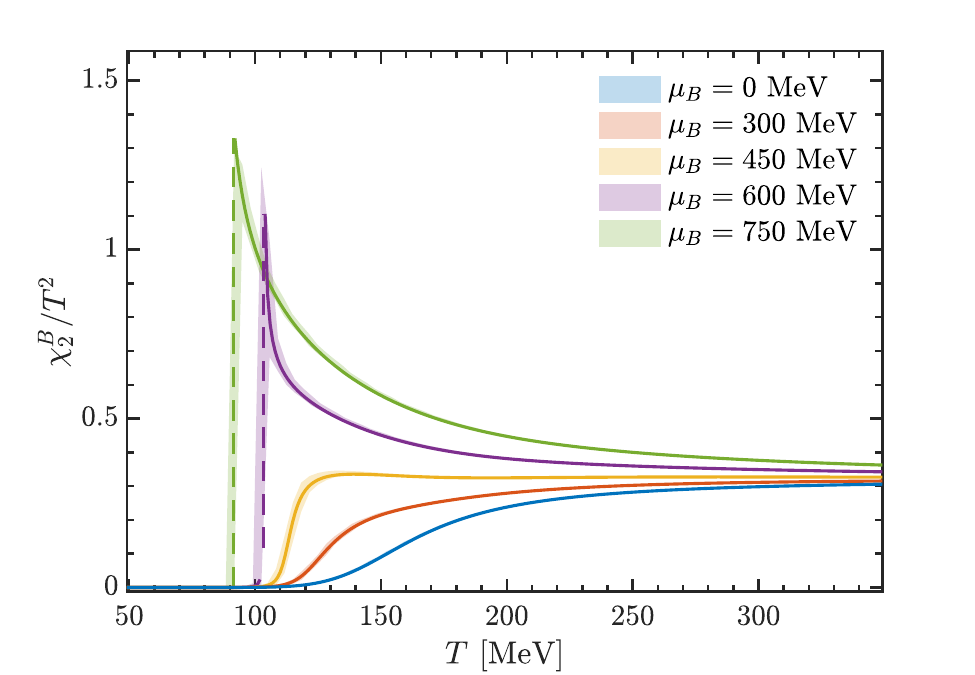}
    \end{subfigure}
    \hfill
    % ===== Right panel =====
    \begin{subfigure}[t]{0.49\linewidth}
        \centering
\includegraphics[width=\linewidth]
{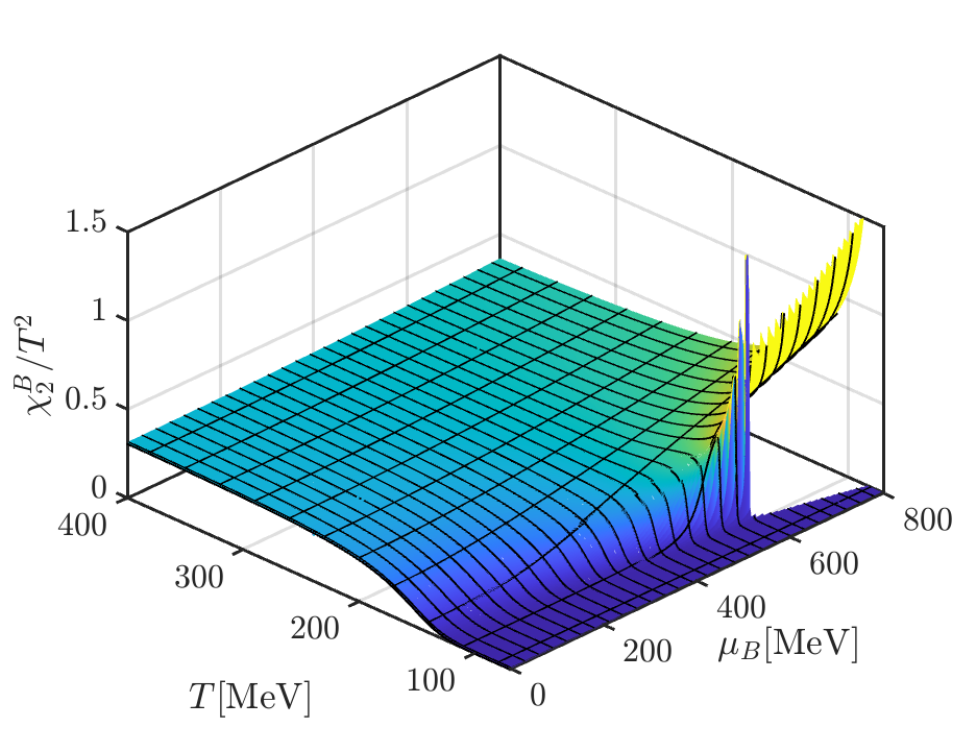}
    \end{subfigure}
\caption{
The left panel shows posterior bands at 95\% confidence level, and corresponding best-fit curves (solid lines) for $\chi^{B}_{2}$ as a function of $T$ for different values of $\mu_B$ while the right panel displays the $\chi^{B}_{2}$ as a function of $T$ and $\mu_B$ computed from the best-fit parameterization.}
    \label{fig:chi2}
\end{figure}

In Fig.~\ref{fig:baryon_conductivity}, we show our Bayesian results for the baryon conductivity. The Bayesian posterior bands at 95\% confidence level are relatively narrow near the crossover. This occurs because the EMD parameters are closely constrained by the lattice data in this temperature range. As a result, there is strong consistency among the posterior samples. The dimensionless ratio $\sigma_B/T$ increases with the temperature, as shown in the left panel of Fig.~\ref{fig:baryon_conductivity}. Within the temperature range $T \sim 150 - 180~\mathrm{MeV}$, the bands and corresponding best-fit lines of $\sigma_B/T$ at fixed $\mu_B$ cross each other. 
%at a temperature slightly above the phase transition. 
The baryon conductivity increases with increasing $\mu_B$ below the temperature range $T \sim 150~\mathrm{MeV}$. However, the opposite behavior is observed above the temperature range $T \sim 180~\mathrm{MeV}$. 
At the critical point, the baryon conductivity remains finite. Across the first-order phase transition line at higher values of $\mu_B$ and lower values of $T$, it develops a small discontinuity gap.

In Fig.~\ref{fig:chi2}, we show our Bayesian results for the dimensionless baryon susceptibility $\hat{\chi}_2^B\equiv\chi_2^B/T^2$ across the phase diagram of the holographic EMD setup at finite temperature and baryon chemical potential. The uncertainty band comprises several different EMD models in the posterior distribution of the Bayesian inference procedure. One can notice $\hat{\chi}_2^B$ diverges at the critical point, as expected for a second-order phase transition point.

\begin{figure}%[H]
    \centering
    % ===== Left panel =====
    \begin{subfigure}[t]{0.49\linewidth}
        \centering
        \includegraphics[width=\linewidth]
        {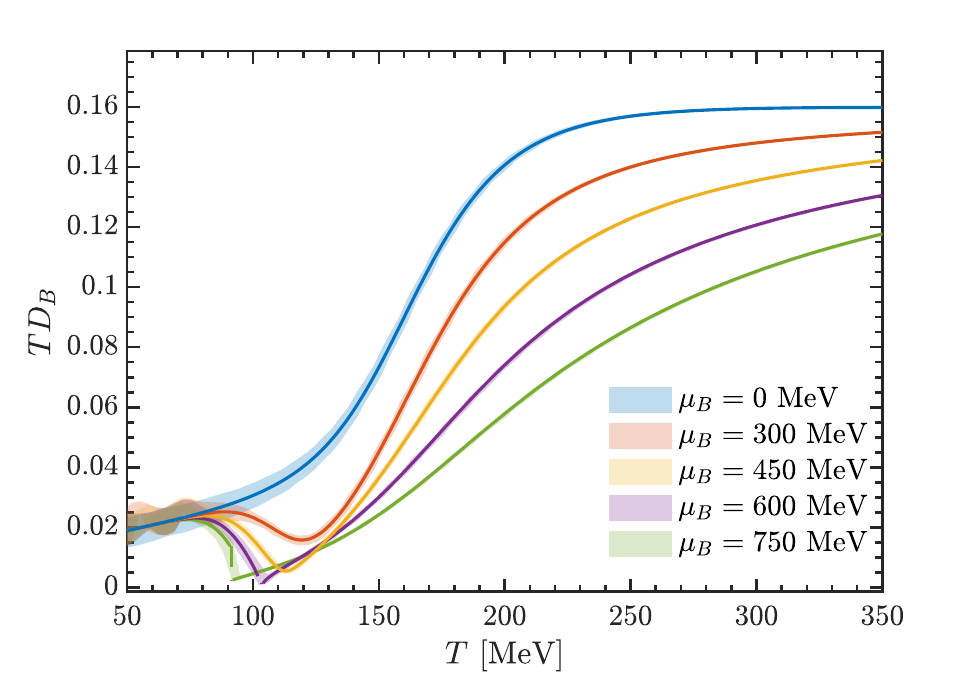}
    \end{subfigure}
    \hfill
    % ===== Right panel =====
    \begin{subfigure}[t]{0.49\linewidth}
        \centering
\includegraphics[width=\linewidth]
{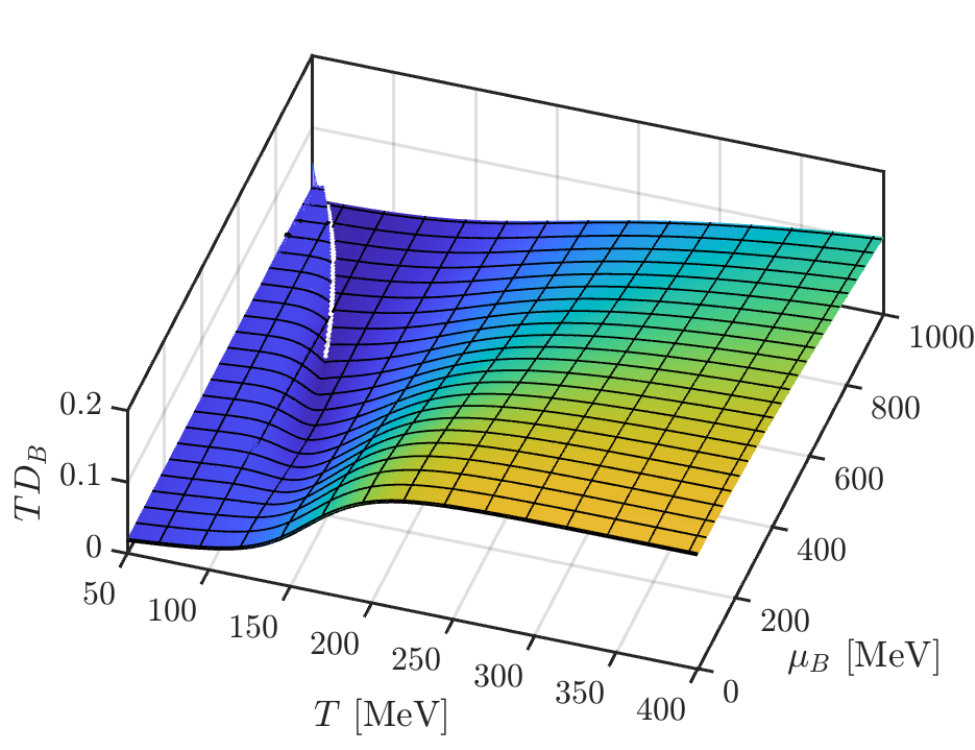}
    \end{subfigure}
\caption{
The left panel shows posterior bands at 95\% confidence level, and corresponding best-fit curves (solid lines) of scaled baryon diffusion coefficient $T D_{B}$ as a function of $T$ for different values of $\mu_B$, while the right panel displays $T D_{B}$ as a function of $T$ and $\mu_B$ computed from the best-fit parameterization.
}
\label{fig:baryon_diffusion}
\end{figure}

Fig.~\ref{fig:baryon_diffusion} presents our Bayesian results for the baryon diffusion coefficient. One can see that by increasing $\mu_B$, the baryon diffusion is generally suppressed at high temperatures, and that it also increases with temperature in the high temperature regime, asymptotically saturating. At the critical point $D_B \rightarrow 0$, because $\chi^B_2$ diverges while $\sigma_B$ remains finite, and $TD_B = \hat{\sigma}_B / \hat{\chi}^B_2$. Across the first-order phase transition, the baryon diffusion shows a small discontinuity gap, which is visible in the best fit line; this gap increases as we increase the baryon chemical potential $\mu_B$.

%\newpage
\begin{figure}%[H]
    \centering
    % ===== Left panel =====
    \begin{subfigure}[t]{0.49\linewidth}
        \centering
        \includegraphics[width=\linewidth]
        {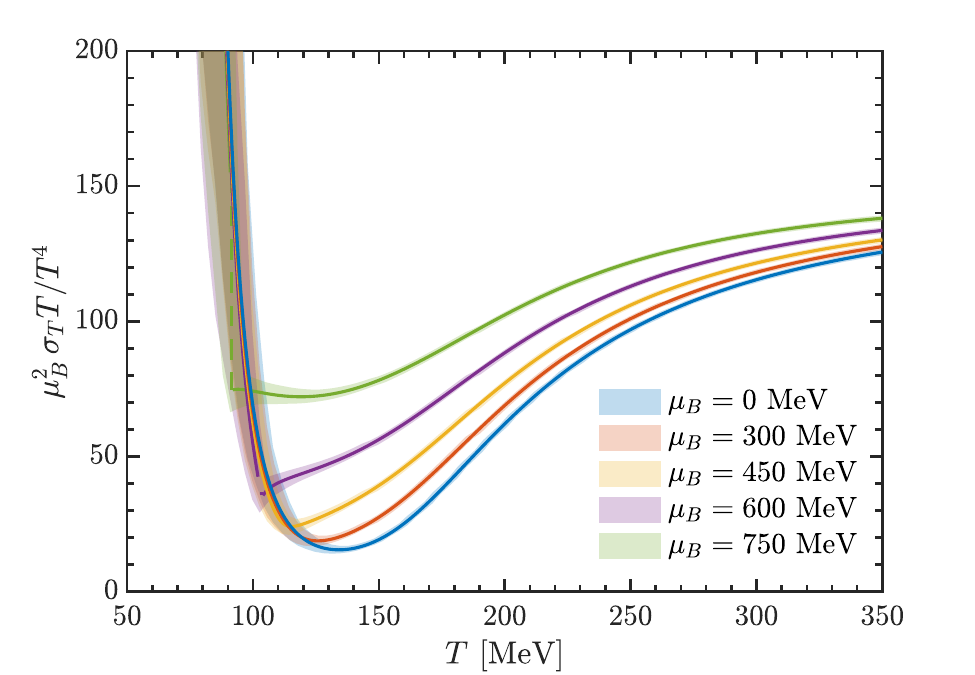}
    \end{subfigure}
    \hfill
    % ===== Right panel =====
    \begin{subfigure}[t]{0.49\linewidth}
        \centering
\includegraphics[width=\linewidth]
{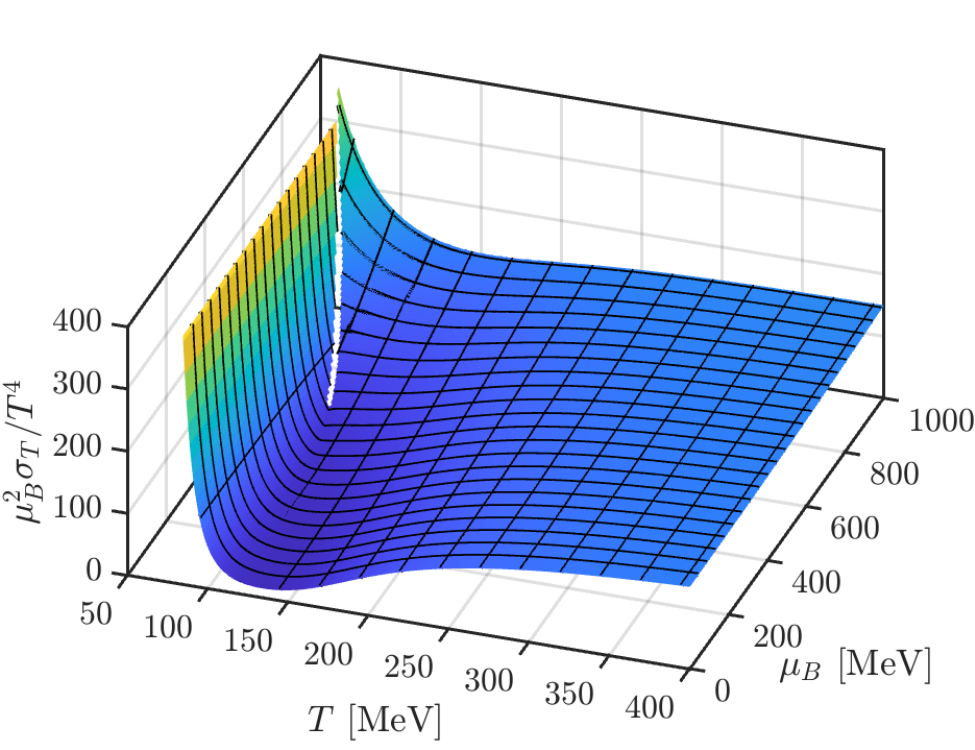}
    \end{subfigure}
\caption{
The left panel shows posterior bands at 95\% confidence level, and corresponding best-fit curves (solid lines) of holographic thermal conductivity as a dimensionless combination given in Eq.~\eqref{eq:sigmaT}  as functions of $T$ for different values of $\mu_B$. The right panel shows the same observable as a function of $T$ and $\mu_B$ computed from the best-fit parameterization.}
    \label{fig:Thermal_conductivity}
\end{figure}

Finally, in Fig.~\ref{fig:Thermal_conductivity}, we show our Bayesian results for the thermal conductivity.
The dimensionless holographic thermal conductivity reveals a minimum within the temperature range $T = 100 - 150~\mathrm{MeV}$
for $\mu_B = 0$. With increasing $\mu_B$ the minima move toward the critical point and turn into a cusp. Beyond the critical point, this observable develops a small discontinuity gap along the first-order phase transition line.

%%%%%%%%%%%%%%%%%%%%%%%%%%%%%%%%%
\subsection{Shear and bulk viscosities}
\label{sec:bulk}

The shear viscosity  represents a measure of the resistance of the fluid to sheared flow. It can be obtained from the analysis of the diffeomorphism and gauge invariant perturbation $Z(r,\omega)$ in the $SO(3)$ quintuplet channel of the EMD model, corresponding to any of the five independent spatial components of the graviton perturbation~\cite{DeWolfe:2011ts}.\footnote{The spatial components of the graviton field, corresponding to a symmetric and traceless perturbation $h_{\mu\nu}$ of the bulk metric field $g_{\mu\nu}$, are given by (in the radial gauge, $h_{\mu r}=0$): $h_{xy}$, $h_{xz}$, $h_{yz}$, $h_{xx}$, $h_{yy}$, and $h_{zz}$. Since the graviton field $h_{\mu\nu}$ is traceless, one may express one of the diagonal components of this perturbation in terms of the others, so that there are just five independent spatial components $h_{ij}$ in five dimensions.} This perturbation satisfies the equation of motion for a massless scalar field on top of the numerical EMD backgrounds. It has been shown in~\cite{DeWolfe:2011ts} that the shear viscosity for the EMD model satisfies $\eta/s = 1/4\pi$ for all values of temperature $T>0$ and chemical potential $\mu_B\ge 0$, as previously stated. However, as mentioned before, the dimensionless combination involving the shear viscosity that naturally appears in hydrodynamics at finite baryon density is $\eta T/(\epsilon+P)$, which reduces to $\eta/s$ in the limit of zero chemical potential \cite{Liao:2009gb,Denicol:2013nua}. This dimensionless combination presents a non-trivial behavior as a function of $(T,~\mu_B)$ in holographic models at finite baryon density. In fact, taking into account that in the EMD model $\eta/s=1/4\pi$, it follows that
\begin{align}
    \left[\frac{\eta\, T}{\epsilon+P}\right](T,\mu_{B})&=\frac{1}{4\pi\left(1+\frac{\mu_{B}\rho_{B}}{Ts}\right)},\label{eq:eta}\\
    &\to \frac{1}{4\pi}=\frac{\eta}{s}, \,\,\, \textrm{as} \,\,\, \mu_B\to 0.\label{eq:eta0}
\end{align}
Note that~\eqref{eq:eta} can be computed solely in terms of the basic thermodynamic quantities given in Eqs.~\eqref{eq:T} ---~\eqref{eq:rhoB}.

The small values of the ratio $\eta/s\sim 1/4\pi\approx 0.08$ --- $0.2$ within the temperature window associated to the QCD deconfinement crossover at $\mu_B\sim 0$ is the main fingerprint of the nearly-perfect fluidity associated to the strongly coupled behavior of the QGP produced at current collider energies. This was recently confirmed by the JETSCAPE Collaboration, through a  Bayesian analysis that uses a hybrid phenomenological model matching several high energy heavy-ion data~\cite{JETSCAPE:2020shq,JETSCAPE:2020mzn}. 

One can see in Fig.~\ref{fig:shear_viscosity} that $\eta T/(\epsilon+P)$ in the EMD model decreases with increasing baryon chemical potential, indicating that the strongly coupled medium further approaches the nearly-perfect fluidity in the baryon dense regime. This behavior is characterized by the presence of a minimum and an inflection point which moves towards the CEP. Beyond the critical end point, there is a discontinuity gap signaling the phase transition taking place in the medium, as already seen in the other observables.

\begin{figure}%[H]
    \centering
    % ===== Left panel =====
    \begin{subfigure}[t]{0.49\linewidth}
        \centering
        \includegraphics[width=\linewidth]{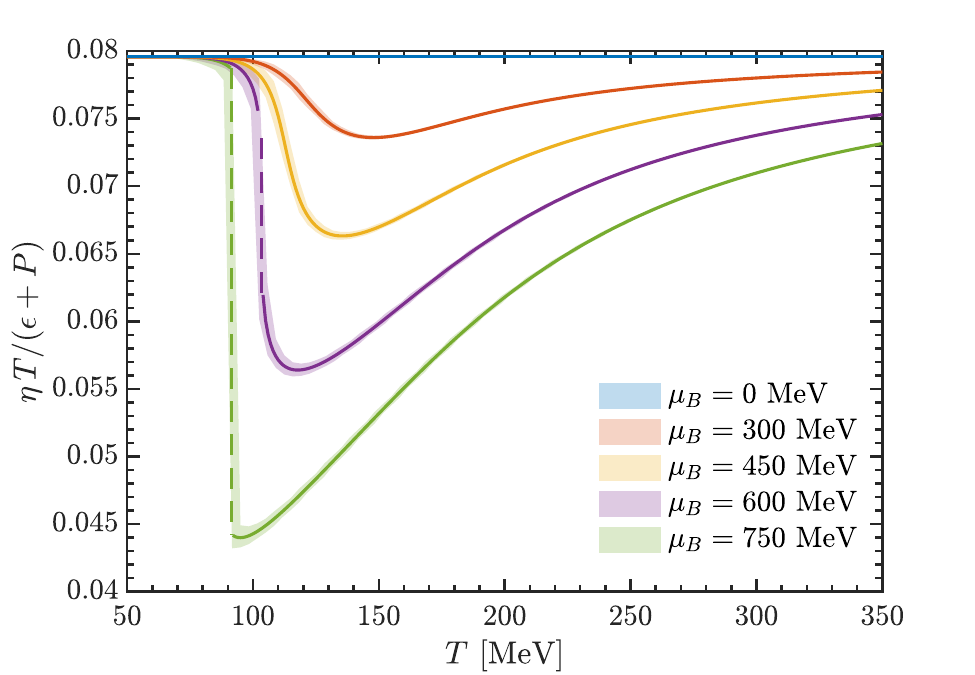}
        \label{fig:shear_viscosity}
    \end{subfigure}
    \hfill
    % ===== Right panel =====
    \begin{subfigure}[t]{0.49\linewidth}
        \centering
\includegraphics[width=\linewidth]{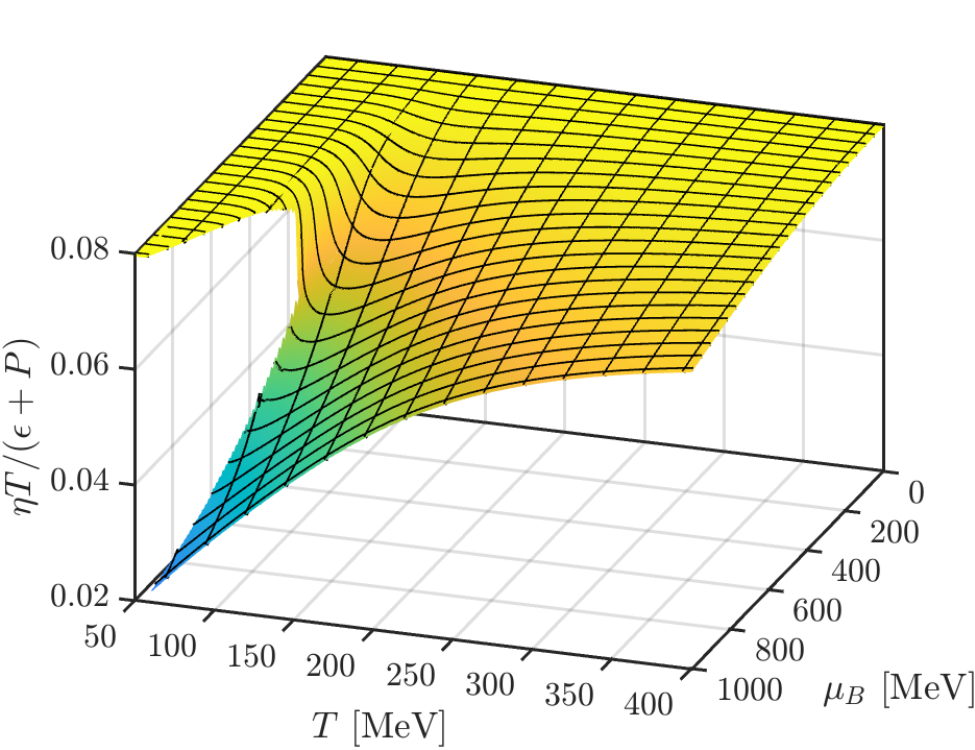}
\label{fig:shear_viscosity3d}
    \end{subfigure}
\caption{The left panel shows posterior bands at 95\% confidence level, and corresponding best-fit curves (solid lines) of holographic shear viscosity times temperature over enthalpy density obtained from Eq.~\eqref{eq:eta} for fixed $\mu_B$ as functions of temperature $T$, whereas the right panel presents a surface plot of the same observable obtained by using the best-fit parameterization.}
\label{fig:shear_viscosity}
\end{figure}

Bulk viscosity measures the resistance of the fluid against compressions or expansions. It can be obtained from the analysis of the diffeomorphism and gauge invariant perturbation $\mathcal{H}(r,\omega)$ in the $SO(3)$ singlet channel of the EMD model, which corresponds to a combination of the spatial trace of the graviton and the dilaton perturbation. This perturbation satisfies the following equation of motion~\cite{DeWolfe:2011ts}, 
\begin{align}
\mathcal{H}''+\left[4A'+\frac{h'}{h}+\frac{2\phi''}{\phi'}-\frac{2A''}{A'}\right]\mathcal{H}'+\left[\frac{e^{-2A}\omega^{2}}{h^{2}}+\frac{h'}{h}\left(\frac{A''}{A'}-\frac{\phi''}{\phi'}\right)+\frac{e^{-2A}}{h\phi'}\left(3A'\partial_\phi f(\phi)-f(\phi)\phi'\right)\Phi'^{2}\right]\mathcal{H}=0,
\label{eq:EOM-H}
\end{align}
which is to be solved with infalling boundary condition at the event horizon, with the solution being normalized to unity at the boundary. These two boundary conditions are implemented via~\cite{Grefa:2022sav},
\begin{align}
\mathcal{H}(r,\omega)\equiv\frac{r^{-i\omega}F(r,\omega)}{r_{\textrm{max}}^{-i\omega}\,F(r_{\textrm{max}},\omega)},
\label{eq:HBdyCond}
\end{align}
where $F(r,\omega)$ must be a regular function at the event horizon, with its equation of motion being obtained by substituting \eqref{eq:HBdyCond} into \eqref{eq:EOM-H}. The holographic formula for the dimensionless ratio between bulk viscosity and entropy density in the EMD model reads as follows~\cite{DeWolfe:2011ts,Rougemont:2017tlu,Grefa:2022sav}
\begin{align}
\left[\frac{\zeta}{s}\right](T,\mu_{B})=-\frac{1}{36\pi}\lim_{\omega\rightarrow0}\frac{1}{\omega}\left(\frac{e^{4A(r)}h(r)\phi'(r)^{2}\,\textrm{Im}[\mathcal{H}^{*}(r,\omega)\mathcal{H}'(r,\omega)]}{A'(r)^{2}}\right)\biggr|_\textrm{on-shell},
\label{eq:HKubo}
\end{align}
where the term between brackets is a radially conserved flux which can then be computed at any value of $r$~\cite{DeWolfe:2011ts}. Similar observations to those below Eq.~\eqref{eq:conduct} also apply here.

From~\eqref{eq:HKubo}, by using the fundamental thermodynamic relation $\epsilon=Ts+\mu_B\rho_B-P$, and the relation $\eta/s=1/4\pi$, which is valid for the holographic EMD model, one obtains the following dimensionless combination
\begin{align}
    \left[\frac{\zeta T}{\epsilon+P}\right](T,\mu_{B})&=\left[\frac{\zeta}{s}\right](T,\mu_{B})\frac{1}{1+\frac{\mu_{B}\rho_{B}}{Ts}},\label{eq:zeta}\\
&\to \frac{\zeta}{s}, \,\,\, \textrm{as} \,\,\, \mu_B\to 0,\label{eq:zeta0}
\end{align}
which is naturally featured in the hydrodynamic expression for the bulk viscous pressure of charged fluids.

Together with the shear viscosity, bulk viscosity plays an important role in the description of the QGP, affecting e.g.~the azimuthal momentum anisotropy, the transverse momentum spectra, and the multiplicity of charged hadrons produced in heavy-ion collisions \cite{Noronha-Hostler:2013gga,Noronha-Hostler:2014dqa,Ryu:2015vwa,Bernhard:2016tnd,Bernhard:2019bmu,JETSCAPE:2020mzn,JETSCAPE:2020shq}. In Fig.~\ref{fig:bulk_viscosity}, we present the results for the best-fit parametrization of the EMD model for the dimensionless combination~\eqref{eq:zeta} along with its Bayesian uncertainty bands from the posterior distribution. Similar to shear viscosity, the overall behavior is characterized by a decrease of bulk viscosity with increasing baryon density in the medium. This prediction is also present in previous holographic EMD setups \cite{DeWolfe:2011ts, Rougemont:2017tlu, Grefa:2022sav}, making it a recurrent aspect of such holographic models. Another feature that appears in this EMD model is the shifting of the peak towards increasing temperature as $\mu_B$ increases, both in the polynomial-hyperbolic Ansatz, as shown in Fig.~\ref{fig:bulk_viscosity}, and in the parametric Ansatz, 
% for the free functions $V(\phi)$ and $f(\phi)$ 
as shown in Fig.~\ref{fig:PA_zeta} in Appendix~\ref{sec:app}. The peak height becomes smaller with increasing chemical potential, exhibiting therefore a suppression of viscosity effects in the baryon dense regime. This aspect is also present in \cite{Rougemont:2017tlu, Grefa:2022sav}, but not in \cite{DeWolfe:2011ts}, where the peak height in the bulk viscosity remains nearly the same as one increases the baryon density. This highlights the model-dependent aspect of the bulk viscosity in the EMD setup.

The normalized bulk viscosity develops a dip at finite baryon density which moves towards the CEP, located at $\mu_B=598\,\text{MeV}$. At the CEP, it acquires an infinite slope, similar to what happens with the shear viscosity. Beyond the critical point, across the first order line, there is a discontinuity gap as well, signaling the phase transition in this observable.

One can also notice that the peak present at $\mu_B=0$ seems to be a robust feature of this observable compared to other Bayesian analyses of the quantity $\zeta/s$, such as those from the JETSCAPE Collaboration~\cite{JETSCAPE:2020mzn,JETSCAPE:2020shq} and the DUKE Group~\cite{Bernhard:2019bmu}, illustrated in Fig.~\ref{fig:JetScape}, where we compare the best-fit curve of the present EMD model and its Bayesian uncertainty band with the Bayesian analyses of those groups. It is important to remark that, while our result for $\zeta/s$ displayed in this comparison is a genuine prediction coming from a microscopic holographic calculation, the Bayesian profiles shown from the JETSCAPE Collaboration and the Duke Group are not microscopic calculations or predictions for $\zeta/s$, being instead adjusted by matching the results from the underlying hybrid phenomenological models to several different heavy-ion experimental data. Therefore, the overall good agreement shown between the EMD prediction for $\zeta/s$ and the hybrid model Bayesian results further strengthens the phenomenological reliability of the holographic EMD setup.

\begin{figure}%[H]
    \centering
    % ===== Left panel =====
    \begin{subfigure}[t]{0.49\linewidth}
        \centering
        \includegraphics[width=\linewidth]
        {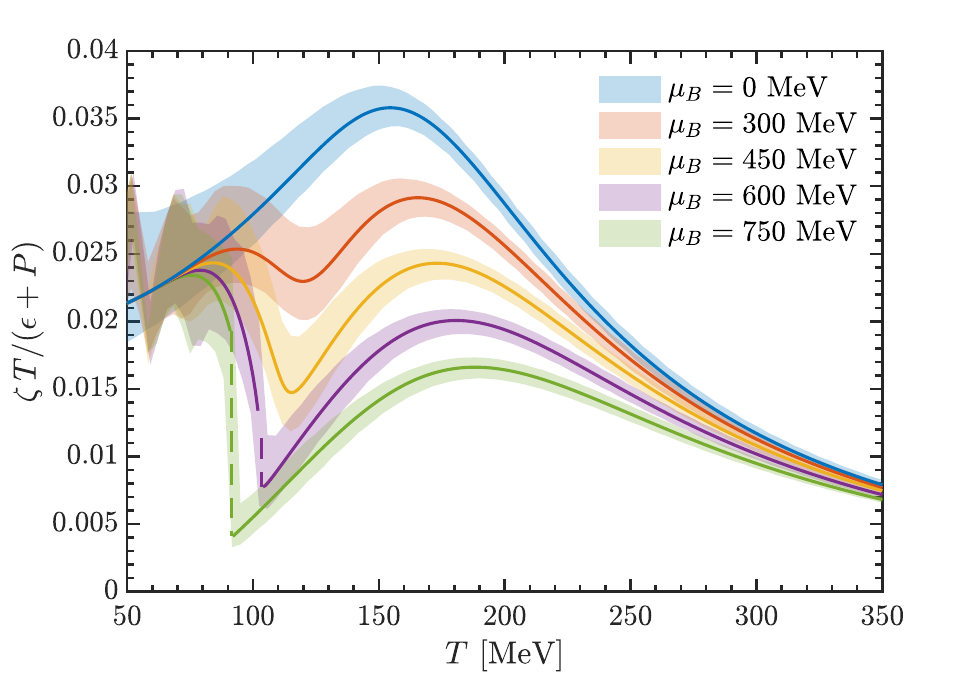}
        \label{fig:bulk_viscosity}
    \end{subfigure}
    \hfill
    % ===== Right panel =====
    \begin{subfigure}[t]{0.49\linewidth}
        \centering
\includegraphics[width=\linewidth]
{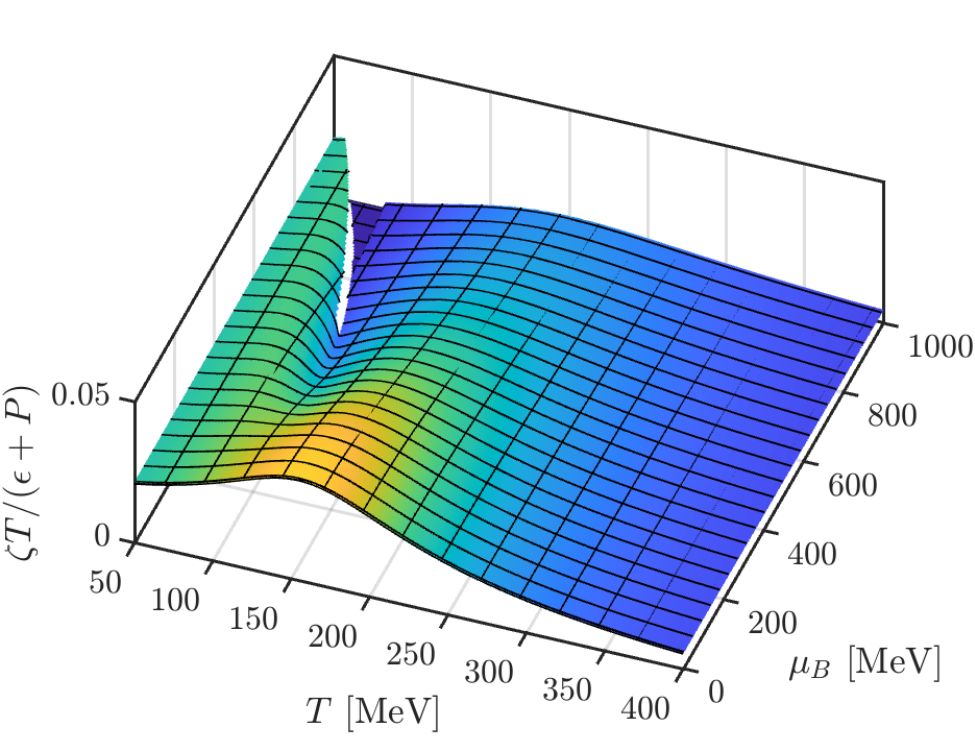}
\label{fig:bulk_viscosity3d}
    \end{subfigure}
\caption{Holographic bulk viscosity, normalized as ${\zeta T}/({\epsilon + P})$. Posterior bands at 95\% confidence level along with best-fit curves (solid lines) for several values of $\mu_B$ (left) and the same observable as a function of $T$ and $\mu_B$ generated from the best-fit parameterization (right).}
\label{fig:bulk_viscosity}
\end{figure}

\begin{figure}%[H]
    \centering
    % ===== Left panel =====
    \begin{subfigure}[t]{0.49\linewidth}
        \centering
        \includegraphics[width=\linewidth]
        {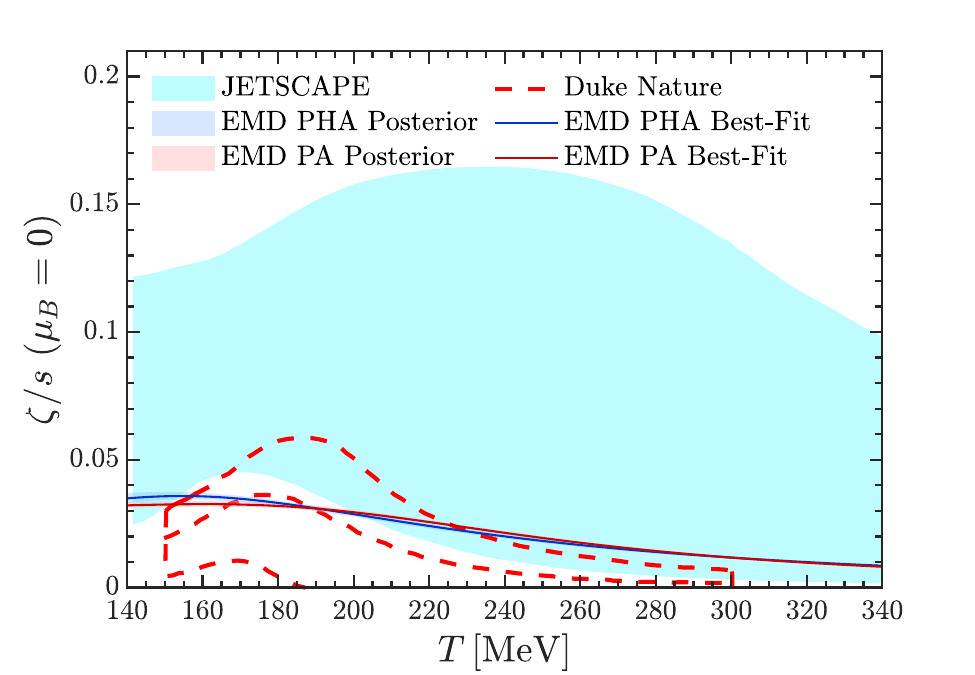}
    \end{subfigure}
    \hfill
    % ===== Right panel =====
    \begin{subfigure}[t]{0.49\linewidth}
        \centering
\includegraphics[width=\linewidth]
{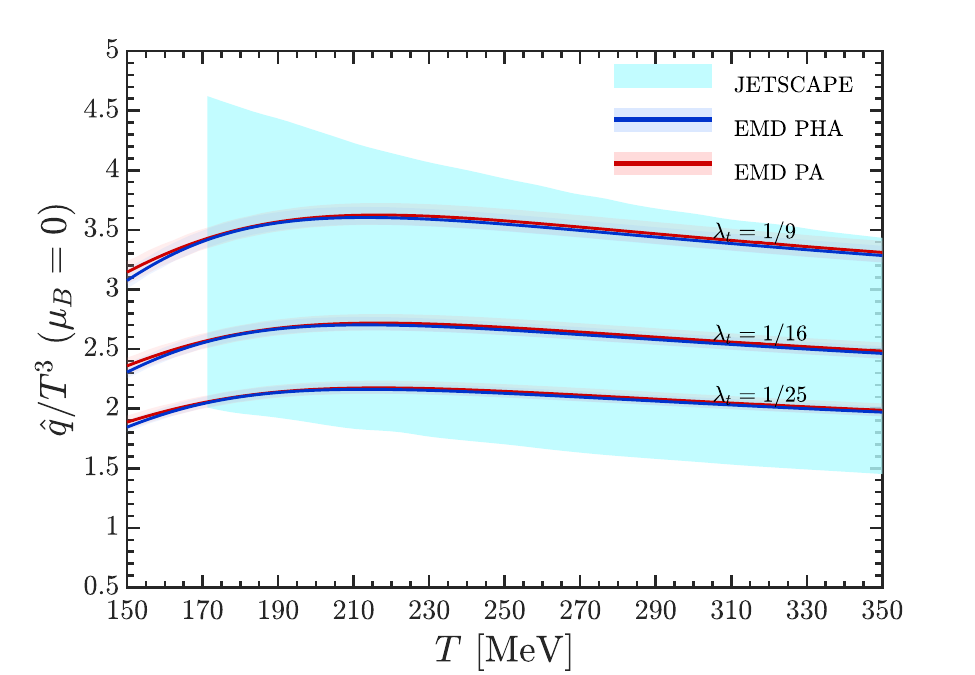}
    \end{subfigure}
\caption{Left panel: Holographic bulk viscosity $\zeta/s$ Bayesian posterior band at 95\% confidence intervals, and best fit (solid line) for PA and PHA ansatze both at $\mu_B$ = 0 compared with the 90\% credible intervals of the JETSCAPE Bayesian
model from \cite{JETSCAPE:2020shq,JETSCAPE:2020mzn}, and the Duke Group results \cite{Bernhard:2019bmu}.
Right panel: Jet-quenching parameter Bayesian posterior band at 95\% confidence level, and best fit (solid line) from EMD model for PA and PHA ansatze at $\mu_B =0$ for three representative values of the `t Hooft parameter $\lambda_t = \{1/25, 1/16, 1/9\}$, compared to the JETSCAPE result from Ref.~\cite{JETSCAPE:2020shq}.
}
    \label{fig:JetScape}
\end{figure}
%\newpage

%%%%%%%%%%%%%%%%%%%%%%%%%%%%%%%%%
\subsection{Energy loss: jet-quenching parameter, heavy quark drag force and Langevin diffusion coefficients}
\label{sec:loss}

The energy loss experienced by a heavy quark traversing a strongly coupled medium can be computed in holography by using the trailing string method, originally proposed for conformal models in~\cite{Gubser:2006bz,Herzog:2006gh} and later generalized for non-conformal dilatonic models in~\cite{Gursoy:2009kk}. In the trailing string method, a probe heavy quark traveling within the medium at a constant velocity $v$ along, say, the $x$ direction, is holographically modeled by an open string where one of the string endpoints is attached to the heavy quark at the boundary, while the rest of the string trails behind it within the higher dimensional bulk, with the other string endpoint being attached to a two-dimensional event horizon formed on top of the string worldsheet~\cite{Gubser:2006nz,Casalderrey-Solana:2007ahi,Gursoy:2010aa}. As the heavy quark traverses the strongly coupled medium at the boundary, it loses energy and momentum due to the drag force $F_\textrm{drag}=dp_{x}/dt$, which is calculated via the energy flow $dE/dx$ from the string endpoint at the boundary towards the string endpoint at the worldsheet event horizon in the interior of the higher dimensional bulk.

\noindent It was shown in~\cite{Rougemont:2015wca} that the heavy quark drag force for the isotropic EMD model can be written as
\begin{equation}\label{eq:Fdrag}
\left[\frac{F_{\textrm{drag}}}{\sqrt{\lambda_{t}}T^{2}}\right](T,\mu_{B};v)=-8\pi \,v\,h_{0}^{\textrm{far}}\,e^{\sqrt{2/3}\,\phi(r_{*})+2A(r_{*})},
\end{equation}
where $r_{*}$ is the radial location of the string worldsheet event horizon given by the solution of the equation \cite{Rougemont:2015wca}
\begin{equation}\label{eq:r_star}
    h(r_{*})=h_{0}^{\textrm{far}}v^{2}.
\end{equation}
In the conformal limit of high temperatures, the EMD background solutions tend to the AdS$_{5}$-Schwarzschild black hole metric. Consequently, the conformal limit of~\eqref{eq:Fdrag} tends to the SYM result~\cite{Gubser:2006bz,Herzog:2006gh},
\begin{equation}\label{eq:FdragCFT}
    \lim_{T\rightarrow\infty} \left[\frac{F_{\textrm{drag}}}{\sqrt{\lambda_{t}}T^{2}}\right](T,\mu_{B};v)\to-\frac{\pi\gamma(v)v}{2}=-\frac{\pi v}{2\sqrt{1-v^{2}}}.
\end{equation}
%==drag force plot v=0.5
\begin{figure}[H]
    \centering
    % ===== Top Row =====
    \begin{subfigure}[t]{0.49\linewidth}
        \centering
        \includegraphics[width=\linewidth]{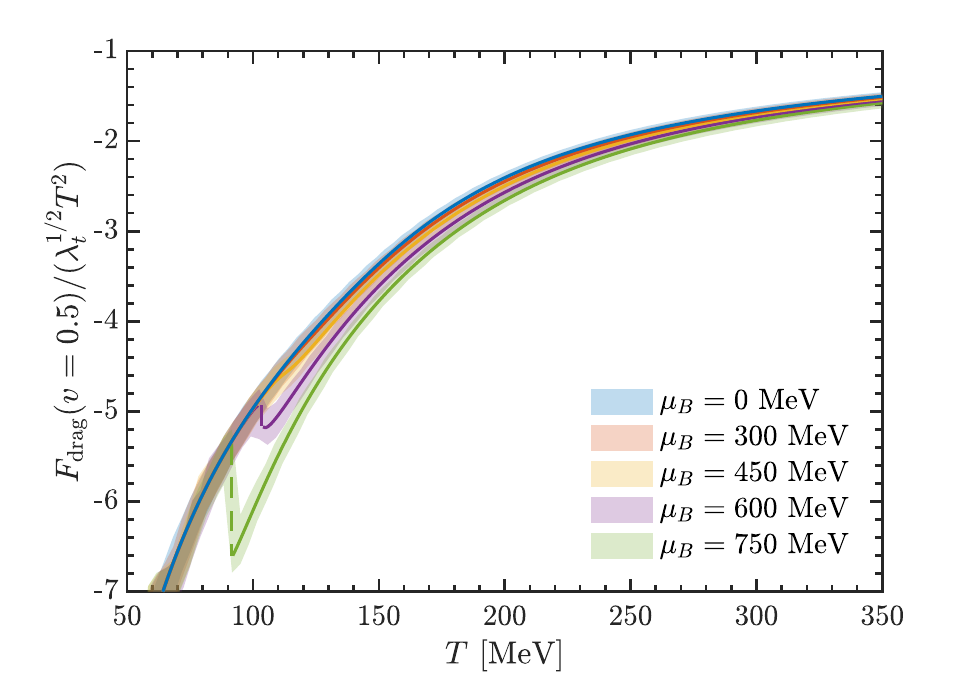}
        \label{fig:Drag_Force}
    \end{subfigure}
    \hfill
    \begin{subfigure}[t]{0.49\linewidth}
        \centering
        \includegraphics[width=\linewidth]{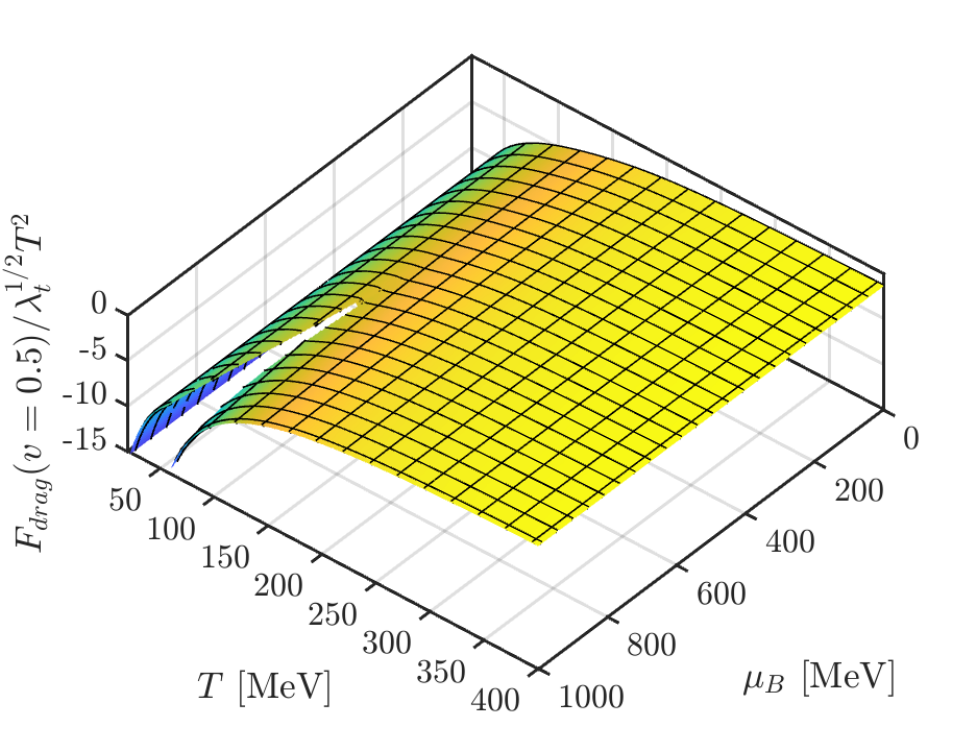}
        \label{fig:Drag_Force3D0.5}
    \end{subfigure}
     \caption{ Heavy quark drag force at $v=0.5$. The
     left panel displays posterior bands at 95\% confidence level, and corresponding best-fit curves (solid lines) as functions of $T$ keeping $\mu_B$ constant and the right panel shows the same observable as a function of $T$ and $\mu_B$ using the best-fit parameterization.}
    \label{fig:BaysDforce0.5}
\end{figure}
%==drag force plot v=0.99
\begin{figure}[b]
    \centering
    \vspace{1em}
    \begin{subfigure}[t]{0.49\linewidth}
        \centering
        \includegraphics[width=\linewidth]{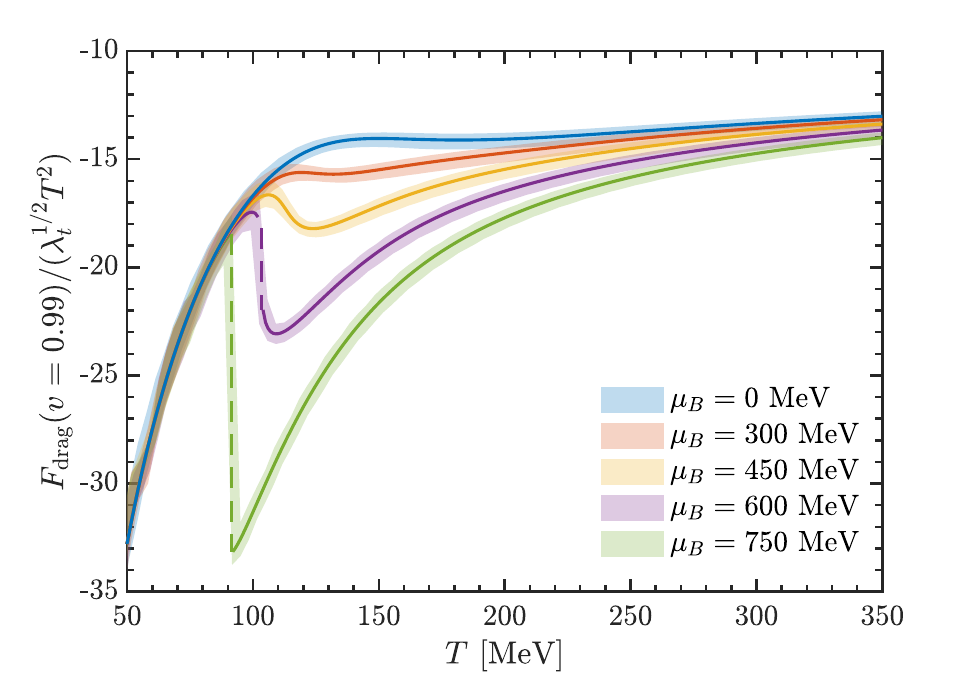}
        \label{fig:Drag_Force099}
    \end{subfigure}
    \hfill
    \begin{subfigure}[t]{0.49\linewidth}
        \centering
        \includegraphics[width=\linewidth]{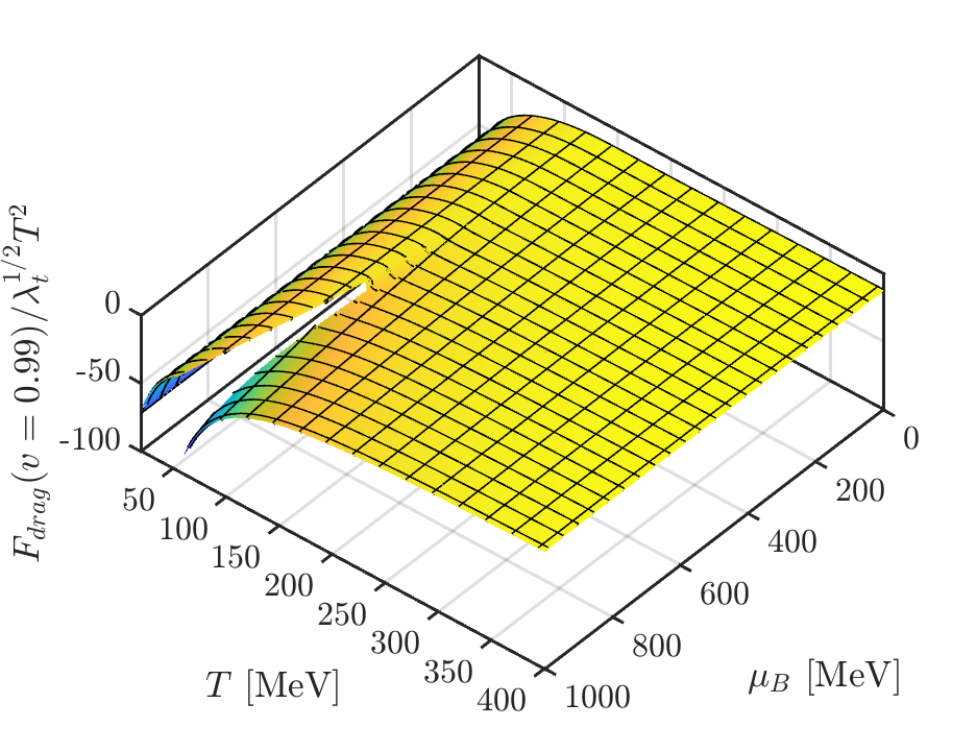}
        \label{fig:Drag_Force3D099}
    \end{subfigure}

    \caption{Heavy quark drag force at $v=0.99$. The
     left panel displays posterior bands at 95\% confidence level, and corresponding best-fit curves (solid lines) as functions of $T$ keeping $\mu_B$ constant and the right panel shows the same observable as a function of $T$ and $\mu_B$ using the best-fit parameterization.}
    \label{fig:BaysDforce0.99}
\end{figure}
%====

In Fig.~\ref{fig:BaysDforce0.5} and Fig.~\ref{fig:BaysDforce0.99} we show our Bayesian results for the heavy quark drag force at finite temperature and baryon chemical potential. 
The magnitude of the drag force, which quantifies the energy loss of a heavy quark moving through the plasma, increases with the medium’s baryon density. This enhancement becomes more pronounced at larger quark velocities. When comparing, for instance, $v=0.5$ and $v=0.99$, one clearly observes that the separation between curves at fixed $\mu_B$ becomes more visible for higher velocities. Although our calculation employs the large-mass (probe) limit, where the quark mass does not explicitly enter, the qualitative trend suggests that quarks capable of reaching higher velocities, thus less massive ones, would experience a stronger sensitivity to the baryon density of the surrounding medium.
In both scenarios, the magnitude of the drag force also increases as the temperature of the medium lowers down. In particular, the predictions of both the Bayesian bands and their best fit curves show that the fixed $\mu_B$ curves develop a dip in the crossover region that increases towards the CEP. Across the first order line, as already seen in the other transport coefficients, there is also a discontinuity gap indicating that the system undergoes a phase transition.

\vspace{15pt}

The holographic method for treating stochastic diffusion processes was originally proposed for conformal models in~\cite{Gubser:2006nz,Casalderrey-Solana:2007ahi}, and later extended for non-conformal dilatonic setups in~\cite{Gursoy:2010aa}. In the trailing string picture, the effects of thermal fluctuations on the propagation of a heavy probe quark are neglected. On the other hand, Refs.~\cite{Gubser:2006nz,Casalderrey-Solana:2007ahi,Gursoy:2010aa} took into account the Brownian motion of the heavy quark as it traverses the medium, which can be approximated by a linearized local Langevin equation describing the thermal fluctuations in the heavy quark trajectory. By the standards of the holographic dictionary, from the minimal coupling interaction for a probe heavy quark at the boundary, $\int dt \,\delta X^\mu(r\to\infty,t)F_\mu(t)$, one can identify the boundary value of the fluctuating trailing string worldsheet embedding, $\delta X^\mu(r\to\infty,t)$, as the source for a gauge invariant QFT operator at the boundary, $F_\mu(t)$, which effectively describes a stochastic force acting upon the heavy quark. The momentum dependent Langevin diffusion coefficients are thus computed from the symmetrized real time two-point correlation function of this boundary QFT operator. Here, we consider the zero frequency limit of these coefficients, describing the long time behavior of the stochastic diffusion process within the hot and baryon dense medium.

The EMD holographic formulas for the heavy quark Langevin diffusion coefficients, perpendicular and parallel to the heavy quark velocity, are given by~\cite{Rougemont:2015wca}
\begin{align}
    \left[\frac{\kappa_{\perp}}{\sqrt{\lambda_{t}}T^{3}}\right](T,\mu_{B};v)&=16\pi\, v\,(h_{0}^{\textrm{far}})^{3/2}\,e^{\sqrt{2/3}\,\phi(r_{*})+3A(r_{*})}\,\sqrt{h'(r_{*})\left[4A'(r_{*})+\sqrt{\frac{8}{3}}\phi'(r_{*})+\frac{h'(r_{*})}{h(r_{*})}\right]},\label{eq:kappaPerp}\\
    \left[\frac{\kappa_{\parallel}}{\sqrt{\lambda_{t}}T^{3}}\right](T,\mu_{B};v)&=16\pi \,v^{3}\,(h_{0}^{\textrm{far}})^{5/2}\,\frac{e^{\sqrt{2/3}\,\phi(r_{*})+3A(r_{*})}}{h'(r_{*})^2}\left(h'(r_{*})\left[4A'(r_{*})+\sqrt{\frac{8}{3}}\phi'(r_{*})+\frac{h'(r_{*})}{h(r_{*})}\right]\right)^{3/2},\label{eq:kappaPar}
\end{align}
and the corresponding conformal limits of high temperatures are given by \cite{Gubser:2006nz,Casalderrey-Solana:2007ahi}
\begin{align}
    \lim_{T\to\infty}\left[\frac{\kappa_{\perp}}{\sqrt{\lambda_{t}}T^{3}}\right](T,\mu_B;v)&\to\pi\gamma(v)^{1/2}=\frac{\pi}{(1-v^{2})^{1/4}},\\
      \lim_{T\to\infty}\left[\frac{\kappa_{\parallel}}{\sqrt{\lambda_{t}}T^{3}}\right](T,\mu_B;v)&\to\pi\gamma(v)^{5/2}=\frac{\pi}{(1-v^{2})^{5/4}}.
\end{align}

As discussed in detail in~\cite{Gursoy:2010aa}, one can define two velocity-dependent jet-quenching parameters as follows
\begin{align}\label{eq:heavyqhats}
\hat{q}_\perp(T,\mu_B;v)=\frac{\langle p_\perp^2 \rangle}{vt}=\frac{2\kappa_\perp(T,\mu_B;v)}{v}, \qquad \hat{q}_\parallel(T,\mu_B;v)=\frac{\langle \Delta p_\parallel^2 \rangle}{vt}=\frac{\kappa_\parallel(T,\mu_B;v)}{v},
\end{align}
where $\langle p_\perp^2 \rangle$ and $\langle \Delta p_\parallel^2 \rangle$ correspond to linear approximations for, respectively, the noise-averaged transverse and longitudinal momentum fluctuations of a probe heavy quark after it traversed a distance $vt$ within the medium. In particular, $\hat{q}_\perp(T,\mu_B;v)$ gives the transverse momentum broadening of a probe heavy quark traveling with velocity $v$ through the medium \cite{Gursoy:2010aa,Gubser:2006nz}. 
%====k_perp
\begin{figure}[h!]
    \centering
    % ===== Row 1 =====
    \begin{subfigure}[b]{0.45\linewidth}
        \centering
        \includegraphics[width=\linewidth]{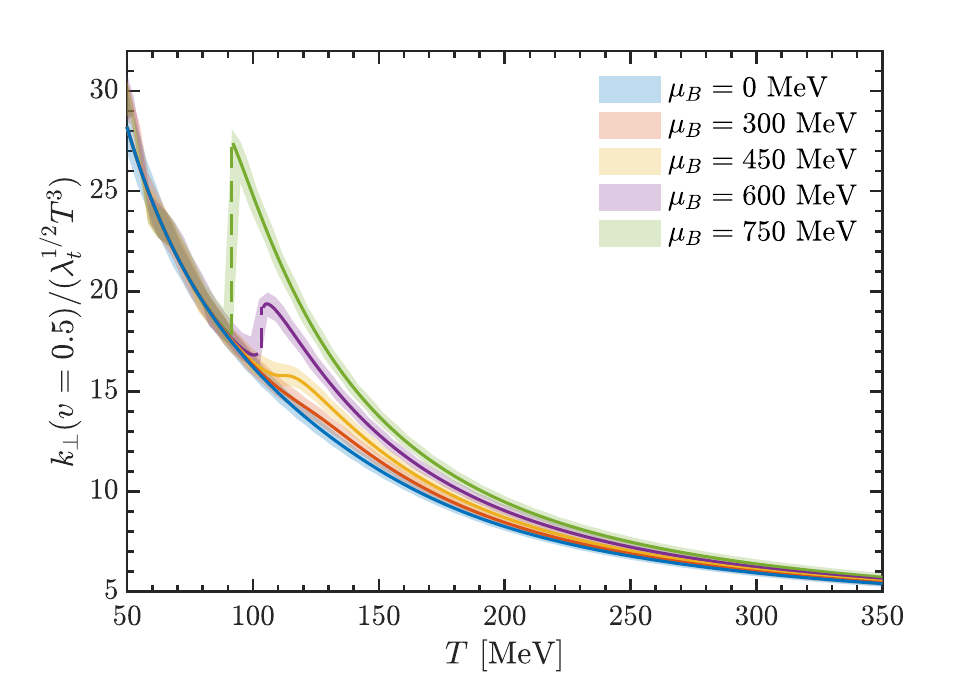}
        \label{fig:kperp05}
    \end{subfigure}
    \hfill
    \begin{subfigure}[b]{0.45\linewidth}
        \centering
        \includegraphics[width=\linewidth]{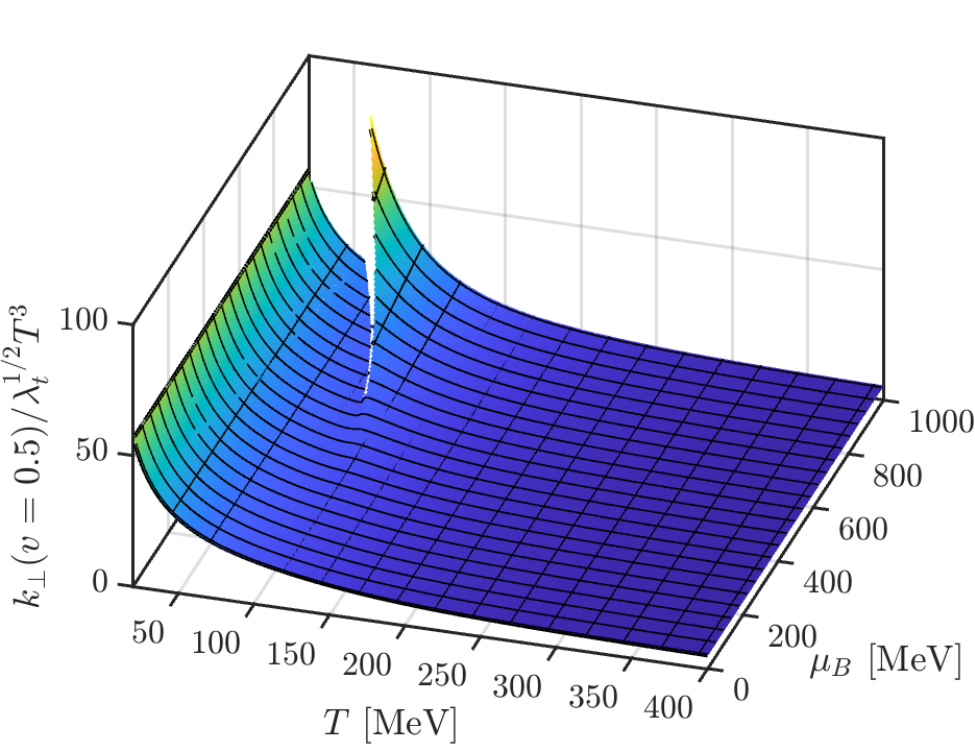}
        \label{fig:kperp3D05}
    \end{subfigure}
    % ===== Row 2 =====
    \begin{subfigure}[b]{0.45\linewidth}
        \centering
        \includegraphics[width=\linewidth]{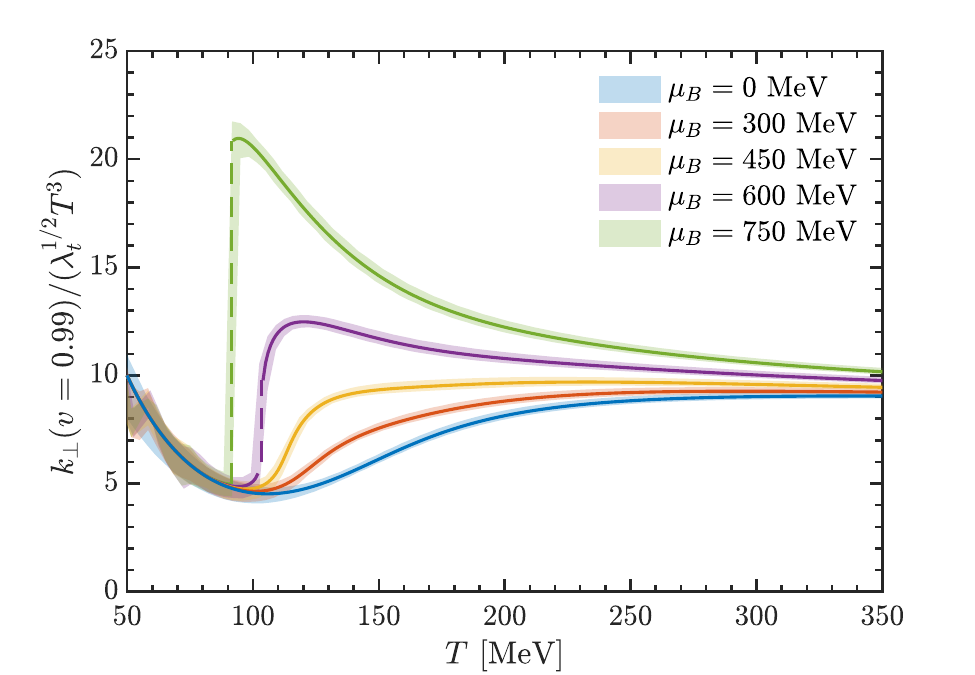}
        \label{fig:kperp099}
    \end{subfigure}
    \hfill
    \begin{subfigure}[b]{0.45\linewidth}
        \centering
        \includegraphics[width=\linewidth]{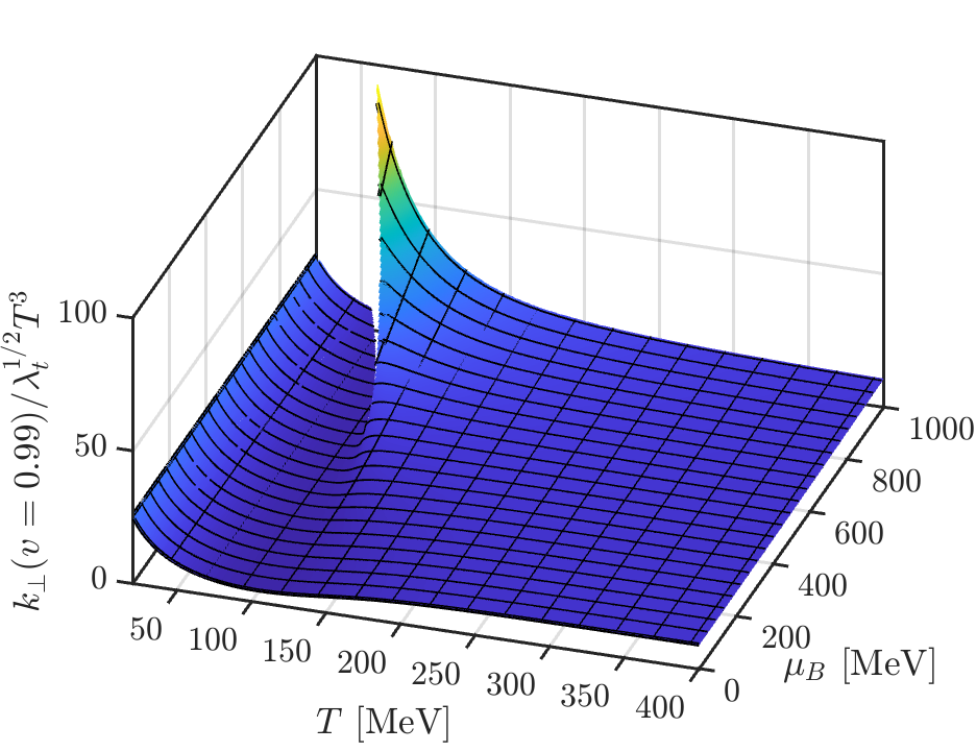}
        \label{fig:kperp3D099}
    \end{subfigure}
    \caption{Perpendicular Langevin diffusion coefficient $\kappa_{\perp}$ obtained by using Eq.~\eqref{eq:kappaPerp}. The panels on the left show posterior bands at 95\% confidence level, and the best-fit curves (solid lines) as functions of $T$ for fixed $\mu_B$, with $v=0.5$ (top left) and $v=0.99$ (bottom left). The panels on the right display $\kappa_{\perp}$ surface plots as a function of $T$ and $\mu_B$ using the best-fit parameterization, with $v=0.5$ (top right) and $v=0.99$ (bottom right).}
    \label{fig:Bayskperp}
\end{figure}
%=====k_||
\begin{figure}[h!]
    \centering
    % ===== Row 1 =====
    \begin{subfigure}[b]{0.45\linewidth}
        \centering
        \includegraphics[width=\linewidth]{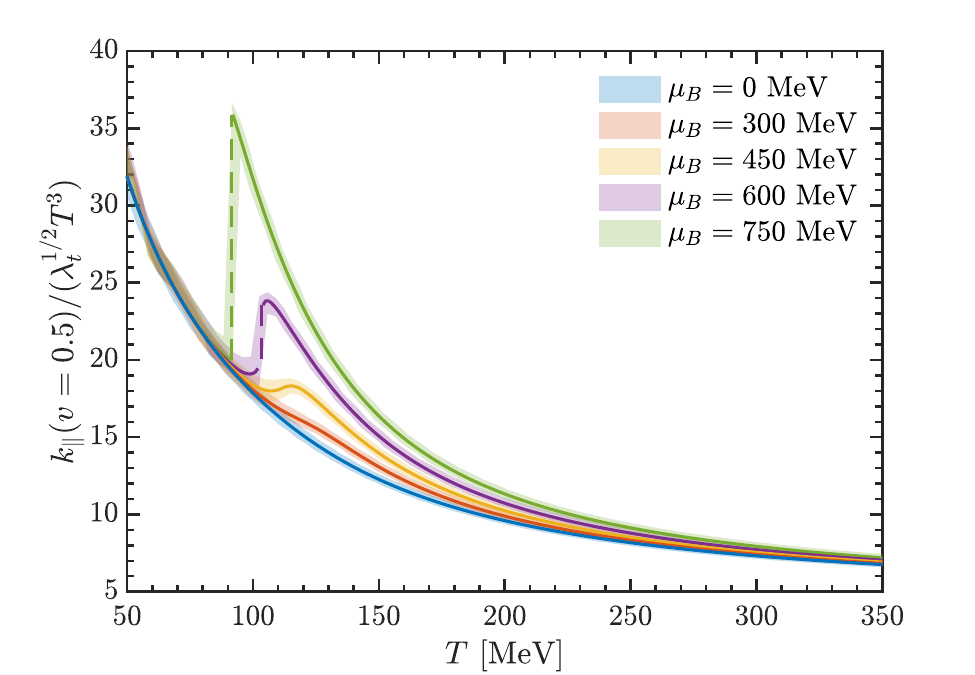}        \label{fig:kparallel05}
    \end{subfigure}
    \hfill
    \begin{subfigure}[b]{0.45\linewidth}
        \centering
        \includegraphics[width=\linewidth]{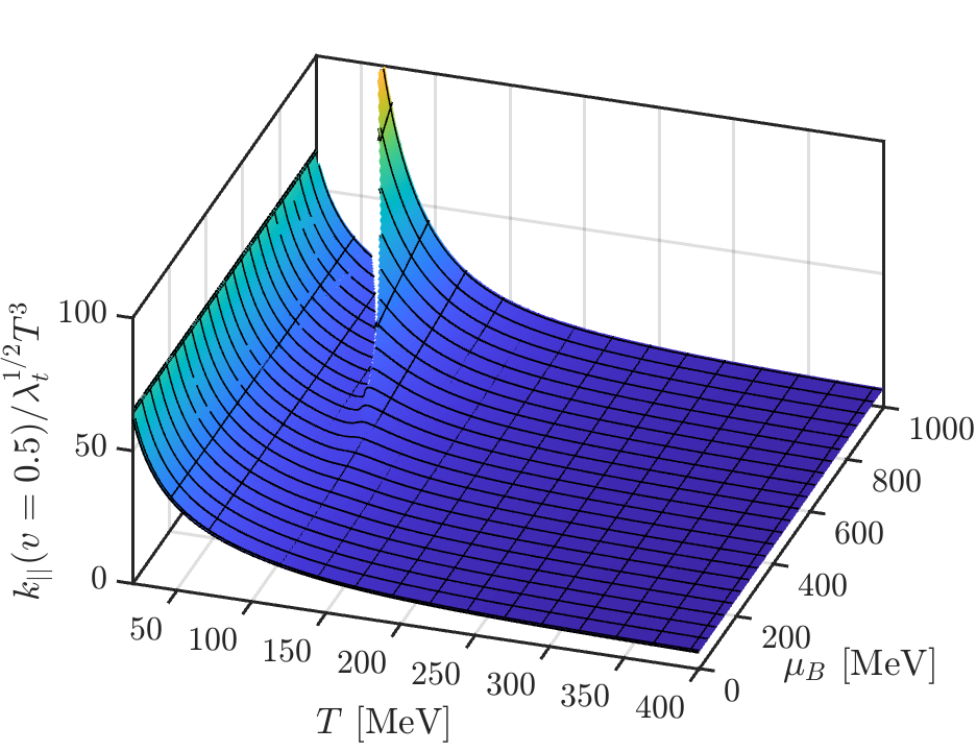}
        \label{fig:kparallel3D05}
    \end{subfigure}

    % ===== Row 2 =====
    \vspace{-1cm}
    \begin{subfigure}[b]{0.45\linewidth}
        \centering
        \includegraphics[width=\linewidth]{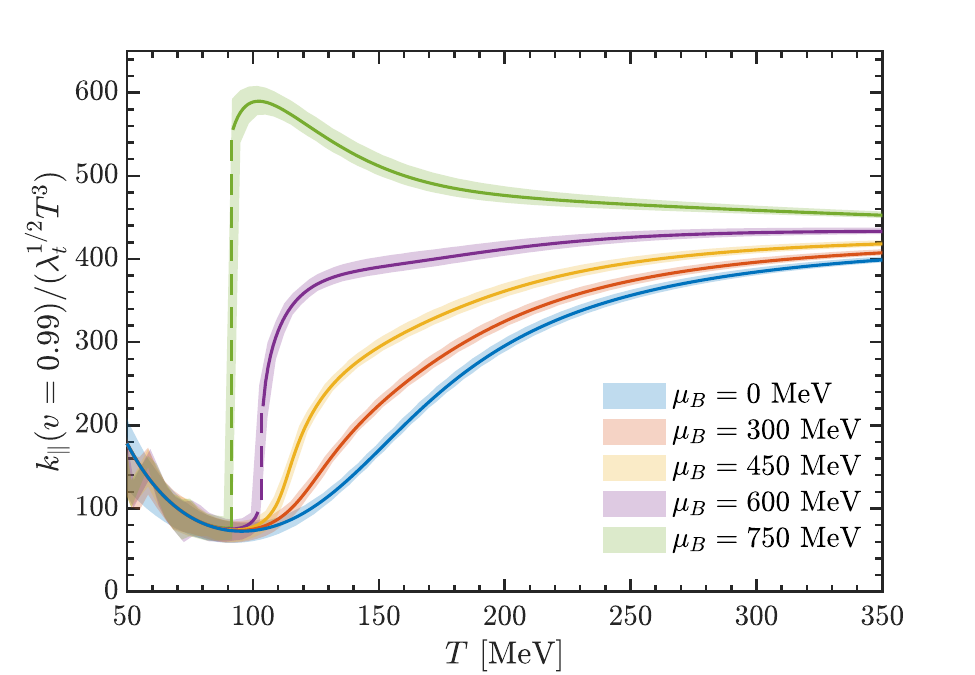}
        \label{fig:kparallel099}
    \end{subfigure}
    \hfill
    \begin{subfigure}[b]{0.45\linewidth}
        \centering
        \includegraphics[width=\linewidth]{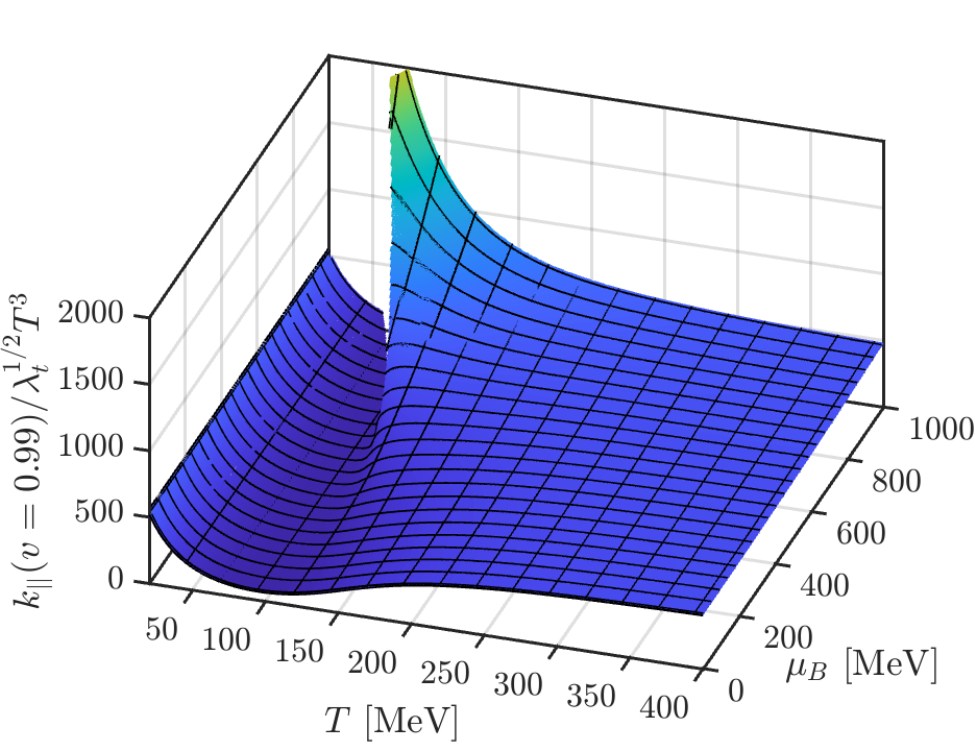}
        \label{fig:kparallel3D099}
    \end{subfigure}

    \caption{Parallel Langevin diffusion coefficient $\kappa_{||}$ obtained by using Eq.~\eqref{eq:kappaPar}. The panels on the left show the posterior bands at 95\% confidence level, and the best-fit curves (solid lines) as functions of $T$ for fixed $\mu_B$, with $v=0.5$ (top left) and $v=0.99$ (bottom left). The panels on the right display the $\kappa_{\perp}$ surface plots as functions of $T$ and $\mu_B$ using the best-fit parameterization, with $v=0.5$ (top right) and $v=0.99$ (bottom right).}
    \label{fig:Bayskparal}
\end{figure}

Figs.~\ref{fig:Bayskperp} and~\ref{fig:Bayskparal} show the behavior of the Langevin diffusion coefficients $\kappa_{\parallel}$ and $\kappa_{\perp}$. Similar to the drag force results, the coefficients increase with the medium’s baryon density, and their sensitivity to $\mu_B$ becomes more pronounced at higher quark velocities. This is reflected in the larger separation between fixed-$\mu_B$ curves as $v$ increases, suggesting that quarks moving at higher velocities experience stronger medium-induced diffusion.
One can also notice how the magnitude of both Langevin diffusion coefficients increases by lowering temperature and/or by increasing the medium's baryon density. In contrast to the drag force, $\kappa_{\parallel}$ and $\kappa_{\perp}$ develop a peak towards the CEP along the crossover region, with a discontinuity gap over the first order phase transition line. One important feature of the predictions of both coefficients is that they are consistent with the universal inequality $\kappa_{\parallel}\geq \kappa_{\perp}$ derived in~\cite{Gursoy:2010aa}.

Finally, we consider the phenomenon of jet quenching in the QGP formed in high-energy heavy-ion collisions. High-momentum partons produced in the initial hard scatterings radiate gluons and develop parton showers prior to hadronization. In a dense deconfined medium, these showers are modified through multiple scatterings with the medium constituents, resulting in parton energy loss and the suppression of high-transverse-momentum hadrons and jets~\cite{Majumder:2010qh,JET:2013cls}. This suppression provides a powerful probe of the QGP, as the initial jet production can be computed perturbatively while its subsequent medium-induced modification encodes information on the properties of the plasma~\cite{PHENIX:2003qdw,STAR:2003pjh,PHOBOS:2003uzz,BRAHMS:2003sns}. The combined collisional and radiative energy loss is commonly characterized by the jet-quenching parameter $\hat{q}$, also defined as the rate of transverse momentum broadening~\cite{Baier:1996sk,Zakharov:1997uu}.

\begin{figure}[H]
    \centering
    % ===== Left panel =====
    \begin{subfigure}[t]{0.49\linewidth}
        \centering
        \includegraphics[width=\linewidth]{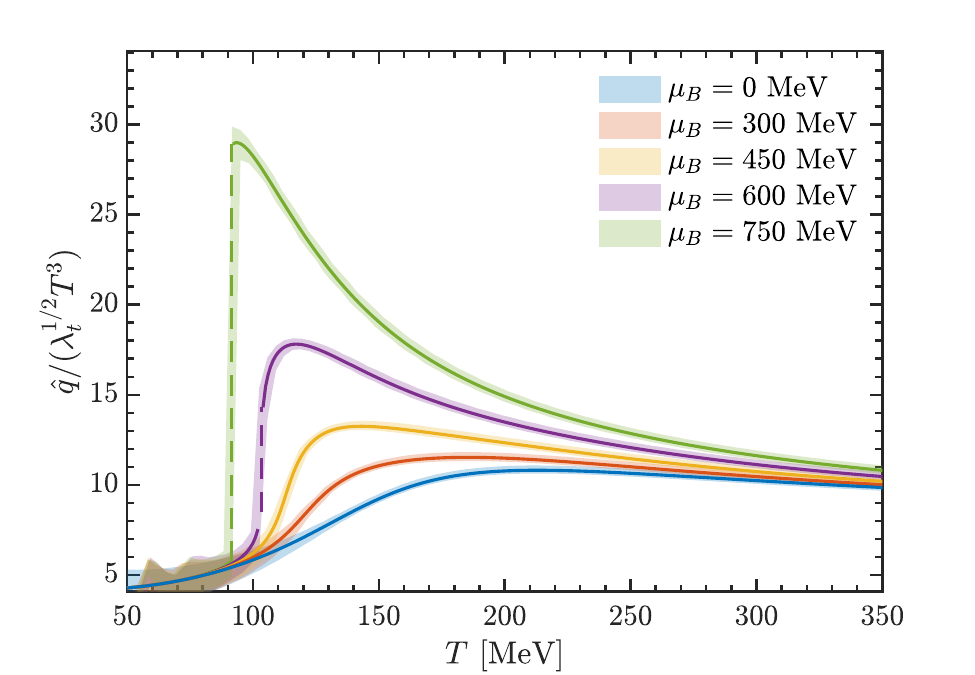}
        \label{fig:qhat}
    \end{subfigure}
    \hfill
    % ===== Right panel =====
    \begin{subfigure}[t]{0.49\linewidth}
        \centering
\includegraphics[width=\linewidth]{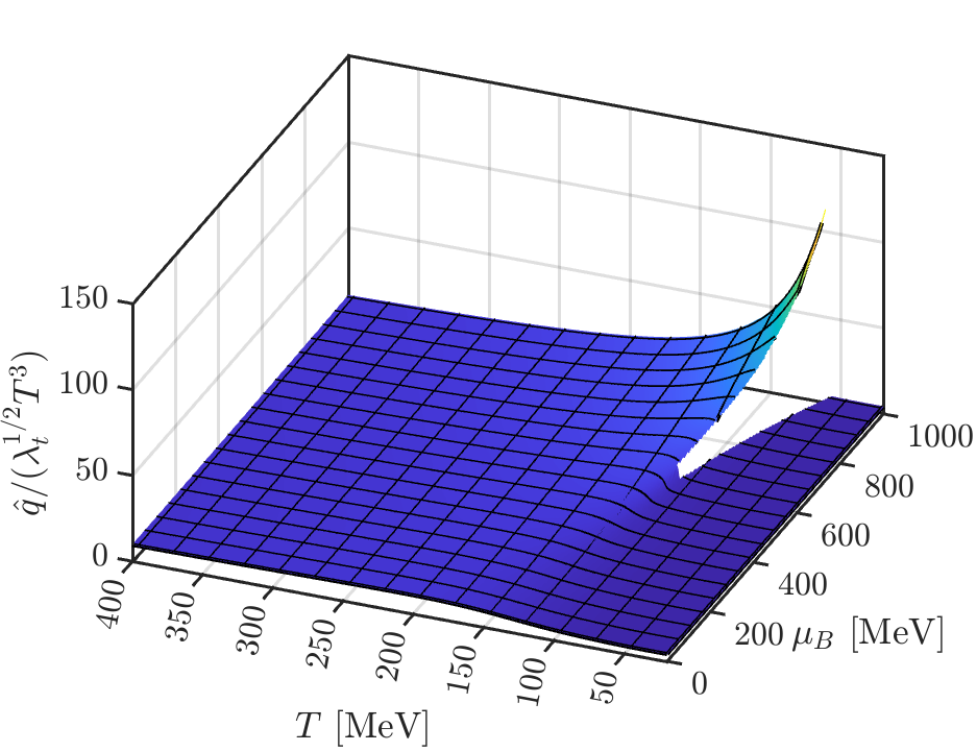}
\label{fig:qhat3d}
    \end{subfigure}
\caption{Left: scaled jet-quenching parameter $\hat{q}/{\lambda^{1/2}_t T^3}$ posterior bands at 95\% confidence level, and the
best-fit curves (solid lines) obtained from Eq.~\eqref{eq:qhat} as  functions of $T$ for several values of $\mu_B$. Right: scaled jet-quenching
parameter as a function of $T$ and $\mu_B$, from the best-fit parameterization.}
    \label{fig:3Dqhat}
\end{figure}

The holographic prescription for computing the jet-quenching parameter $\hat{q}$ associated with the transverse momentum broadening of light partons moving at the speed of light within the QGP~\cite{Baier:1996sk}, originally proposed for conformal holographic models in~\cite{Liu:2006ug,Liu:2006he}, was later adapted to non-conformal EMD models in~\cite{Rougemont:2015wca}, leading to the following integral formula,
\begin{equation}\label{eq:qhat}
    \left[\frac{\hat{q}}{\sqrt{\lambda_{t}}T^{3}}\right](T,\mu_{B})=\frac{64\pi^{2}\,h_{0}^{\textrm{far}}}{\int_{r_{\textrm{start}}}^{r_{\textrm{max}}}dr\frac{e^{-\sqrt{2/3}\,\phi(r)-3A(r)}}{\sqrt{h(r)\left[h_{0}^{far}-h(r)\right]}}},
\end{equation}
with the corresponding conformal limit of high temperatures being given by~\cite{Liu:2006ug},
\begin{equation}
    \lim_{T\to\infty}\left[\frac{\hat{q}}{\sqrt{\lambda_{t}}T^{3}}\right](T,\mu_B)=\frac{\pi^{3/2}\,\Gamma(3/4)}{\Gamma(5/4)}\approx7.52814.
\end{equation}
\noindent In Fig.~\ref{fig:3Dqhat} we display our Bayesian results for the jet-quenching parameter. One notes that the energy loss also grows in magnitude as the medium temperature decreases and its baryon density increases, indicating more jet suppression in the baryon-dense regime. A peak is also developed towards the CEP around the crossover region, with a discontinuity gap over the first-order transition line.
\vspace{-1cm}
In the right panel of Fig.~\ref{fig:JetScape}, we show the holographic EMD prediction for the jet-quenching parameter at zero baryon chemical potential compared to the Bayesian profile obtained with the hybrid phenomenological model of the JETSCAPE Collaboration \cite{JETSCAPE:2020shq}, which is adjusted to describe heavy ion data on partonic energy loss. As in any holographic calculation involving the NG action, there is an extra free parameter corresponding to the `t Hooft coupling, $\lambda_t$. In the right panel of Fig.~\ref{fig:JetScape}, we adjusted the height of the holographic jet-quenching parameter by choosing three representative values for $\lambda_t = \{1/25, 1/16, 1/9\}$, all of them lying within the broad JETSCAPE uncertainty band. Since $\hat{q} \propto \sqrt{\lambda_t}$, the overall normalization of the holographic result is controlled by the choice of $\lambda_t$. Therefore, the holographic model can only make a prediction for the shape of the jet quenching parameter as a function of $(T,\mu_B)$, but not for its magnitude. In this sense, the aforementioned values for $\lambda_t$, employed to plot the holographic result for the jet quenching parameter on the right panel of Fig.~\ref{fig:JetScape}, should be regarded as a phenomenological normalization used for comparison with the JETSCAPE result, rather than as a determination of the fundamental QCD `t Hooft coupling.

Indeed, we remark that such small values for $\lambda_t=g_\textrm{QCD}^2 N_c$ are in tension with one of the basic assumptions of the standard holographic gauge-gravity dictionary, which prescribes that the `t Hooft coupling should be very large. Such a tension is also similar to what was discussed at the end of Section~\ref{sec:holo} concerning the value of the gravitational constant in the present bottom-up EMD model, $\kappa_5^2\equiv 8\pi G_5=11.4$. By the standards of the holographic gauge-gravity dictionary, one should expect the gravitational constant $\kappa_5^2$ to be very small, in association with the large $N_c$ limit of the dual gauge theory at the boundary. However, in our bottom-up phenomenological approach, we have never matched any of our results to large $N_c$ QCD, but used instead inputs from the physical $SU(N_c=3)$ QCD. Although the approach followed here provides in practice holographic predictions in good agreement with first principles LQCD results, as shown in Fig.~\ref{fig:Lattice}, and also in reasonable agreement with the JETSCAPE results for the bulk viscosity and the jet quenching parameter inferred from heavy-ion experimental data, as displayed in Fig.~\ref{fig:JetScape}, from a purely theoretical perspective, the resolution of the aforementioned tensions with holographic top-down expectations is still something to be understood.

%\vspace{-20pt}
\section{Comparison Between PHA and PA ansatze}
\label{sec:comparison}

\begin{figure}%[H]
    \centering
    % ===== Left panel =====
    \begin{subfigure}[t]{0.49\linewidth}
        \centering
        \includegraphics[width=\linewidth]{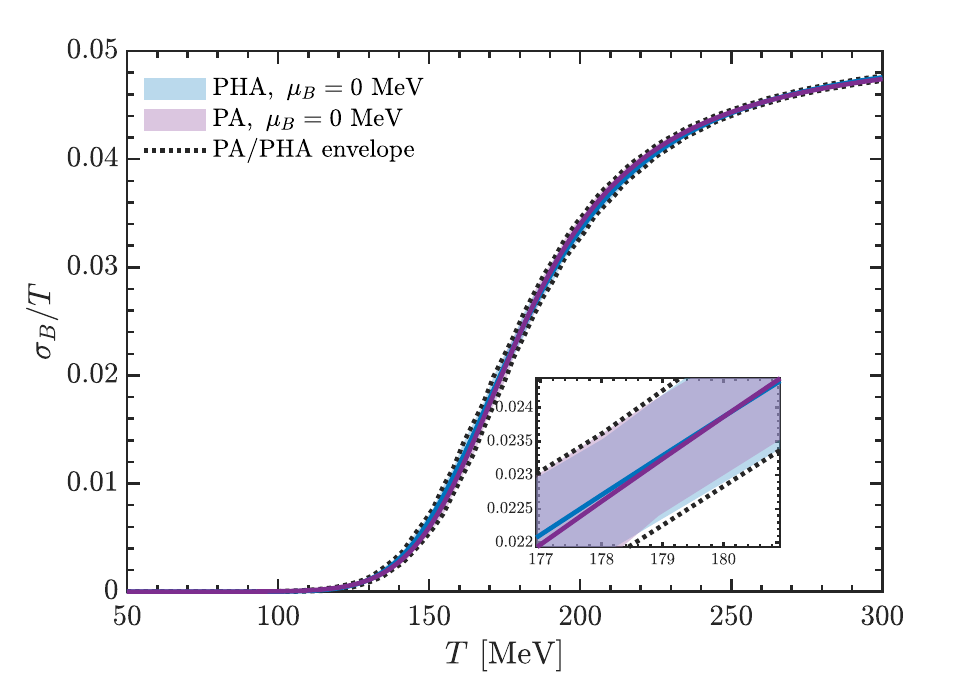}
        \label{fig:sigmaB_PHAvsPAmu0}
    \end{subfigure}
    \hfill
    % ===== Right panel =====
    \begin{subfigure}[t]{0.49\linewidth}
        \centering
\includegraphics[width=\linewidth]{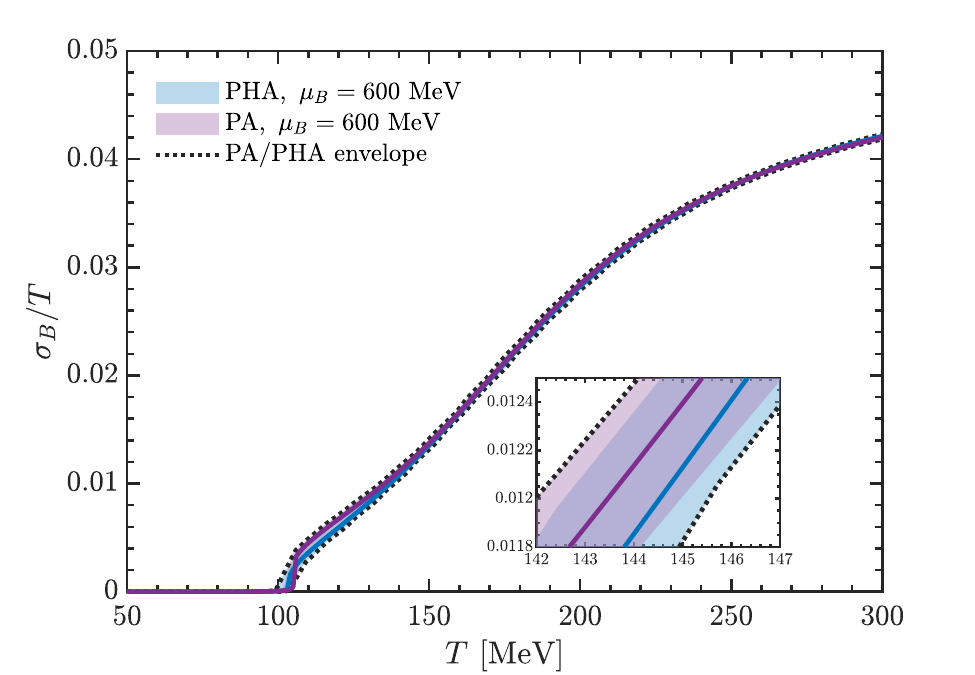}
\label{fig:sigmaB_PHAvsPAmu600}
    \end{subfigure}
\caption{Panels display normalized baryon conductivity, $\sigma_{B}/T$, obtained from PA and PHA ansatze. Bands denote the 95 \% credible intervals, while solid curves represent the corresponding best-fit predictions. The black dotted curves indicate the combined envelope of the PA and PHA results. Left panel: $\mu_B=0$ MeV. Right panel : $\mu_B=600$ MeV.}
    \label{fig:sigmaBPHAvsPA}
\end{figure}

\begin{figure}%[H]
    \centering
    % ===== Left panel =====
    \begin{subfigure}[t]{0.49\linewidth}
        \centering
        \includegraphics[width=\linewidth]{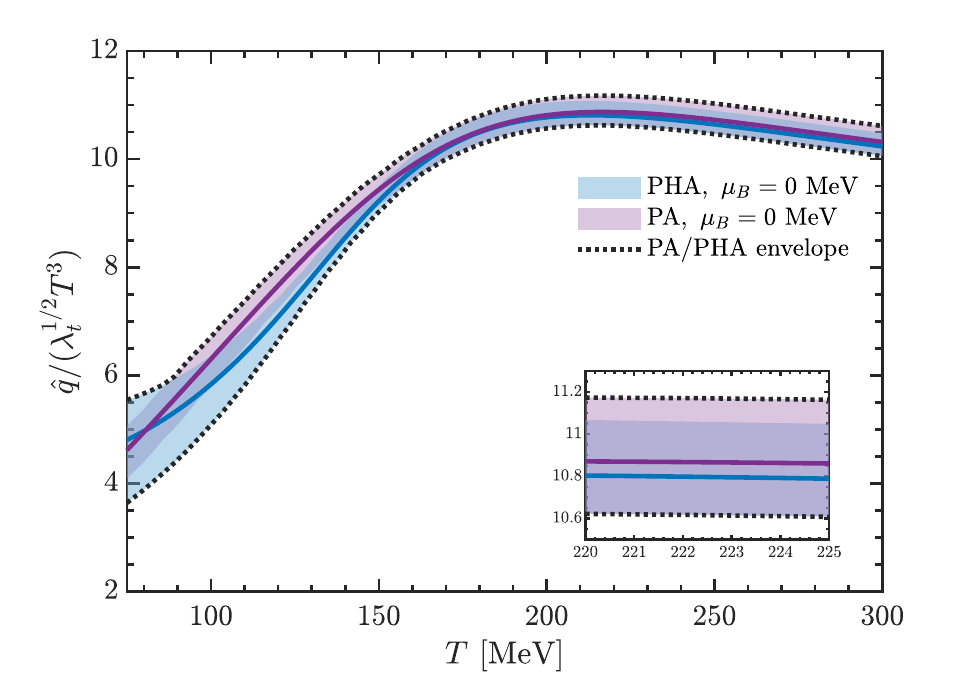}
        \label{fig:qhat_PHAvsPAmu0}
    \end{subfigure}
    \hfill
    % ===== Right panel =====
    \begin{subfigure}[t]{0.49\linewidth}
        \centering
\includegraphics[width=\linewidth]{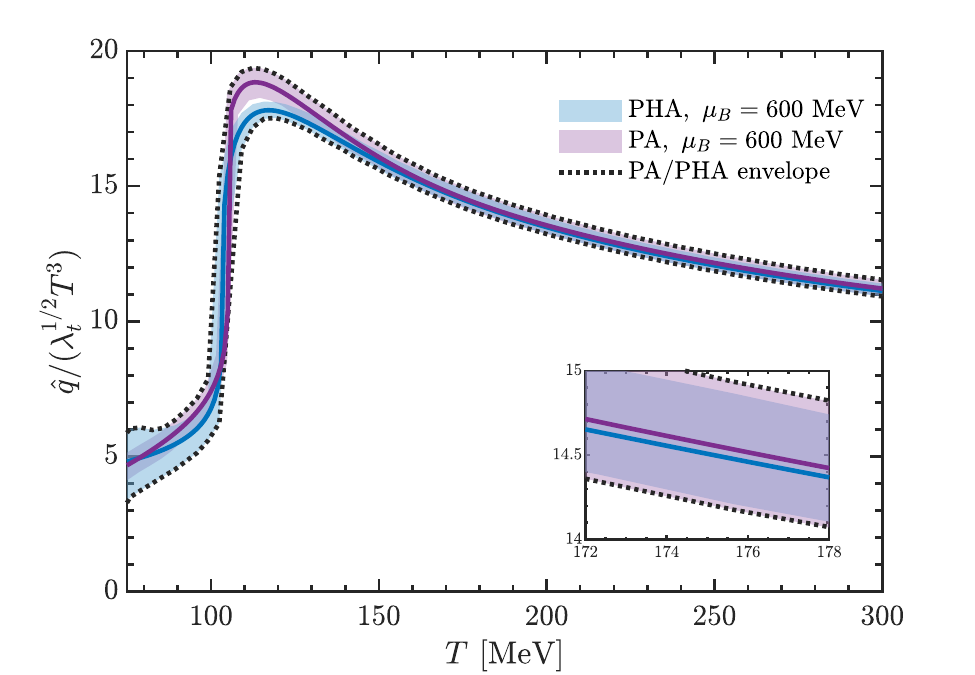}
\label{fig:qhat_PHAvsPAmu600}
    \end{subfigure}
\caption{Comparison of the scaled jet-quenching parameter $\hat q/(\lambda_t^{1/2}T^3)$ obtained using the PA and PHA ansatze. Shaded bands represent 95\% credible intervals while solid lines correspond to the best-fit curves. The dotted black lines indicate the combined PA/PHA envelope. Left panel: $\mu_B=0$ MeV. Right panel: $\mu_B = 600$ MeV.}
    \label{fig:qhatPHAvsPA}
\end{figure}

\begin{figure}%[H]
    \centering
    % ===== Left panel =====
    \begin{subfigure}[t]{0.49\linewidth}
        \centering
        \includegraphics[width=\linewidth]{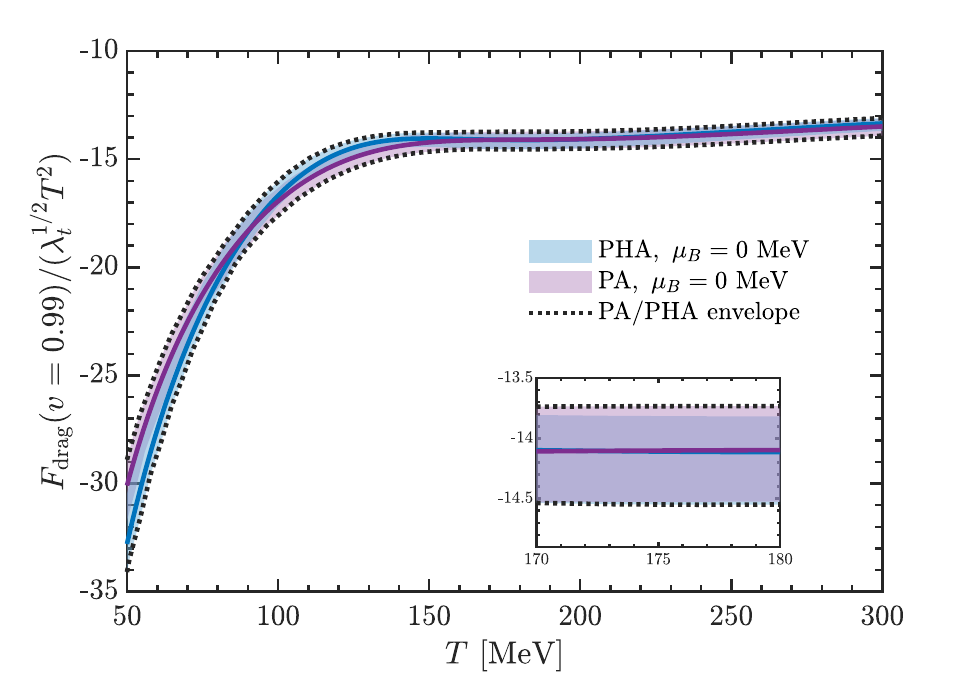}
        \label{fig:DF_99_PHAvsPAmu0}
    \end{subfigure}
    \hfill
    % ===== Right panel =====
    \begin{subfigure}[t]{0.49\linewidth}
        \centering
\includegraphics[width=\linewidth]{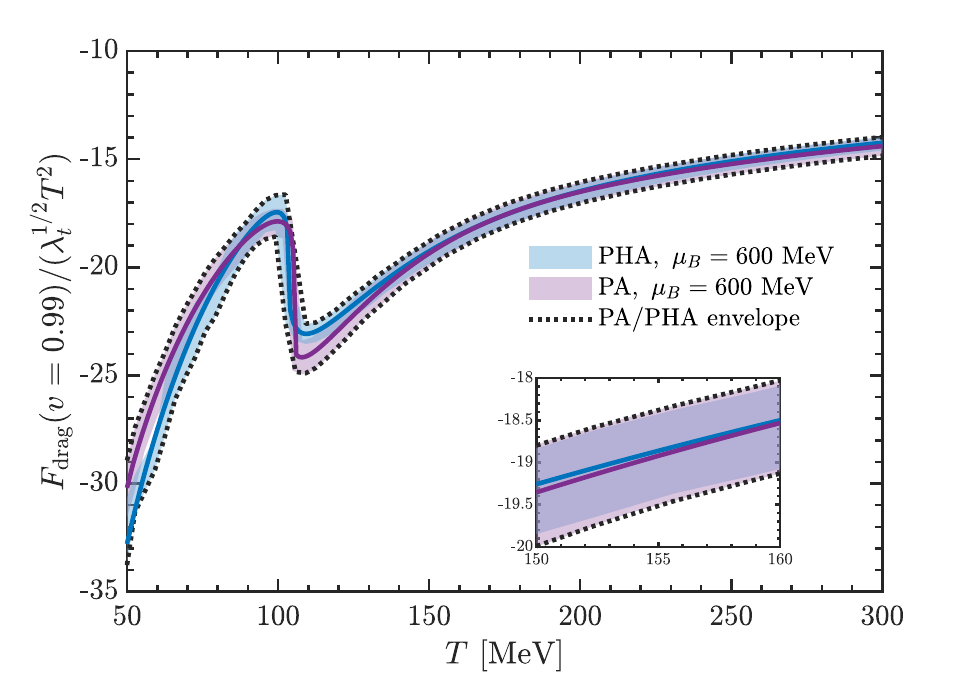}
\label{fig:DF_PHAvsPAmu600}
    \end{subfigure}
\caption{Comparison of the heavy quark drag force with $v = 0.99$ obtained using the PA and PHA ansatze. Bands represent 95\% credible intervals while solid lines correspond to the best-fit curves. The dotted black lines indicate the combined PA/PHA envelope. Left panel: $\mu_B=0$ MeV. Right panel: $\mu_B=600$ MeV.}
    \label{fig:DFPHAvsPA}
\end{figure}

\begin{figure}%[H]
    \centering
    % ===== Left panel =====
    \begin{subfigure}[t]{0.49\linewidth}
        \centering
        \includegraphics[width=\linewidth]{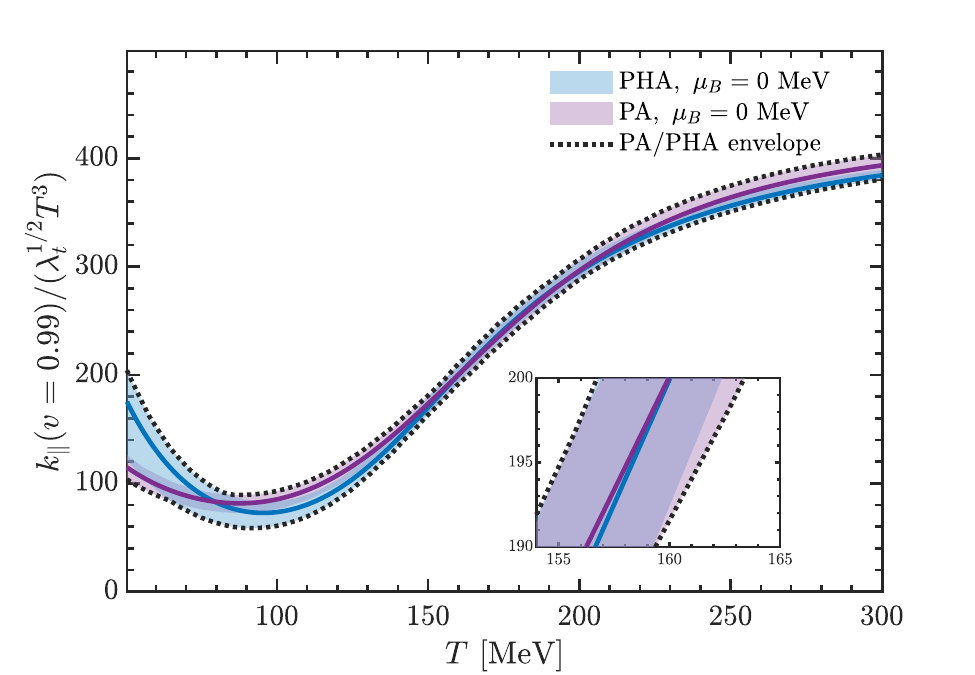}
        \label{fig:KII_PHAvsPAmu0}
    \end{subfigure}
    \hfill
    % ===== Right panel =====
    \begin{subfigure}[t]{0.49\linewidth}
        \centering
\includegraphics[width=\linewidth]{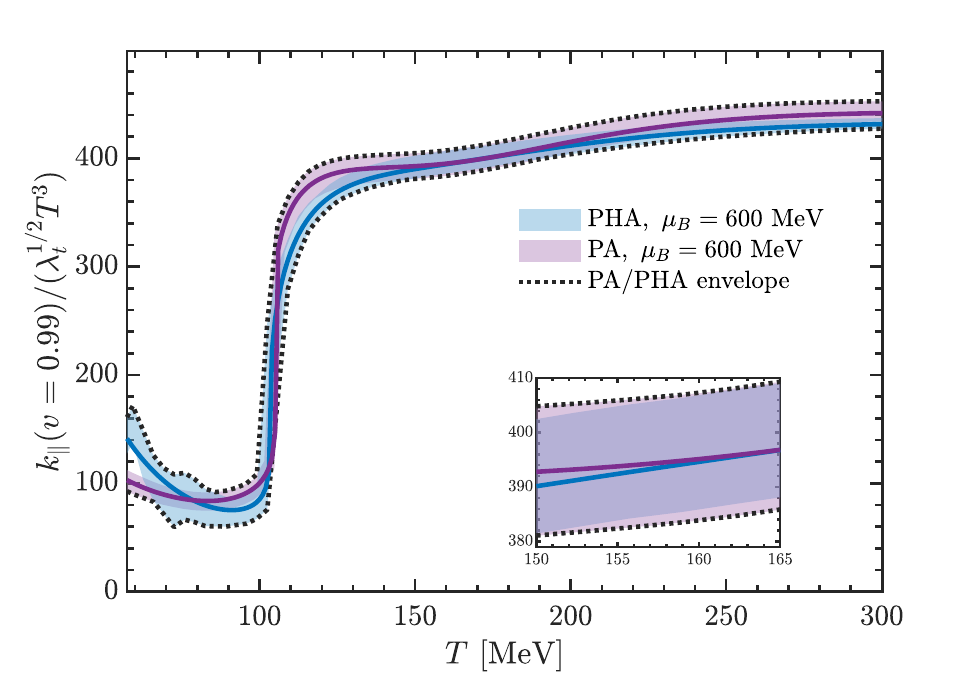}
\label{fig:KII_PHAvsPAmu600}
    \end{subfigure}
\caption{Comparison of the Parallel Langevin diffusion coefficient $\kappa_{\parallel}$ with $v = 0.99$ obtained using the PA and PHA ansatze. Bands represent 95\% credible intervals while solid lines correspond to the best-fit curves. The dotted black lines indicate the combined PA/PHA envelope. Left panel: $\mu_B=0$ MeV. Right panel: $\mu_B=600$ MeV.}
    \label{fig:kIIPHAvsPA}
\end{figure}

\begin{figure}%[H]
    \centering
    % ===== Left panel =====
    \begin{subfigure}[t]{0.49\linewidth}
        \centering
        \includegraphics[width=\linewidth]{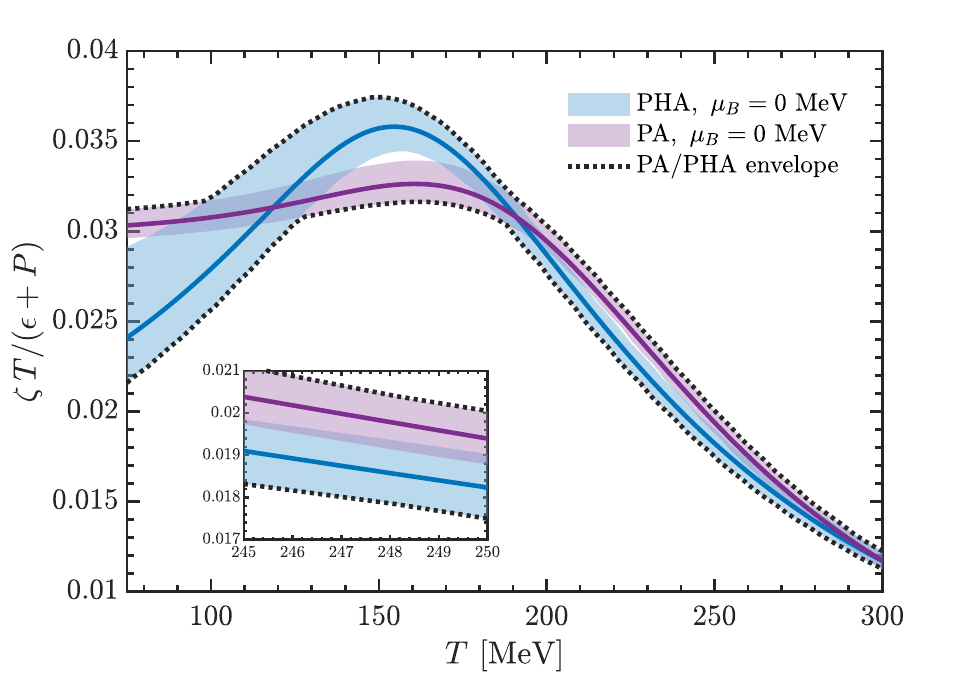}
        \label{fig:zeta_PHAvsPAmu0}
    \end{subfigure}
    \hfill
    % ===== Right panel =====
    \begin{subfigure}[t]{0.49\linewidth}
        \centering
\includegraphics[width=\linewidth]{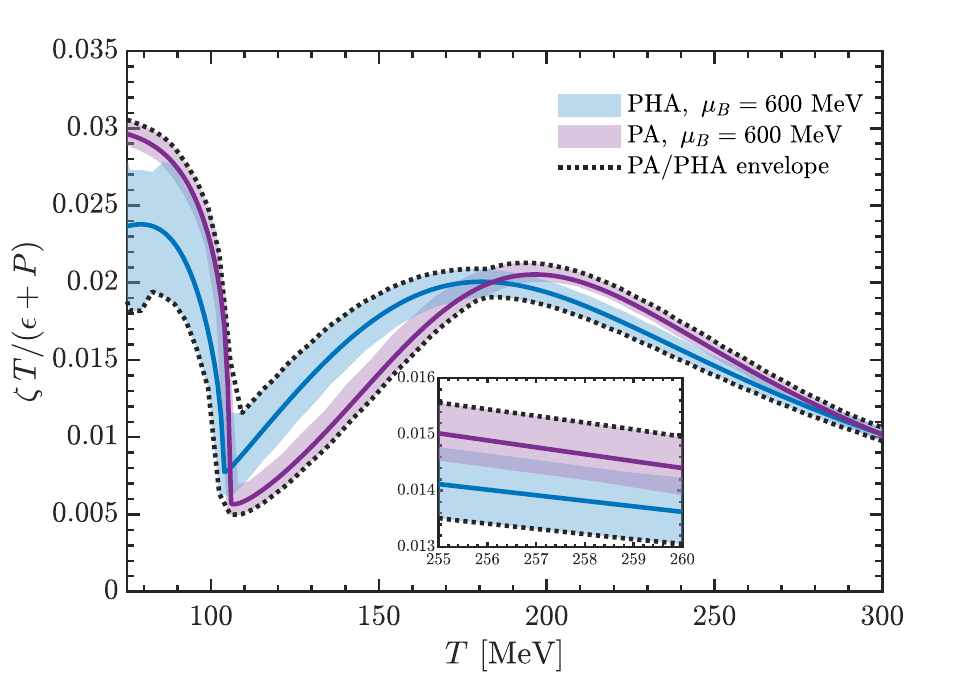}
\label{fig:zeta_PHAvsPAmu600}
    \end{subfigure}
\caption{Comparison of the bulk viscosity, normalized as $\zeta T/(\epsilon +P )$. Obtained using the PA and PHA ansatze. Shaded bands represent 95\% credible intervals while solid lines correspond to the best-fit curves. The dotted black lines indicate the combined PA/PHA envelope. Left panel: $\mu_B=0$ MeV. Right panel: $\mu_B = 600$ MeV.}
    \label{fig:zetaPHAvsPA}
\end{figure}

The results for transport coefficients and energy loss presented in Section~\ref{sec:results}, were obtained using the polyhyperbolic ansatz (PHA), while the corresponding results obtained with the parametric ansatz (PA) are shown in Appendix~\ref{sec:app}. Since both PHA and PA ansatze were constrained by the same lattice QCD data \cite{Borsanyi:2021sxv} through Bayesian analysis in \cite{Hippert:2023bel}, a direct comparison of PA and PHA $95\%$ credible bands offers a quantitative measure of how sensitively the transport coefficient predictions depend on the choice of ansatz. To assist this comparison, we show in Figs.~\ref{fig:sigmaBPHAvsPA} ---~\ref{fig:zetaPHAvsPA} the PA--PHA envelope as the dotted black lines, which were constructed by taking the smallest lower limit and the largest upper limit of the two credible bands. Its width provides a visual measure of the ansatz dependence. We discuss such a comparison below.

Fig.~\ref{fig:sigmaBPHAvsPA} compares the bands for the normalized baryon conductivity, $\sigma_{B}/T$, as a function of temperature obtained from PA and PHA ansatze at $\mu_{B}=0$ MeV and $\mu_{B}=600$ MeV. The two bands overlap almost completely across the temperature range, resulting in a very narrow PA-PHA envelope. This near-perfect agreement demonstrates that predictions for $\sigma_{B}/T$ are remarkably robust to the choice of ansatz.

Fig.~\ref{fig:qhatPHAvsPA} compares the posterior bands for the scaled jet-quenching parameter, $\hat q/(\lambda_t^{1/2}T^3)$, obtained from PA and PHA ansatze at fixed chemical potential. The 95\% credible bands show significant overlap throughout the temperature range, indicating that the predicted behavior of $\hat q/(\lambda_t^{1/2}T^3)$ is largely insensitive to the ansatz choice. A modest difference does emerge around the peak region at $\mu_{B}=600$ MeV, where the PA-PHA envelope somewhat broadens, signaling a slightly stronger ansatz dependence around the peak.

Fig.~\ref{fig:DFPHAvsPA} compares the posterior bands for the scaled heavy quark drag force at $v=0.99$ obtained using the PA and PHA ansatze for $\mu_B=0$ and $\mu_B=600$ MeV as function of $T$. The $95 \%$ credible bands overlap almost completely across the entire temperature range, indicating negligible dependence on the choice of ansatz.

Fig.~\ref{fig:kIIPHAvsPA} compares the scaled parallel component of the Langevin diffusion coefficient at $v=0.99$ obtained from PA and PHA ansatze for $\mu_B=0$ and $600$ MeV. The $95\%$ credible bands overlap almost completely across the entire temperature range, indicating that the heavy-quark drag force is largely insensitive to the choice of holographic ansatz.

Fig.~\ref{fig:zetaPHAvsPA} compares the posterior bands for the normalized bulk viscosity $\zeta T/(\epsilon+P)$  at $\mu_B=0$ and $600$ MeV. In contrast to other observables considered in this work, the bulk viscosity exhibits stronger dependence on the choice of holographic ansatz. At $\mu_B=0$, both PA and PHA bands show qualitatively different behavior for $T\lesssim150$ MeV. The PHA band decreases towards lower temperature, whereas the PA band remains nearly flat. Moreover, the PHA credible band becomes noticeably wider in this region, implying that low-temperature bulk viscosity is less constrained. This low-temperature sensitivity is 
similar to the larger differences between the PA and PHA trace anomaly, $(\epsilon-3P)/T^4$, compared to the other thermodynamic observables shown in Fig.~8 of Ref.~\cite{Hippert:2023bel}. 
At $\mu_B=600$ MeV, both ansatze show the same qualitative features, such as a clear dip at low temperatures and a broad peak at intermediate temperatures. The PA–PHA envelope is wider around this temperature region than for the other transport observables, but the two posterior bands overlap, especially at high temperatures.

For each observable, we overlay the PA and PHA 95\% credible bands and quantify their separation through the PA--PHA envelope. The envelope is defined by the  lower bound and  upper bound of the two bands at each temperature. The width of the PA--PHA envelope quantifies ansatz dependence of the observables. A narrow envelope shows that the observable is largely insensitive to the choice of ansatz. Whereas a broader envelope shows greater model dependence. This comparison also provides an estimate of the uncertainty associated with the choice of holographic parametrization.
Note also from the comparison done in Figs.~\ref{fig:sigmaBPHAvsPA} ---~\ref{fig:zetaPHAvsPA} that the PA ansatz exhibits smaller posterior uncertainty than the PHA ansatz (which is also the case for thermodynamic observables, as originally disclosed in Ref.~\cite{Hippert:2023bel}).

%SigmaB

%df
%%

%KP
%%

%%

%%%%%%%%%%%%%%%%%%%%%%%%%%%%%%%%%
%%%%%%%%%%%%%%%%%%%%%%%%%%%%%%%%%
\section{Conclusions}
\label{sec:conclusion}

In the present work, we employed the holographic Einstein-Maxwell-Dilaton model  supplemented with Bayesian inference tools~\cite{Hippert:2023bel} to evaluate several transport coefficients for the QGP with $N_f=2+1$ flavors throughout the QCD phase diagram. The model employed in this work shows good overall agreement with state-of-the-art $(2+1)$-flavor lattice QCD thermodynamics and with the finite-density equation of state obtained via the $T'$-expansion \cite{Borsanyi:2021sxv,Bellwied:2015lba}.

We introduced a new relaxational numerical strategy for extracting ultraviolet asymptotics in holographic models, leading to substantial gains in stability and performance, 
which in turn enabled the Bayesian sampling of lattice QCD–constrained posterior distributions in the analysis of Ref.~\cite{Hippert:2023bel}. 

A publicly available \texttt{C++} implementation of the EMD model incorporating this method was extended to include the calculation of several transport coefficients: 
 baryon and thermal conductivities, the baryon diffusion coefficient, shear and bulk viscosities, the heavy-quark drag force and Langevin diffusion coefficients, and the jet-quenching parameter \cite{yang_2025_14695243}.

Combining this improved \texttt{C++} code with Bayesian samples from \cite{Hippert:2023bel,Hippert2024_ZenodoDataset}, we obtained predictions for these transport coefficients across the phase diagram, including Bayesian uncertainty bands constrained by lattice QCD. 
Our results go beyond previous work in the literature (see e.g.~\cite{Chen:2025fpd}) by presenting lattice QCD-constrained uncertainty bands at finite baryon chemical potential, and by considering baryon and thermal conductivities.
Moreover, by exploring two different sets of functional forms for the holographic potentials,  we were able to confirm the robustness of our predictions in the crossover and phase-transition regions. 
Within the regime of validity of the model, the lattice-constrained posterior leads to tightly bounded transport coefficients. 
The Bayesian uncertainty bands obtained in the present work are typically broader at low temperatures, as expected, since this region of the phase diagram lies beyond the regime of applicability of the model, and was not constrained by lattice QCD data at zero baryon density. 

We found an overall suppression of the diffusion of baryon charge, shear viscosity, and bulk viscosity with increasing baryon density, indicating that the strongly interacting QGP becomes even closer to the perfect-fluid limit at large baryon chemical potentials. On the other hand, the jet-quenching parameter and the heavy-quark energy loss and Langevin momentum-diffusion coefficients were found to be enhanced with increasing baryon density.
Furthermore, we found that our prediction for the bulk viscosity is in good agreement with the uncertainty band obtained from the Bayesian analysis of heavy-ion data by the JETSCAPE Collaboration~\cite{JETSCAPE:2020shq,JETSCAPE:2020mzn}. This provides further support for the holographic EMD description of dissipative transport in the quark-gluon plasma. For the jet-quenching parameter, we found that the predicted temperature dependence is in agreement with the JETSCAPE band after fixing the overall holographic normalization for this observable by phenomenologically choosing $\lambda_t \in [1/25,1/9]$. Since $\hat{q}/T^3 \propto \sqrt{\lambda_t}$, the choice of $\lambda_t$ only fixes the overall magnitude of the holographic jet quenching parameter, while the $(T,\mu_B)$-dependence remains unchanged.

%%%%%%%%%%%%%%%%%%%%%%%%%%%%%%%%%
%%%%%%%%%%%%%%%%%%%%%%%%%%%%%%%%%
\acknowledgments
We thank J. Jahan for his assistance in running all posterior samples on the cluster. This material is based upon work supported by the National Science Foundation under grants No. PHY-2208724, PHY-2116686 and PHY-2514763, and within the framework of the MUSES collaboration, under Grant No. OAC-2103680. This material is also based upon work supported by the U.S. Department of Energy, Office of Science, Office of Nuclear Physics, under Award Numbers DE-SC0022023, DE-SC0026065, and DE-
SC0023861, as well as by the National Aeronautics and Space Agency (NASA) under Award Number 80NSSC24K0767.
A.N. acknowledges the financial support by the Brazilian National Council for Scientific and Technological Development (CNPq) under process number 176343/2025-3.
M.H. was supported by the Brazilian National Council for Scientific and Technological Development (CNPq) under process No. 313638/2025-0. 
R.R. acknowledges financial support by the Brazilian National Council for Scientific and Technological Development (CNPq) under grants number 407162/2023-2 and 305466/2024-0.

%%%%%%%%%%%%%%%%%%%%%%%%%%%%%%%%%
%%%%%%%%%%%%%%%%%%%%%%%%%%%%%%%%%
\newpage
\appendix
\section{Holographic Bayesian transport with the Parametric Ansatz}
\label{sec:app}

\begin{figure}[H]
\begin{subfigure}[b]{0.45\textwidth}
    \centering
    \includegraphics[width=\textwidth]{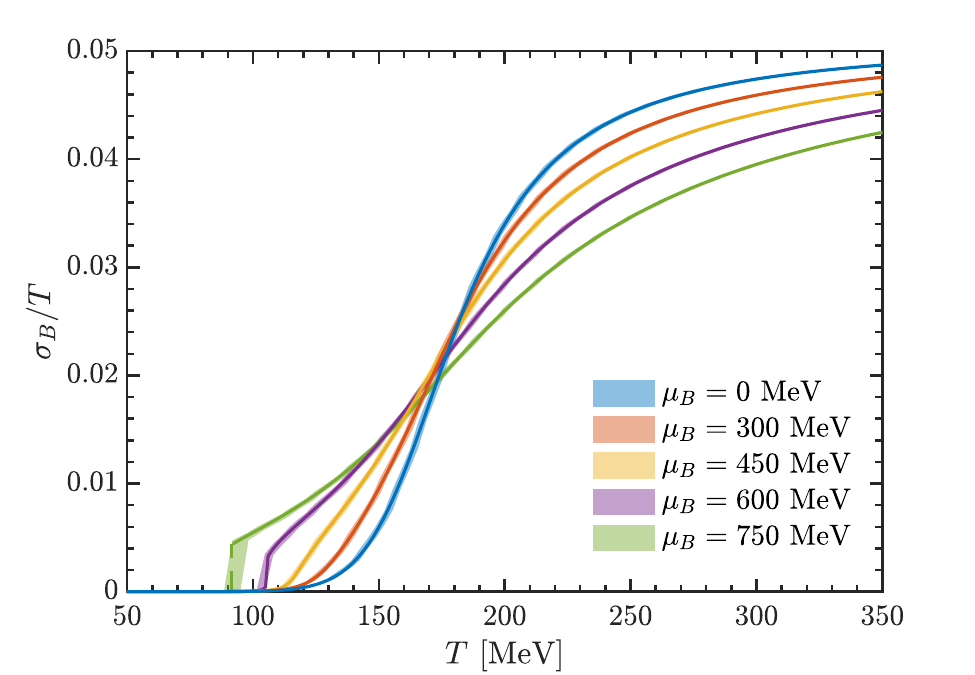}
    \caption{}
\end{subfigure}
\hfill
\begin{subfigure}[b]{0.45\textwidth}
    \centering
    \includegraphics[width=\textwidth]{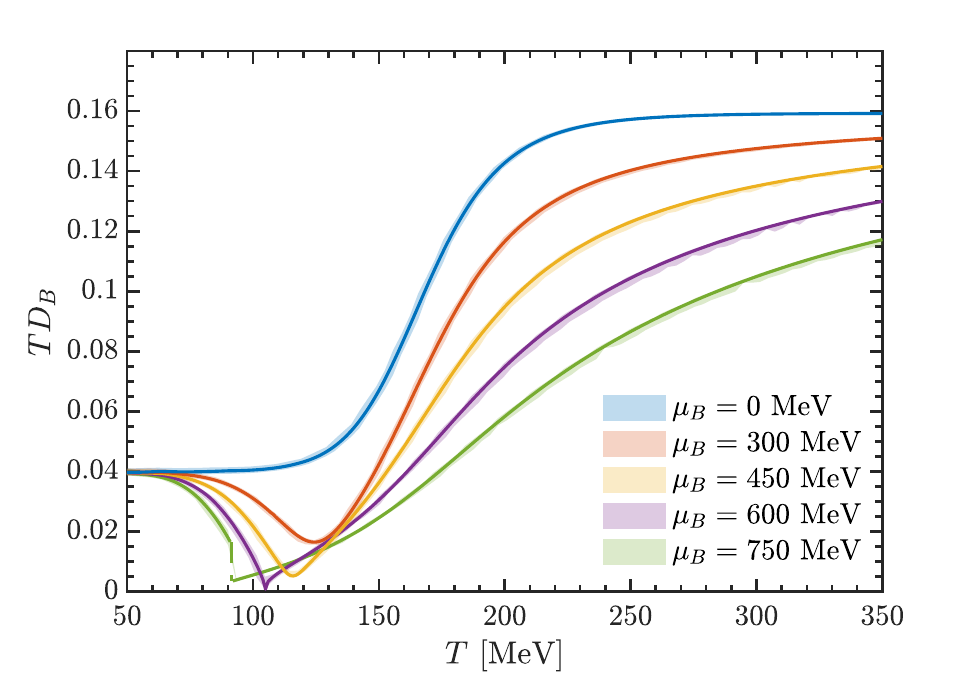}
    \caption{}
\end{subfigure}
\vspace{0.5cm}
\begin{subfigure}[b]{0.45\textwidth}
    \centering
    \includegraphics[width=\textwidth]{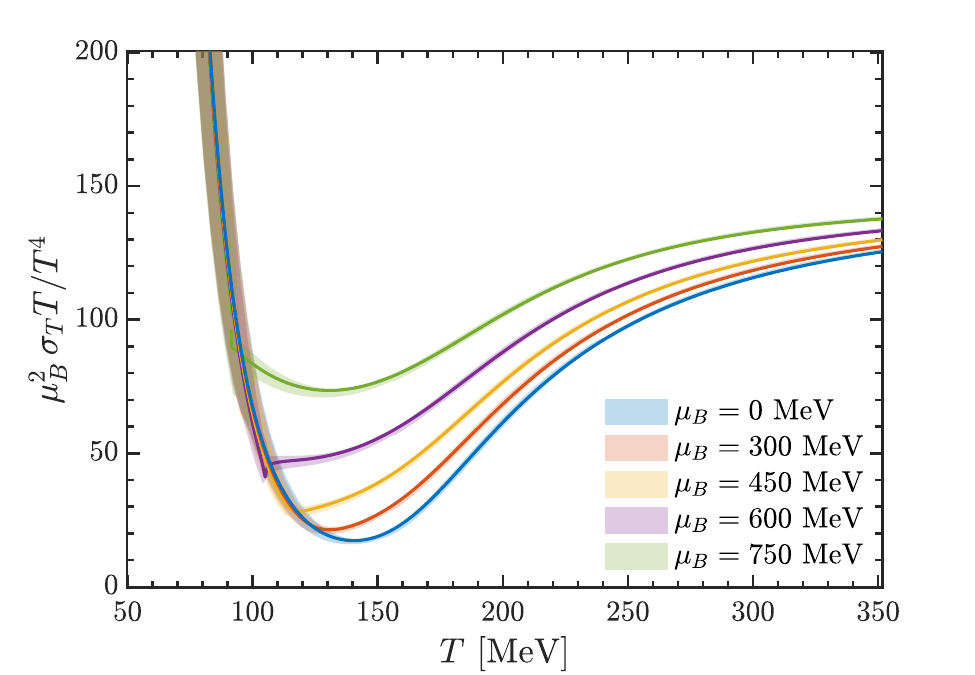}
    \caption{}
\end{subfigure}
\hfill
\begin{subfigure}[b]{0.45\textwidth}
    \centering
    \includegraphics[width=\textwidth]{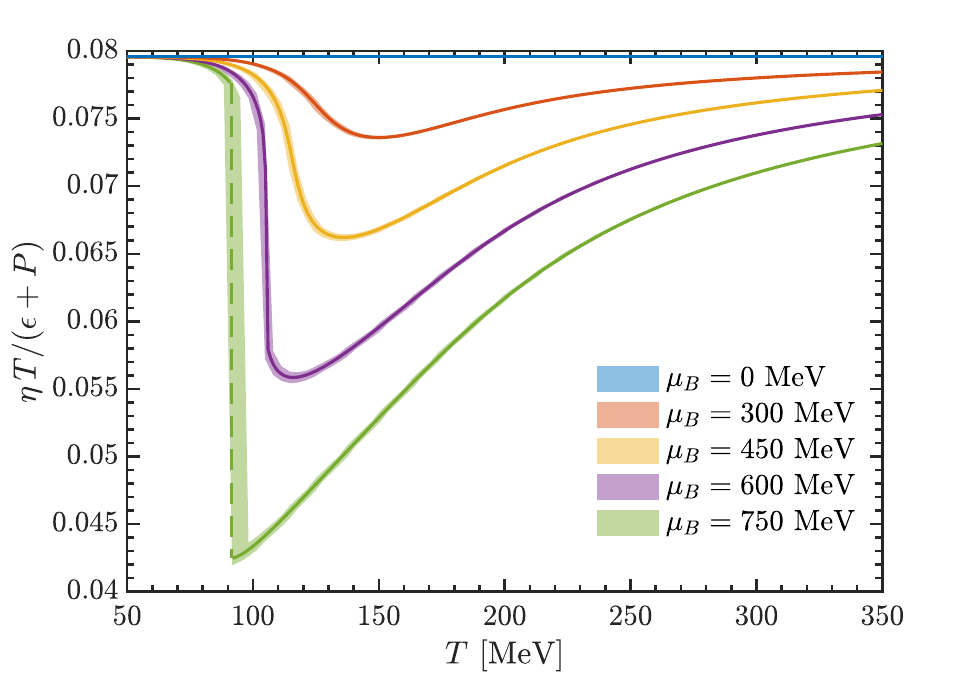}
    \caption{}
\end{subfigure}
\caption{Panel (a) display posterior bands at 95\% confidence level and corresponding best-fit curves (solid lines) for $\sigma_B/T$ as a function of $T$ for different values of $\mu_B$. Panel (b) presents posterior bands at 95\% confidence level and corresponding best-fit curves (solid lines) for the scaled baryon diffusion coefficient $TD_B$ as a function of $T$ for different values of $\mu_B$. Panel (c) shows posterior bands for the holographic thermal conductivity at 95\% confidence level and corresponding best-fit curves (solid lines) as a function of $T$ for different values of $\mu_B$. Panel (d) reveals holographic shear viscosity times temperature over enthalpy density. Posterior bands at 95\% confidence level and the corresponding best-fit curves (solid lines) as a function of $T$ for several values of $\mu_B$.}
\label{fig:sigmaB_DB_sigmaT_eta}
\end{figure}

\begin{figure}[H]
    \centering
    \begin{subfigure}[t]{0.49\linewidth}
        \centering
        \includegraphics[width=\linewidth]{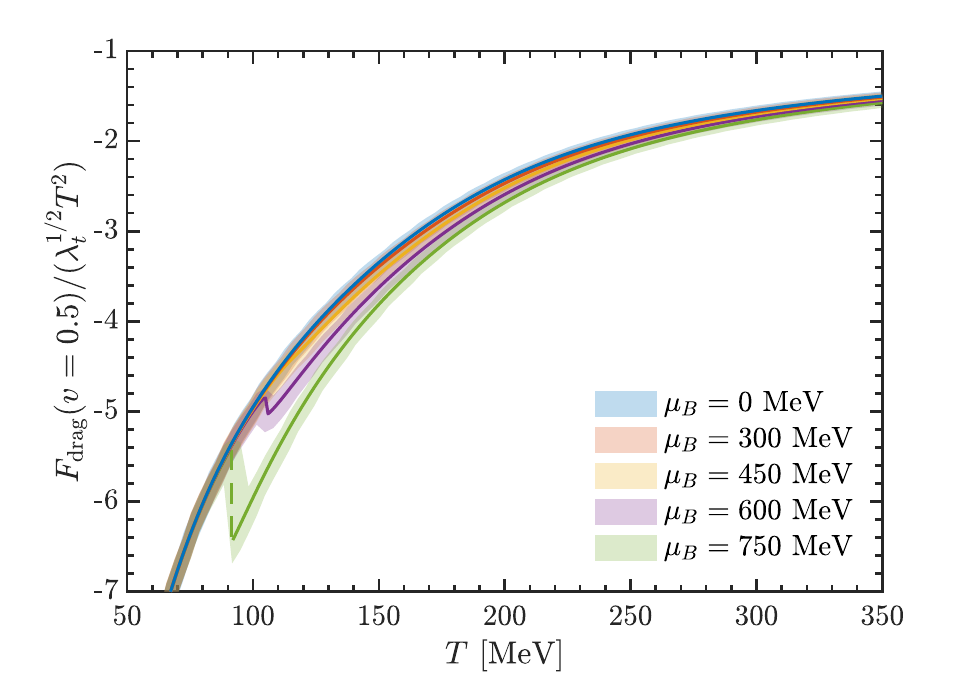}
        \label{fig:Drag_Force099}
    \end{subfigure}
    \hfill
    \begin{subfigure}[t]{0.49\linewidth}
        \centering
        \includegraphics[width=\linewidth]{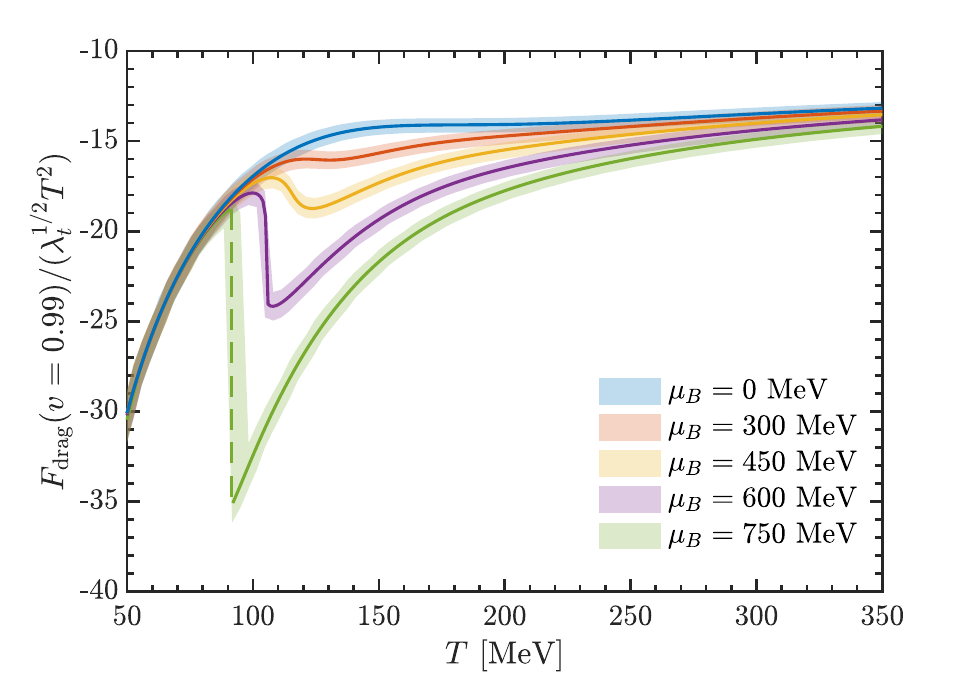}
        \label{fig:Drag_Force3D099}
    \end{subfigure}
    \caption{Heavy quark drag force at v = 0.5 (left panel) and v=0.99 (right panel). Both figures show posterior band at 95\% confidence level and corresponding best-fit curves (solid lines).}
    \label{fig:PA_DrageForce}
\end{figure}

\begin{figure}[H]
    \centering
    \begin{subfigure}[t]{0.49\linewidth}
        \centering
        \includegraphics[width=\linewidth]{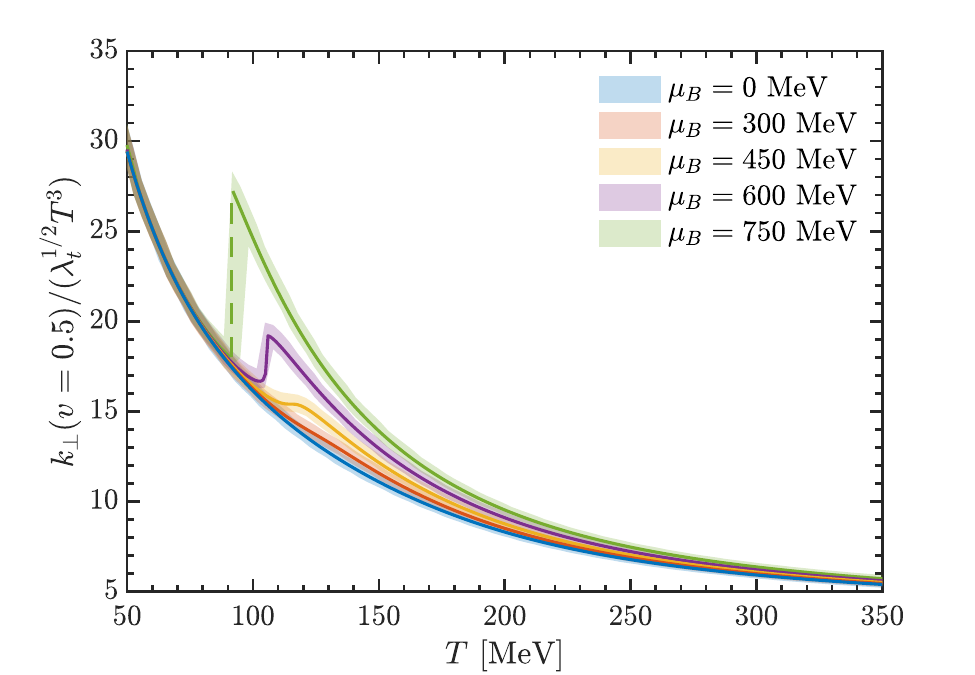}
        \label{fig:Drag_Force099}
    \end{subfigure}
    \hfill
    \begin{subfigure}[t]{0.49\linewidth}
        \centering
        \includegraphics[width=\linewidth]{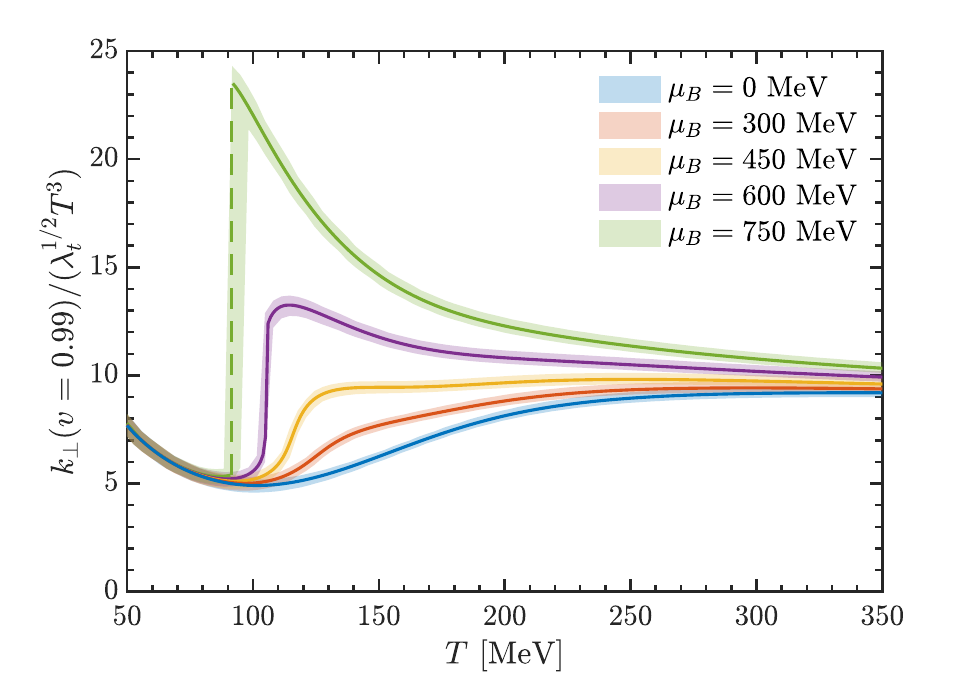}
        \label{fig:Drag_Force3D099}
    \end{subfigure}
    \caption{Perpendicular Langevin diffusion coefficient $\kappa_\perp$ at v = 0.5 (left panel) and v=0.99 (right panel). Both figures show posterior band at 95\% confidence level and the corresponding best-fit curves (solid lines).}
    \label{fig:PA_kappa_perp}
\end{figure}

\begin{figure}[H]
    \centering
    % LEFT
    \begin{minipage}[t]{0.48\textwidth}
        \centering
        \includegraphics[width=\linewidth]{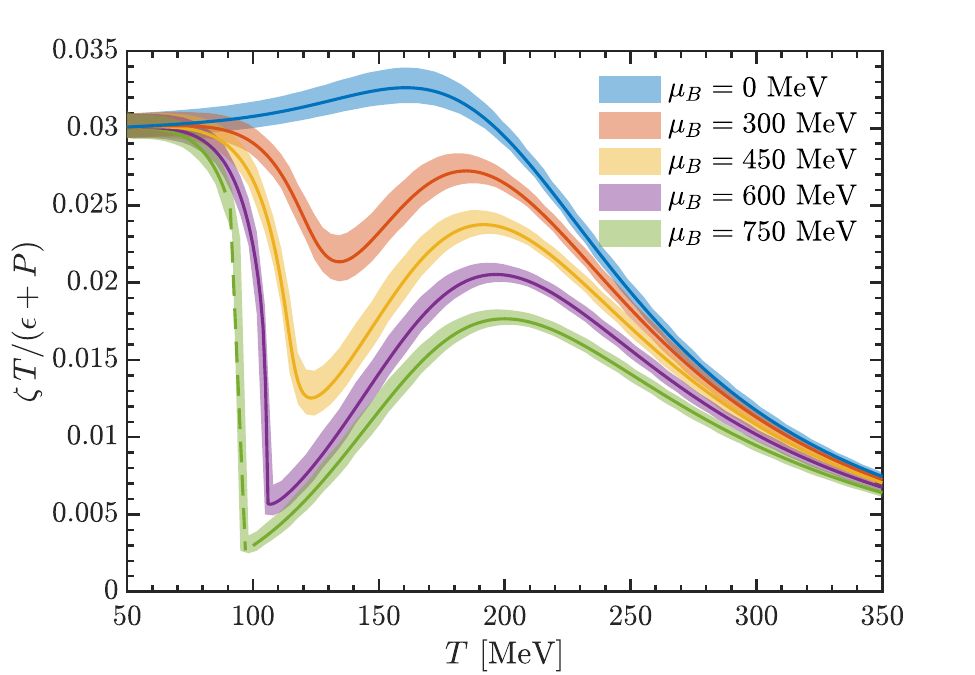}
        \caption{
        Posterior bands for the holographic bulk viscosity at 95\% confidence level and thecorresponding best-fit curves (solid lines) as a function of $T$ for different values of $\mu_B$.}
        \label{fig:PA_zeta}
    \end{minipage}
    \hfill
    % Right
    \begin{minipage}[t]{0.48\textwidth}
        \centering
        \includegraphics[width=\linewidth]{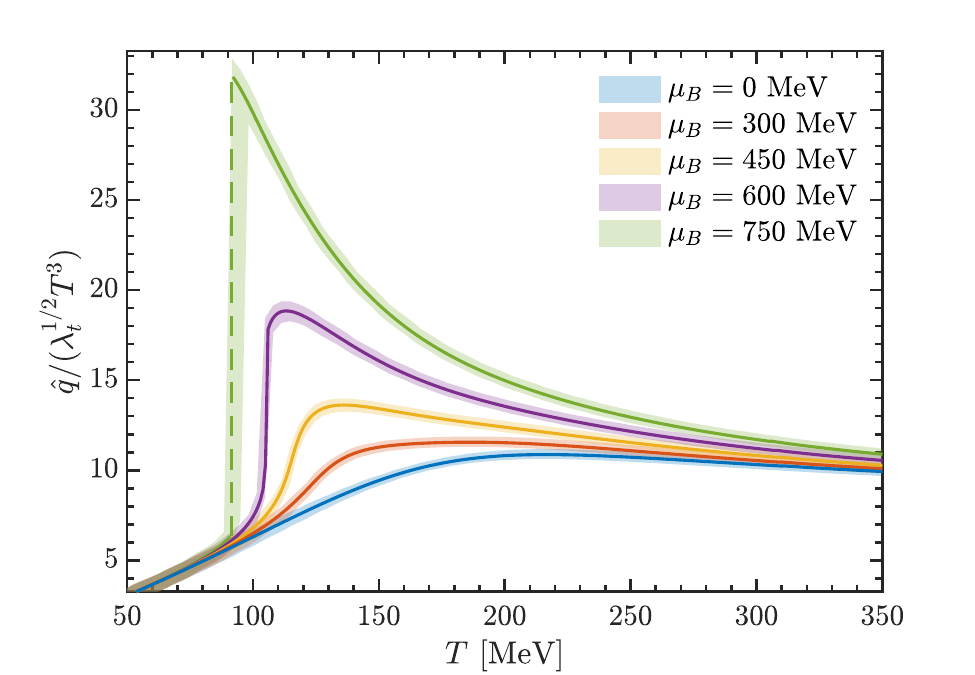}
    \caption{
Scaled jet-quenching parameter. Posterior bands at 95\% confidence level, and the corresponding best-fit curves (solid lines) of $\sigma_B/T$ as a function of $T$ for different values of $\mu_B$.}
    \label{fig:PA_qhat}
    \end{minipage}
\end{figure}

%\vspace{-0.5cm}
In the main section, we discussed the results obtained using the polyhyperbolic Ansatz (PHA) given in Eqs.~\eqref{PHA_fofphi} and \eqref{PHA_Vofphi}. Here, we present the corresponding transport coefficient and energy loss results computed using the parametric Ansatz (PA), Eqs.~\eqref{PA_fphi} and \eqref{PA_Vphi}. The baryon conductivity, baryon diffusion, and thermal conductivity, along with the shear viscosity, are shown in Fig.~\ref{fig:sigmaB_DB_sigmaT_eta}. Drag force and perpendicular Langevin diffusion coefficient are given in Fig.~\ref{fig:PA_DrageForce} and Fig.~\ref{fig:PA_kappa_perp}, respectively. Finally, the bulk viscosity is shown in Fig.~\ref{fig:PA_zeta}, while the jet-quenching parameter is presented in Fig.~\ref{fig:PA_qhat}.

%\newpage
\vspace{-1cm}

\vspace{-1cm}
\vspace{-0.9cm}

%%%%%%%%%%%%%%%%%%%%%%%%%%%%%%%%%
%%%%%%%%%%%%%%%%%%%%%%%%%%%%%%%%%
\bibliographystyle{unsrturl}
\bibliography{main.bib,noninspire.bib}

\end{document}